\documentclass[a4paper,fleqn,usenatbib]{mnras}
\usepackage[T1]{fontenc}
\usepackage{times}

\usepackage{graphicx}	% Including figure files
\usepackage{amsmath}	% Advanced maths commands
\usepackage{amssymb}	% Extra maths symbols
\graphicspath{{./}{Figures/}}

\newcommand{\ep}{e_\mathrm{p}}
\newcommand{\Mp}{M_\mathrm{p}}
\newcommand{\ap}{a_\mathrm{p}}

\newcommand{\Ap}{A_\mathrm{p}}
\newcommand{\Bp}{B_\mathrm{p}}
\newcommand{\md}{\mathrm{d}}
\newcommand{\asd}{\bar\Sigma}

\newcommand{\tsec}{t_\mathrm{sec}}

\newcommand{\ef}{e_\mathrm{f}}
\newcommand{\efree}{e_\mathrm{free}}
\newcommand{\ecmin}{e_\mathrm{min}}
\newcommand{\ffree}{f_\mathrm{free}}

\newcommand{\ecmax}{e_\mathrm{max}}

\def\ba{\begin{eqnarray}}
\def\ea{\end{eqnarray}}

\title[Debris Disc Substructures]{Debris Disc Substructures Induced by Secular Planetary Perturbations}

\author[R. R. Rafikov, D. Akansoy, \& A. A.  Sefilian]{
  Roman R. Rafikov$^{1,2}$, Deniz Akansoy$^{1}$ and Antranik A.  Sefilian$^{3}$\\
$^1$Department of Applied Mathematics and Theoretical Physics, University of Cambridge, Wilberforce Road, Cambridge CB3 0WA, UK\\
$^2$Institute for Advanced Study, Einstein Drive, Princeton, NJ 08540, USA
\\
$^{3}$Department of Astronomy and Steward Observatory, University of Arizona, 933 N Cherry Ave, Tucson, AZ 85721, USA
}

\date{Accepted XXX. Received YYY; in original form ZZZ}

\pubyear{2020}

\begin{document}
\label{firstpage}
\pagerange{\pageref{firstpage}--\pageref{lastpage}}
\maketitle

\begin{abstract}
Observations of debris discs have the potential to provide valuable information about massive planets perturbing them gravitationally. In this work, we explore the evolution of the azimuthally-averaged (or axisymmetric) surface density (ASD) --- a characteristic routinely derived from observations --- in a disc that is secularly perturbed by an inner planet. We develop a detailed analytical understanding of ASD evolution and verify it using a novel numerical framework {\sc debrispy}, which we make publicly available. With these tools we show that in a secularly evolving disc ASD develops a set of sharp features --- sharp peaks at eccentricity nulls and finite discontinuities at caustic points where particle periastra or apoastra pile up --- marching out through the disc as it ages. We probe the dependence of these features on planetary eccentricity, ratio of the free to forced particle eccentricity, and underlying radial mass distribution, showing in particular that more eccentric planets produce more prominent ASD features. Convolution with the PSF of realistic observations (as well as the non-zero random free eccentricity of debris) smooths out these features, but they can still be detectable in high-resolution observations. The radial locations of secular ASD peaks follow a particular, well-defined pattern, which should unambiguously point to their secular nature in observations. We illustrate how both detection and non-detection of such secular features in observed discs can be used (in combination with other constraints) to measure or constrain key parameters of perturbing planets (even those not yet detected) --- mass, semi-major axis and eccentricity. 
%Substructures in ASD secularly induced by planets in debris disks
\end{abstract}
%%%%%%%%%%%%%%%%%%%%%%%%%%%%%%%%%%%%%%%%%%%%%%%%%%


%%%%%%%%%%%%%%%%%%%%%%%%%%%%%%%%%%%%%%%%%%%%%%%%%%

%%%%%%%%%%%%%%%%% BODY OF PAPER %%%%%%%%%%%%%%%%%%

%%%%%%%%%%%%%%%%%%%%%%%%%%%%%%%%%%%%%%%%
%%%%%%%%%%%%%%%%%%%%%%%%%%%%%%%%%%%%%%%%
\section{Introduction}
\label{sec:Introduction}
%%%%%%%%%%%%%%%%%%%%%%%%%%%%%%%%%%%%%%%%
%%%%%%%%%%%%%%%%%%%%%%%%%%%%%%%%%%%%%%%%

Debris discs are a common by-product of planet formation and are observed around $\sim 20-30 \%$ of nearby main sequence stars spanning a wide range of ages and spectral types \citep[for reviews, see e.g.][]{Hughes2018,Wyatt2018}. 
Consisting of dust continually replenished through collisions among larger planetesimals, these systems provide a direct window into the reservoirs of planetesimals that survive long after the dispersal of protoplanetary gas \citep{Wyatt2008}. As a result, debris discs offer valuable insights into the architecture, evolution, and dynamical state of planetary systems. 
Their spatial structure is particularly informative, especially in the context of multi-wavelength observations, which probe different dust populations and thereby reveal the interplay between gravitational dynamics, radiation forces, and collisional evolution. Very importantly, in some cases debris discs can also reveal signs of gravitational coupling to more massive objects in the system --- planets. 

Gravitational effects of planets are often inferred via the non-axisymmetric or non-coplanar features exhibited by debris discs. For example, \citet{Wyatt1999} \& \citet{Kennedy2020}, among many others, discussed the possibility of driving eccentricity in debris discs by unseen eccentric planets, whereas \citet{Mou1997}, \citet{Au2001}, \citet{Dawson2011} explored the possibility of a warp in the debris disc of $\beta$ Pic as being driven by a planet on an inclined orbit \citep[see also][]{Sefilian2025}. This type of inference naturally requires high-resolution observations of debris discs, sensitive to often subtle non-axisymmetric features.   

Such observations have become possible in the recent decade thanks to {\it ALMA} and {\it JWST}. However, in many cases the signal-to-noise ratio (SNR) of these observations is still not high enough to make definitive claims about the spatial disc structure via the surface brightness distribution in the plane of the sky. In such cases one often resorts to constructing (upon an appropriate de-projection of observations) an azimuthally-averaged profile of the disc brightness, which can then be translated into the azimuthally-averaged (or axisymmetric) surface density \citep[ASD, as defined in][]{Rafikov2023} profile of emitting debris particles \citep[e.g.][]{Marino2021,Imaz2023,ARKS2}. This disc metric, while lacking information on the full two-dimensional disc structure, has an elevated SNR and can still be directly connected to the key dynamical characteristics of the disc.  

Indeed, \citet{Rafikov2023} has shown that given the knowledge of the mass in debris particles and their eccentricity distribution at every semi-major axis, one can (semi-)analytically compute the ASD without any approximations or uncertainties. This framework provides a natural and efficient way of forward modeling the ASD of observed systems and will be used in the present study. 

The goal of this work is to provide a description of the evolution of ASD in a debris disc subject to secular perturbations due to a planet. We seek to identify the features in the ASD that can be considered as telltale signs of a perturbing planet and used to constrain planetary properties --- mass and orbital characteristics. In the process, we develop an analytical understanding of the ASD behavior in discs of secularly evolving debris particles using the formalism of \citet{Rafikov2023} and its numerical implementation in the form of a novel software package {\sc debrispy}, which we make publicly available as part of this work. 

This work is organized as follows. After describing our physical setup and basic facts about secular evolution in Section \ref{sec:SecularFramework}, we proceed to describe our calculation of ASD in Section \ref{sec:ASD}, including our main numerical tool, {\sc debrispy}. In Section \ref{sec:ASD-behavior} we describe the main features exhibited by ASD in a secularly perturbed disc, and in Section \ref{sec:sec-evolve} we investigate the evolution of these features in time. In Section \ref{sec:par-var} we explore how ASD appearance and evolution depend on system parameters such as planetary eccentricity (Section \ref{sec:ep-dep}), amplitude of the free eccentricity component of debris particles (Section \ref{sec:free-ec}), and disc mass distribution (Section \ref{sec:complex-discs}). We then discuss the appearance of ASD structures in observations with finite resolution in Section \ref{sec:obs}, and in the presence of random free eccentricity component in Section \ref{sec:free-ec-rnd}. We discuss our results in Section \ref{sec:disc}, with a particular emphasis on their application to interpreting debris disc observations in Sections \ref{sec:apps} and \ref{sec:non-detect}. We summarize our key findings in Section \ref{sec:Summary}. Appendices contain various details of this work, including the technical description of {\sc debrispy} capabilities in Appendix \ref{sec:debrispy} and mathematical results on ASD features in Appendix \ref{sec:features_math}.

%%%%%%%%%%%%%%%%%%%%%%%%%%%%%%%%%%%%%%%%
%%%%%%%%%%%%%%%%%%%%%%%%%%%%%%%%%%%%%%%%%

\section{Basic Setup and General Secular Framework}
\label{sec:SecularFramework}

%%%%%%%%%%%%%%%%%%%%%%%%%%%%%%%%%%%%%%%%

We consider a debris disc composed of particles on bound Keplerian orbits around a star of mass $M_\star$. This disc is perturbed by a massive planet of mass $\Mp\ll M_\star$. The planet orbits the star on an eccentric orbit with a semi-major axis $\ap$, an eccentricity $\ep$ and an apsidal angle $\varpi_\mathrm{p}$. To be more specific, we assume an {\it inner} planet, 
i.e., one orbiting
%%%%%%%%%%%%%%%%%%%%%%%%%%%%i.e. that $\ap$ is 
{\it interior} to the inner edge of the debris disc. There are multiple natural generalizations of this setup, e.g. a planet (i) embedded within the disc or (ii) exterior to it, (iii) a stellar companion instead of a planet, or any combination thereof, but for simplicity we will not consider these situations here. 

We restrict ourselves to a planar geometry, so that the orbit of each debris particle comprising the disc can be characterized by its semi-major axis $a$, eccentricity $e$, and apsidal angle $\varpi$ (relative to a fixed reference direction). Very importantly, for the purposes of this work, specifically, the calculations made in \S\ref{sec:ASD}, we will assume that eccentricity $e$ is a unique function of $a$, i.e. $e=e(a)$. This assumption is appropriate for a disc secularly evolving under the planetary gravity, see Section \ref{sec:EccEvol}.

Next we outline the details of the long-term dynamical evolution of debris disc particles that follow from this general setup.

%%%%%%%%%%%%%%%%%%%%%%%%%%%%%%%%%%%%%%%%%

\subsection{Secular framework}
\label{sec:sec}

%%%%%%%%%%%%%%%%%%%%%%%%%%%%%%%%%%%%%%%%

We consider evolution of orbital elements of individual particles comprising the debris disc in a secular approximation, which assumes averaging over the short-period orbital motion of particles \citep{Murray1999}. Without loss of generality, we adopt the following simple model for the disturbing function (due to a planet) describing the secular evolution of a particle with the semi-major axis $a$, eccentricity $e$ and apsidal angle $\varpi$:
\ba  
{\cal R}=na^2\left[\frac{1}{2}A(a)e^2+B(a)e_p e\cos(\varpi-\varpi_p)\right],
\label{eq:R}
\ea  
where $n=\sqrt{GM_\star/a^3}$ is the particle mean motion and $A(a)$ and $B(a)$ are the functions of $a$ that characterize the axisymmetric and non-axisymmetric parts of the planetary perturbing potential to lowest order in $e$. In particular, if the inner planet is the only source of a perturbation in the system, then one has $A=\Ap$ and $B=\Bp$ with \citep{Murray1999}
\ba  
\Ap(a)=\frac{n}{4}\frac{M_p}{M_\star}\alpha b_{3/2}^{(1)}(\alpha),~~~~ \Bp(a)=-\frac{n}{4}\frac{M_p}{M_\star}\alpha b_{3/2}^{(2)}(\alpha),
\label{eq:ABp}
\ea  
where $\alpha=\ap/a<1$ and $b_s^{(m)}(\alpha)$ are the Laplace coefficients 
%%%%%%%%%%%%%%%%%%%%%%%%%%%%%%%%%%%%%%
\begin{equation}
b_{s}^{(m)}(\alpha) = \frac{2}{\pi} \int\limits_{0}^{\pi} \cos(m\theta) \bigg[1+\alpha^2-2\alpha\cos\theta   \bigg]^{-s} d\theta\, .
\label{eq:bsm}
\end{equation}
%%%%%%%%%%%%%%%%%%%%%%%%%%%%%%%%%%%%%%
Since ${\cal R}$ does not explicitly depend on mean anomaly, the particle semi-major axis $a$ stays fixed in the secular limit. 

For simplicity, we will also assume that the debris disc is far removed from the planet, i.e. that $\alpha=a_p/a \ll 1$ for all disc particles. In this limit the asymptotic behavior of the Laplace coefficients   $b_{3/2}^{(1)}(\alpha)\to 3\alpha$, $b_{3/2}^{(2)}(\alpha)\to (15/4)\alpha^2$ in the expressions (\ref{eq:ABp}) implies that 
\ba  
&& \Ap(a)\approx \frac{3}{4}n_p\frac{M_p}{M_\star}\left(\frac{a_p}{a}\right)^{7/2}\,,
\label{eq:Ap}
\\
&& \Bp(a)\approx -\frac{15}{16}n_p\frac{M_p}{M_\star}\left(\frac{a_p}{a}\right)^{9/2},
\label{eq:Bp}
\ea  
where $n_p = \sqrt{G M_{\star}/a_p^3}$ is the planetary mean motion. We will use equations (\ref{eq:Ap})-(\ref{eq:Bp}) to characterize the `inner planet' scenario in this work.

%%%%%%%%%%%%%%%%%%%%%%%%%%%%%%%%%%%%%%%%%

\subsection{Eccentricity evolution}
\label{sec:EccEvol}

%%%%%%%%%%%%%%%%%%%%%%%%%%%%%%%%%%%%%%%%

Introducing the eccentricity vector of a debris particle ${\bf e}=(k,h)\equiv(e\cos\varpi,e\sin\varpi)$, rewriting ${\cal R}$ in terms of $h$ and $k$, and using Lagrange's equations \citep{Murray1999}, one finds the usual general solution for the particle eccentricity behavior: 
\ba  
{\bf e}(t)={\bf e}_{\mathrm{free}}(t)+{\bf e}_{\mathrm{f}}     ,
\label{eq:e_sol_gen}
\ea
with the free eccentricity vector
\ba  
{\bf e}_{\mathrm{free}}(t)=\efree\left(\cos\left(At+\varpi_0\right),\,\sin\left(At+\varpi_0\right)\right),
\label{eq:e_free}
\ea  
where $\efree$ and $\varpi_0$ are constants set by the initial conditions, and the forced eccentricity vector given by  
\ba  
{\bf e}_{\mathrm{f}}=\ef\left(\cos\varpi_p,\sin\varpi_p\right),~~~~\ef=-\frac{\Bp}{\Ap}\ep\approx \frac{5}{4}\frac{\ap}{a}\ep,
\label{eq:e_forced}
\ea  
where the last approximation assumes $\ap\ll a$ and uses equations (\ref{eq:Ap}), (\ref{eq:Bp}). Without loss of generality, we will subsequently set $\varpi_p=0$ so that ${\bf e}_{\mathrm{f}}=(\ef,0)$.

According to equations (\ref{eq:e_sol_gen})--(\ref{eq:e_forced}), scalar particle eccentricity $e=|{\bf e}|$ is given by
\begin{align}  
e(t) %&=& 
&=\sqrt{\ef^2+\efree^2}
%\nonumber\\
%&\times & 
\left[1-\frac{2\ef\efree}{\ef^2+\efree^2}\cos\left(\Ap t+\varpi_0\right)\right]^{1/2}
\nonumber\\
&= \ef\sqrt{1+\ffree^2}
\left[1-\frac{2\ffree}{1+\ffree^2}\cos\left(\Ap t+\varpi_0\right)\right]^{1/2},
\label{eq:e_full}
\end{align}  
where we introduced a free-to-forced eccentricity ratio,
\begin{equation}
\ffree= \frac{ \efree } {\ef},
\label{eq:f_free}
\end{equation}
which we refer to simply as `free fraction' in the rest of the paper. One can see that $e$ varies between $\ecmin=|\ef-\efree|=\ef(1-\ffree)$ and $\ecmax=|\ef+\efree|=\ef(1+\ffree)$.

The amplitude of the free eccentricity is specified by the initial conditions and depends on the dynamical state of the debris disc at the onset of its secular evolution, as well as on the subsequent evolution of the system often driven by dissipative forces. The simplest but astrophysically very important assumption is that initially the debris disc was dynamically cold, i.e. that particle eccentricities were zero (what happens when a disc is born eccentric is briefly discussed in Section \ref{sec:free-ec}). In this case, adopting the initial condition ${\bf e}=0$ at $t=0$, we find that $\efree=\ef$ (i.e. $\ffree=1$), $\varpi_0=\pi$, and 
\begin{align}  
e(a,t) &= 2\ef(a)\left|
\sin \bigg( \frac{\Ap(a) t}{2} \bigg) \right|,
\label{eq:e_fr-fr=1}\\
\tan\varpi(a,t) &= \frac{\sin\left(\Ap t\right)}{\cos\left(\Ap t\right)-1}=\tan\left(\frac{\Ap(a) t}{2}-\frac{\pi}{2}\right),
\label{eq:varpi_fr-fr=1}
\end{align} 
such that $\varpi(a,t)$ lies in the interval $[-\pi/2,\pi/2]$. The eccentricity profile (\ref{eq:e_fr-fr=1}) will be used in most of our calculations, although we will also consider the situation with $\ffree\neq 1$ in Section \ref{sec:free-ec}. Profiles of $e(a,t)$ will be illustrated throughout the paper, in particular in Figure \ref{fig:overview}, where we also show the corresponding $\varpi(a,t)$ profile. 

%%%%%%%%%%%%%%%%%%%%%%%%%%%%%%%%%%%%%%%%
%%%%%%%%%%%%%%%%%%%%%%%%%%%%%%%%%%%%%%%%%

\section{Calculation of axisymmetric surface density (ASD)}
\label{sec:ASD}

%%%%%%%%%%%%%%%%%%%%%%%%%%%%%%%%%%%%%%%%

Debris discs are composed of numerous particles orbiting in a central Newtonian potential that are gradually ground down to smaller sizes through collisions. Given the slow pace of collisional evolution \citep[compared to the orbital frequencies, e.g.][]{Wyatt2008}, for our purposes we can assume particles to be collisionless and affected only by the gravity of massive objects in the system. This assumption should be justified at least for the more massive objects that provide the mass reservoir in the system \citep[e.g.,][]{wyattdent2002}. 

Observations of debris discs inform us (upon suitable de-projection) about a two-dimensional distribution of the surface density of debris particles $\Sigma(r,\phi)$ in polar $(r,\phi)$ coordinates in the disc plane. One would like to connect the observed $\Sigma(r,\phi)$ to the underlying distribution of orbital elements of debris particles, which sheds light on the dynamical history of the system. To characterize mass distribution in the disc, we define $\Sigma_a(a)$ such that the amount of mass $d\mu(a)$ in objects with semi-major axes in the interval $(a,a+da)$ is
\ba  
d\mu(a)=2\pi a\Sigma_a(a)\md a.
\label{eq:sigma_a}
\ea  
In problems involving secular interactions $\Sigma_a$ is an important quantity, since the semi-major axis of a particle subject to secular gravitational effects is conserved. As a result, both the mass distribution $\mu(a)$ and $\Sigma_a(a)$ stay unchanged in the course of secular evolution. For a disk composed of objects on circular orbits one naturally has  $r = a$ and $\Sigma_a(a)=\Sigma(a)$.

We can also define {\it Axisymmetric (or Azimuthally-averaged) Surface Density} $\asd$ (ASD) as the zeroth-order azimuthal Fourier component of $\Sigma(r,\phi)$,  i.e. its azimuthal average: 
\ba  
\asd(r)=\frac{1}{2 \pi}\int_0^{2\pi}\Sigma(r,\phi)\md\phi.
\label{eq:asd}
\ea  
As alluded to earlier, this is a commonly used disc characteristic that has a  higher SNR in observations than $\Sigma$ itself. In this work, we will use ASD as the key metric that can be directly connected to the dynamical state of the debris disc. The way in which this is done is described next.

%%%%%%%%%%%%%%%%%%%%%%%%%%%%%%%%%%%%%%%%%

\subsection{Analytical framework}
\label{sec:ASD-theory}

%%%%%%%%%%%%%%%%%%%%%%%%%%%%%%%%%%%%%%%%

\citet{Rafikov2023} has demonstrated that in a disc with particle eccentricities having a unique value $e(a)$ at every $a$, the relation between the ASD and $\Sigma_a$ has the form
\ba  
\asd(r) = \pi^{-1}\int_{r/2}^\infty \frac{a^{-1}\Sigma_a(a)~\theta(r,a)}{\sqrt{e^2(a)-\kappa^2(r,a)}}  \md a,
\label{eq:asd-rel-gen}
\ea  
where we introduced a simple but very important function  
\ba
\kappa(r,a)=\left|1-\frac{r}{a}\right|.
\label{eq:kap}
\ea
Also, $\theta(r,a)=\Theta\left(e(a)-\kappa(r,a)\right)$, where $\Theta(z)$ is a Heaviside step-function. The significance of the function $\theta$ can be appreciated by noticing that, for a given $r$, particles at semi-major axes $a$ satisfying the condition $e(a)=\kappa(r,a)$ have their periastra $r_p(a)=a[1-e(a)]$ or apoastra $r_a(a)=a[1+e(a)]$ equal to $r$. Thus, this condition delineates particles that can or cannot reach a specific radius $r$ in the course of their orbital motion: $\theta=1$ for all $a$ such that $r_p(a)\le r\le r_a(a)$, while $\theta=0$ if this condition is violated. Introduction of the indicator function $\theta(r,a)$ forces one to count in the integral (\ref{eq:asd-rel-gen}) the contribution to the ASD only from particles with orbits crossing the radius $r$. It also  constrains the expression inside the root in (\ref{eq:asd-rel-gen}) to be positive.

Equations (\ref{eq:asd-rel-gen})-(\ref{eq:kap}) are derived without any simplifying assumptions (as long as debris particles move on Keplerian orbits) and allow us to directly relate the behavior of $e(a)$ and $\Sigma_a(a)$ to the radial profile of ASD $\asd(r)$. Note that $\asd(r)$ is independent of the particle apsidal angles $\varpi$ since the orientations of particle orbits do not change ASD. This considerably simplifies our task.

%%%%%%%%%%%%%%%%%%%%%%%%%%%%%%%%%%%%%%%%%

\subsection{Numerical implementation: {\sc debrispy}}
\label{sec:ASD-num}

%%%%%%%%%%%%%%%%%%%%%%%%%%%%%%%%%%%%%%%%

To efficiently compute ASD using equations (\ref{eq:asd-rel-gen}), (\ref{eq:kap}), we developed a dedicated software package 
{\sc debrispy}, which is made publicly available\footnote{Available on GitHub: \href{https://github.com/DenizAkansoy/DebrisPy}{github.com/DenizAkansoy/DebrisPy}.}. Given the profiles of $\Sigma_a(a)$ and $e(a)$, this package allows a high-accuracy, point-by-point evaluation of $\asd(r)$ on an adaptive grid, focusing resolution at the locations where $\asd$ exhibits sharp gradients. This capability is particularly important since, as we will see next, ASD of a secularly evolving disc typically features numerous sharp features. Our method is more flexible and computationally efficient than the traditional Monte Carlo sampling techniques, although the latter is also implemented in {\sc debrispy} for verification and calculation of the full 2D maps of particle distribution. A more detailed description of {\sc debrispy}'s capabilities can be found in Appendix \ref{sec:debrispy}. 

%%%%%%%%%%%%%%%%%%%%%%%%%%%%%%%%%%%%%%%%%
%%%%%%%%%%%%%%%%%%%%%%%%%%%%%%%%%%%%%%%%%

\section{Typical structure of ASD in a secularly-evolving disc}
\label{sec:ASD-behavior}

%%%%%%%%%%%%%%%%%%%%%%%%%%%%%%%%%%%%%%%%%

We can now illustrate the behavior of ASD in a disc that evolves under the secular perturbations from an inner planet. According to equation (\ref{eq:asd-rel-gen}), to do this we need to provide two inputs, namely the radial profiles of $\Sigma_a(a)$ and $e(a)$ of debris particles as functions of semi-major axis $a$.

\begin{figure}
	\begin{center}
	\includegraphics[width=0.49\textwidth]{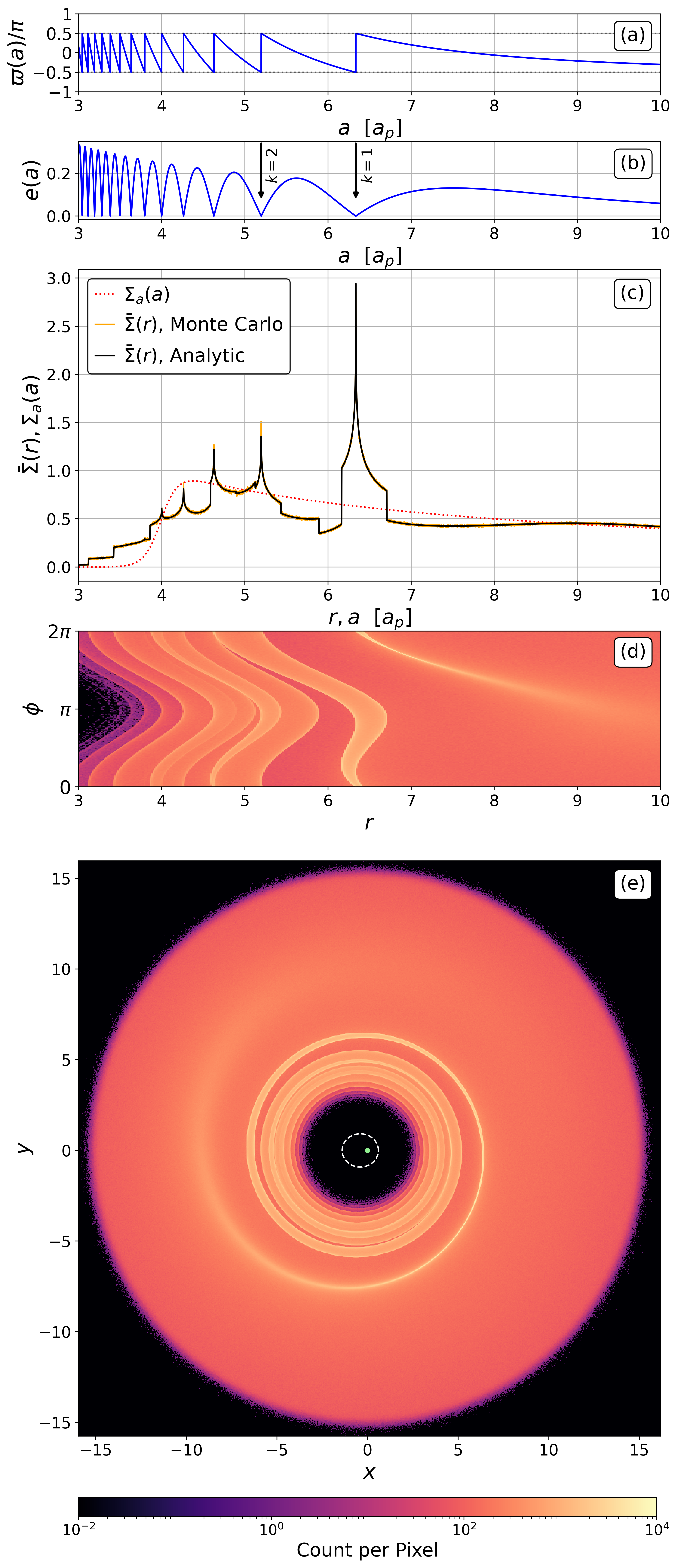}
	\caption{
    Snapshot of various characteristics of a debris disc secularly perturbed by an inner planet. This calculation assumes $\ep=0.4$, $\ffree=1$ Equation \ref{eq:f_free}), and $t=5\tsec$ (as defined in Equation \ref{eq:sec-t}). Different panels show: (a) apsidal angle of debris particles $\varpi(a)$ as a function of their semi-major axis $a$, (b) particle eccentricity $e(a)$ as a function of $a$, (c) radial profile of the ASD $\asd(r)$ and of the underlying $\Sigma_a(a)$, (d) a Cartesian $r-\phi$ map of the two-dimensional particle density $\Sigma(r,\phi)$ obtained by Monte Carlo sampling, and (e) $\Sigma(r,\phi)$ in polar coordinates (dashed white ellipse is the planetary orbit). In panel (c) we show the ASD computed via the Monte Carlo sampling (orange) and analytical framework (black) of \citet{Rafikov2023}, showing their excellent agreement. 
}
	\label{fig:overview}
	\end{center}
\end{figure}

For $\Sigma_a(a)$ we adopt a (smoothly) truncated power-law profile with  inner and outer semi-major axes $a_\mathrm{in}=4\ap$ and  $a_\mathrm{out}=15\ap$ near which $\Sigma_a$ smoothly tapers off to zero. In the bulk of the disc, for $a_\mathrm{in}\lesssim a\lesssim a_\mathrm{out}$, $\Sigma_a$ follows a power-law profile, $\Sigma_a(a)\propto a^{-1}$. An explicit mathematical form of this $\Sigma_a(a)$ profile is provided in Appendix \ref{sec:siga}, see equations (\ref{eq:siga})-(\ref{eq:fout}); this profile is also shown in Figure \ref{fig:overview}(c) and in other figures. The smooth behavior of $\Sigma_a$ in the bulk of the disc allows us to better illustrate the sharp features of the ASD arising due to the secularly-driven $e(a)$ profile. We explore other forms of $\Sigma_a(a)$ in Section \ref{sec:complex-discs}. 

For the eccentricity profile, we adopt $e(a)$ in the form (\ref{eq:e_fr-fr=1}) resulting from secular evolution of disc particles with $\ffree=1$. The assumption of $\efree=\ef$ is implicit through most of the paper, except Section \ref{sec:free-ec} where we explicitly relax it. This $e(a)$ profile features multiple nulls (or zeros) of eccentricity, i.e. the locations where $e(a)=0$, which are very important for ASD behavior, as we show further. 

To judge the progress of secular disc evolution, we measure time $t$ since disc formation (and start of its evolution) in units of the secular time $t_\mathrm{sec}$ at the inner disc edge $a=a_\mathrm{in}$. More explicitly, we set 
\begin{align}  
t_\mathrm{sec}=\frac{2\pi}{\Ap(a_\mathrm{in})} & \approx 5.7\,\mathrm{Myr}\, \left(\frac{\ap}{10\,\mathrm{AU}}\right)^{-2}\left(\frac{a_\mathrm{in}}
{40\,\mathrm{AU}}\right)^{7/2}
\nonumber\\
&\times\left(\frac{\Mp}{M_\mathrm{J}}\right)^{-1}\left(\frac{M_\star}{M_\odot}\right)^{1/2},
\label{eq:sec-t}
\end{align}
where the expression for $\Ap(a)$ is given by equation (\ref{eq:Ap}). Expressed through $\tsec$, the eccentricity profile (\ref{eq:e_fr-fr=1}) becomes
\begin{align}  
e(a,t) = 2\ef(a)\left|
\sin\left[\pi\frac{t}{\tsec}
\left(\frac{a_\mathrm{in}}{a}\right)^{7/2}\right]\right|.
\label{eq:e_fr-fr=2}
\end{align}
Specifying time in units of $\tsec$ frees us from the need to explicitly specify planetary mass $\Mp$ in what follows. Note that the local secular timescale $\tsec(a)$ at a given semi-major axis $a$ is given by equation (\ref{eq:sec-t}) with $a_\mathrm{in}$ replaced by $a$.

%%%%%%%%%%%%%%%%%%%%%%%%%%%%%%%%%%%%%%%%%%%%%%%%%%%%%%%%%%%
%%%%%%%%%%%%%%%%%%%%%%%%%%%%%%%%%%%%%%%%%%%%%%%%%%%%%%%%%%%

%%%%%%%%%%%%%%%%%%%%%%%%%%%%%%%%%%%%%%%%%

\subsection{Example of ASD profile in a secularly-driven disc}
\label{sec:ASD-patterns}

%%%%%%%%%%%%%%%%%%%%%%%%%%%%%%%%%%%%%%%%%

Before focusing solely on the ASD behavior for the rest of the paper, we first provide an illustration of various general debris disc characteristics that ultimately determine $\asd$ profile. Figure \ref{fig:overview} presents a snapshot of ASD and other disc characteristics at $t=5t_\mathrm{sec}$ when secular effects had sufficient time to develop in the inner part of the disc. All these characteristics were computed with {\sc debrispy}. This calculation assumes planetary eccentricity $\ep=0.4$, `free fraction' $\ffree=1$ (i.e. eccentricity profile in the form (\ref{eq:e_fr-fr=1}) or (\ref{eq:e_fr-fr=2})), and we express $r$ and $a$ in units of planetary semi-major axis $\ap$.

Panel (b) shows $e(a)$ profile at that time, with multiple eccentricity nulls being obvious. Their radial density --- number per unit interval of $a$ --- rapidly increases in the inner disc, as expected. Indeed, according to equation (\ref{eq:e_fr-fr=1}) the $e$ nulls occur when
\ba
\frac{\Ap(a_k)t}{2}=k\pi,~~~k=1,2,\dots \,,
\label{eq:null_cond}
\ea
where the integer $k$ numbers eccentricity nulls, with $k=1$ corresponding to the outermost null (in Figure \ref{fig:overview}b we mark $k=1$ and $k=2$ nulls with arrows). Using the expression (\ref{eq:Ap}) for $\Ap$ one then determines the semi-major axis $a_k$ of $k$-th null to be
\ba
a_k=\ap\left(\frac{3}{8\pi}\frac{M_p}{M_\star}\frac{n_\mathrm{p}t}{k}\right)^{2/7}=a_\mathrm{in}\left(\frac{t}{\tsec}k^{-1}\right)^{2/7}.
\label{eq:a_k}
\ea
Expressed in physical units, 
\begin{align}
a_k  \approx 91\,\mathrm{AU}\, &
\left(\frac{\Mp}{M_\mathrm{J}}\frac{t}{100\,\mathrm{Myr}}k^{-1}
\right)^{2/7}
\nonumber\\
&\times\left(\frac{\ap}{10\,\mathrm{AU}}\right)^{4/7}
\left(\frac{M_\star}{M_\odot}\right)^{-1/7}.
\label{eq:a_k_units}
\end{align}
Clearly, the higher-$k$ nulls are located at a progressively smaller $a_k$, and the distance between the adjacent nulls decreases as $k$ goes up. As $t$ increases, a null with a given $k$ shifts to larger radii. These properties are universal for secularly-driven $e(a)$ profiles. 

Although the apsidal angle of disc particles plays no role in setting ASD, we still show $\varpi(a)$ in panel (a) of Figure \ref{fig:overview}. Knowledge of $\varpi(a)$ in addition to $e(a)$ allows us to understand the full 2D distribution of the debris particle density $\Sigma(r,\phi)$, which is shown in the bottom of the Figure in two projections: Cartesian and polar, in panels (d) and (e) correspondingly. These maps of $\Sigma(r,\phi)$ are computed via Monte Carlo sampling of particle orbital elements, which is included in {\sc debrispy}, see Appendix \ref{sec:debrispy}. 

One can see that the rapid variation of $\varpi$ in the inner parts of the disc leads to a tightly wrapped structure in that region; see panels (a) and (e). By contrast, outside the outer $k=1$ null the apsidal angles vary much slower with $r$, resulting in a one-armed spiral forming outside the outer null at $a=a_1$. This spiral is essentially a one-armed kinematic density wave forming in a Keplerian potential \citep{Kalnajs1973,Wyatt2005,Sefilian2021}  Also, the inner edge of the disc exhibits a characteristic non-axisymmetric shape in panel (d), sitting noticeably closer to the star at $\varpi\approx 0$ than at $\varpi\approx \pi$. This is because under secular perturbations the most eccentric particles have $\varpi\approx \varpi_\mathrm{p}$ (which we have set to zero) as can be seen in panels (a),(b) and also deduced from equations (\ref{eq:e_fr-fr=1})-(\ref{eq:varpi_fr-fr=1}). Periastra of these particles are also clustered around $\phi\approx 0$, and this is the direction where their distance to the star is smallest. Their apoastra cluster near $\phi\approx \pi$ and this is where these particles are furthest from the star, explaining the non-circular shape of the inner cavity. This $\phi\approx 0/\pi$ alignment tendency can be traced also in the bulk of the disc, at $a\lesssim a_1$, see panel (d).  

Finally, panel (c) of Figure \ref{fig:overview} illustrates the radial profile of ASD $\asd(r)$. It is computed in two ways: via the analytical approach based on equation (\ref{eq:asd-rel-gen}, black curve) and via the Monte Carlo sampling of particle orbits (orange curve). The Monte Carlo calculation is performed over the full radial domain of the sampled particle distribution, using $N_p=10^8$ particles, and $N_r=1.5\times10^4$ radial bins. The comparison with the analytical solution is restricted to the interval $r\in [3, 10]$ as shown in Figure \ref{fig:overview}, which contains 7314 bins. One can see that the two curves match perfectly, but the one based on Monte Carlo sampling exhibits random dispersion around the analytically-computed curve, which is caused by the inherent Poisson noise of the Monte Carlo method. Over the comparison interval, the root-mean square fractional deviation of the Monte Carlo profile from the analytical curve is 2.88\%. For reference, the dotted red curve shows the underlying $\Sigma_a(a)$ profile given by equations (\ref{eq:siga})-(\ref{eq:fout}).  

One can see that while $\Sigma_a(a)$ is smooth and well-behaved, the ASD curve exhibits a number of localized, sharp features: tall and narrow peaks, for example at $r\approx (4.7, 5.2, 6.3)\ap$, etc., and discontinuous but finite jumps or steps, for example at $r\approx (3.9, 5.9, 6.2, 6.7)\ap$, etc. We investigate the origin of these features next. 

%%%%%%%%%%%%%%%%%%%%%%%%%%%%%%%%%%%%%%%%%%%%%%%%%%%%%%%%%%%
\begin{figure} 
	\begin{center}
	\includegraphics[width=0.49\textwidth]{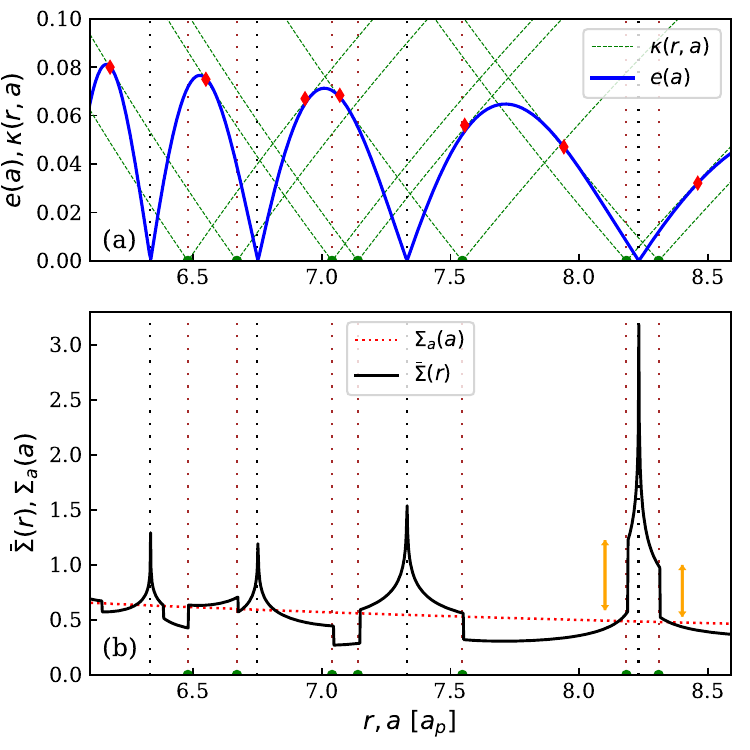}
	\caption{
    Illustration of the origin of ASD features --- sharp peaks and discontinuous jumps. Shown over a particular radial range are (a) the eccentricity profile $e(a)$ and (b) the ASD profile $\asd(r)$. The calculation is done at $t=25\tsec$ for $\ep=0.2$, $\ffree=1$ and $\Sigma_a(a)$ given by equations (\ref{eq:siga})-(\ref{eq:fout}), also shown as red dotted curve in panel (b). Vertical dotted brown and black lines mark the locations of ASD discontinuities and peaks, respectively. The outer peak at $r=8.23\,\ap$ corresponds to the second ($k=2$) eccentricity null, with inner peaks emerging at progressively higher $k$. Thin dashed green curves in panel (a) show $\kappa(r_t,a)$ for $r_t$ corresponding to ASD discontinuities, marked with green points. Comparing the upper and lower panels, one can see that sharp peaks stand at eccentricity nulls, where $e(a)=0$, while the locations $r_t$ of discontinuous jumps are such that $\kappa(r_t,a)$ and $e(a)$ curves are tangent to each other at points $a_t$, marked with red diamonds. Orange arrows in panel (b) illustrate the theory-based amplitude of $\asd$ discontinuities adjacent to them, computed using equation (\ref{eq:dSig2}). 
    }
	\label{fig:features_ill}
	\end{center}
\end{figure}
%%%%%%%%%%%%%%%%%%%%%%%%%%%%%%%%%%%%%%%%%%%%%%%%%%%%%%%%%%%

%%%%%%%%%%%%%%%%%%%%%%%%%%%%%%%%%%%%%%%%%

\subsection{Origin of ASD features}
\label{sec:ASD-features}

%%%%%%%%%%%%%%%%%%%%%%%%%%%%%%%%%%%%%%%%%

We now provide a brief phenomenological description of the nature of singular features exhibited by ASD in Figure \ref{fig:overview}(c). A  more detailed, mathematical characterization of their origin and properties is provided in Appendix \ref{sec:features_math}, and we will refer to it when necessary. Our presentation will be aided by Figure \ref{fig:features_ill}, in which we zoom in on a smaller radial interval than in Figure \ref{fig:overview} and focus on $e(a)$ and $\asd(r)$ behavior. This figure is also made for system parameters  different from Figure \ref{fig:overview}, namely for $\ep=0.2$ and at $t=25\tsec$. The fact that very similar features --- peaks and discontinuities --- appear in both figures asserts the universality of these features in secularly-evolving discs. 

%%%%%%%%%%%%%%%%%%%%%%%%%%%%%%%%%%%%%%%%%

\subsubsection{ASD features: peaks}
\label{sec:ASD-features-peaks}

The first thing one notices by examining Figures \ref{fig:overview}b,c and \ref{fig:features_ill} is that the sharp, narrow peaks of ASD always appear at radii $r$ where $e(r)=0$, i.e. at the nulls of eccentricity. This correspondence is present at every null of $e(a)$. We explain this behavior mathematically in Appendix \ref{sec:peaks}, where we show that these peaks represent a true but weak singularity of ASD, as $\asd$ diverges logarithmically at these locations, see Figure \ref{fig:log}a. As logarithmic divergence is integrable, there is still always a finite number of particles in any region around the peak. 

Physically, the increase in particle density around the nulls can be understood as follows. Near the $k$-th null, particles with $a\approx a_k$ have $e\approx 0$ and move on essentially circular orbits thus fully contributing to ASD at $r\approx a_k$. In addition, there are also eccentric particles with semi-major axes somewhat further from $a_k$ that may reach $r=a_k$ as well and additionally contribute to ASD there. This combined effect of the `local' and `non-local' particles is what drives the ASD increase near $r=a_k$. On the other hand, at $r$ far from the nulls the effect of local particles is strongly diluted as they now contribute to ASD over a finite radial interval due to their epicyclic motion. As a result, $\asd$ can go down there relative to $\Sigma_a$.

There is one important detail in this picture: the non-local particles with semi-major axes different from $a_k$ must be able to reach $r=a_k$ to enhance the density there resulting in a singular peak. For that to be the case, $e(a)$ near the null should grow faster with $|a-a_k|$ than the function $\kappa(r,a)$ at $r=a_k$, i.e. the condition 
\ba
\left|\frac{\partial e(a)}{\partial a}\Big|_{a\to a_k}\right|>\left|\frac{\partial\kappa(a_k,a)}{\partial a}\Big|_{a\to a_k}\right|
\label{eq:peak_cond}
\ea
must be fulfilled. This is indeed the case in Figure \ref{fig:features_ill}, where in panel (a) the dashed green lines show $\kappa(r,a)$ for different $r$: one can see that the slope of $\kappa(r,a)$ is steeper than that of $e(a)$ near the nulls everywhere in the shown region. As a result, peaks appear at every null in this figure. 

Introducing a characteristic parameter $\zeta_k$ as 
\ba
\zeta_k=\left|\frac{\partial e(a)/\partial a}{\partial\kappa(r,a)/\partial a}\right|_{r,a\to a_k},
\label{eq:zt_k}
\ea
we can re-cast the condition (\ref{eq:peak_cond}) for a sharp peak to exist at $k$-th $e$-null as $\zeta_k>1$, see Appendix \ref{sec:zg1}. In Appendix \ref{sec:peaks} we show that for a secularly-evolving debris disc with eccentricity profile (\ref{eq:e_fr-fr=1})
\ba
\zeta_k \approx \frac{35\pi k}{4}\ep \frac{\ap}{a_k},
\label{eq:zeta1}
\ea 
where we used equation (\ref{eq:a_k}). This expression will be used later to understand ASD evolution. 

In the opposite case, when $\zeta_k<1$, the non-local particles do not reach $r=a_k$ and cannot strongly enhance the ASD there. In this case the singularity at $r=a_k$ does not form  and instead only a moderate or weak enhancement of $\asd$ is found at this radius, see Appendix \ref{sec:zl1}. This situation does not appear in Figures \ref{fig:overview} \& \ref{fig:features_ill}, but will be found in many subsequent figures.

%%%%%%%%%%%%%%%%%%%%%%%%%%%%%%%%%%%%%%%%%%%%%%%%%%%%%%%%%%%
\begin{figure*}
	\begin{center}
	\includegraphics[width=0.99\textwidth]{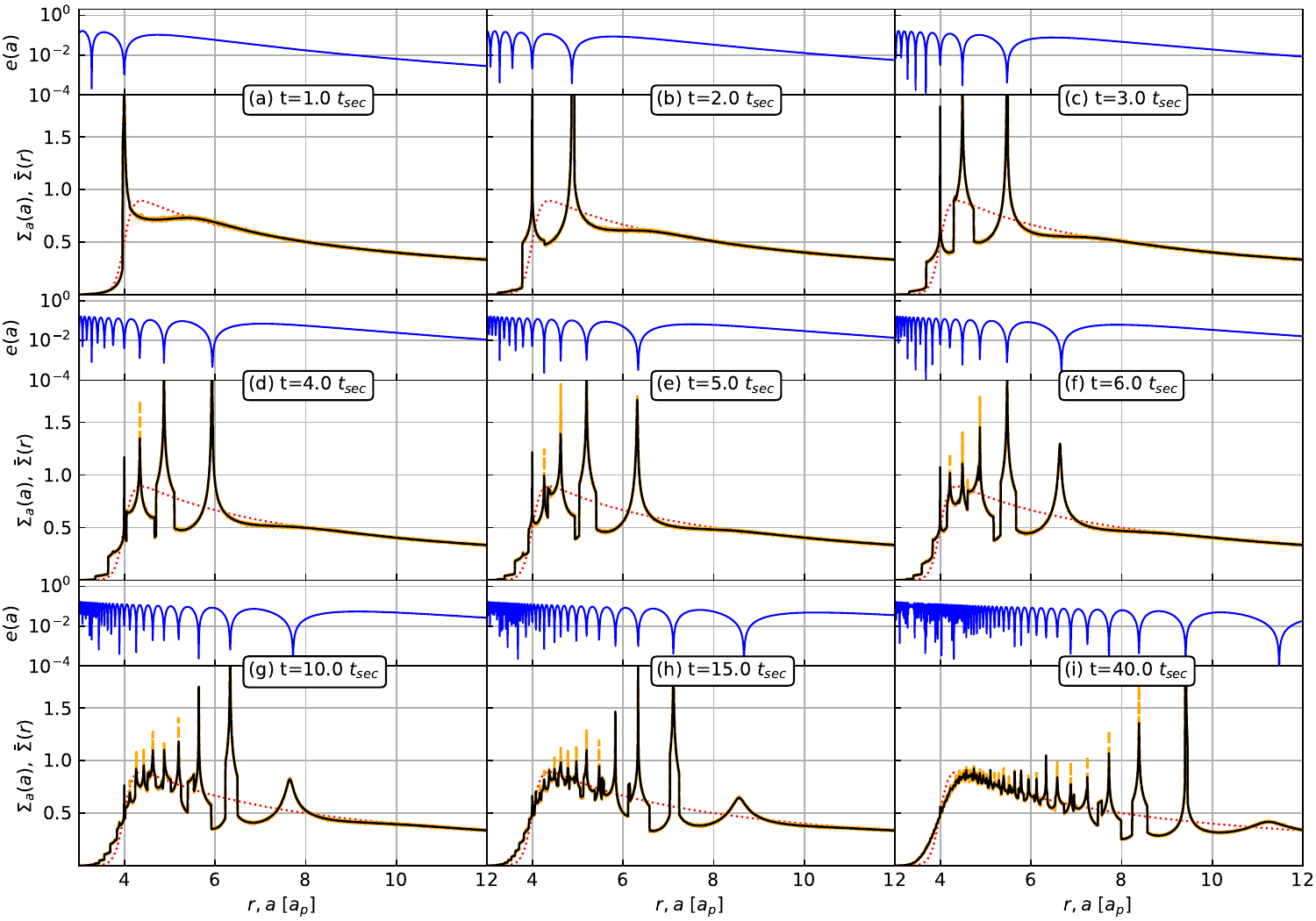}
	\caption{
    ASD evolution secularly driven by a planet with eccentricity $\ep=0.2$, with $\efree=\ef$. Solid black curve is the analytical ASD $\asd(r)$ calculation, dashed orange is the Monte Carlo sampling, dotted red is the underlying $\Sigma_a(a)$ profile. Both $r$ and $a$ are expressed in units of $\ap$. Solid blue curves in subpanels above show the corresponding $e(a)$ at every moment of time (indicated for both subpanels). ASD snapshots are shown at times $t$ expressed in units of $t_\mathrm{sec}$.}
	\label{fig:time_ev}
	\end{center}
\end{figure*}
%%%%%%%%%%%%%%%%%%%%%%%%%%%%%%%%%%%%%%%%%%%%%%%%%%%%%%%%%%%

%%%%%%%%%%%%%%%%%%%%%%%%%%%%%%%%%%%%%%%%%

\subsubsection{ASD features: jumps}
\label{sec:ASD-features-jumps}

Next we address the origin of discontinuous but finite jumps (or steps) of ASD clearly visible in Figures \ref{fig:overview}c and \ref{fig:features_ill}. Sometimes, but not always, such jumps accompany ASD peaks by forming a 'pedestal' on which the peak stands, see e.g. the ASD jumps around two outermost peaks in Figure \ref{fig:features_ill}. 

Careful inspection shows that ASD jumps emerge at radii $r=r_t$ such that the functions $\kappa(r_t,a)$ and $e(a)$ happen to be {\it tangent} to each other at some $a=a_t$. At this semi-major axis one thus has 
\ba
\kappa(r_t,a_t)=e(a_t)~~~~~\mbox{and} ~~~~~\frac{\partial\kappa(r_t,a_t)}{\partial a}=\frac{\partial e(a_t)}{\partial a}.
\label{eq:jump_cond}
\ea
We illustrate this statement in Figure \ref{fig:features_ill}, where the dashed green curves in panel (a) show $\kappa(r,a)$ for several values of $r$ corresponding to the locations of the ASD jumps in panel (b). One can see that either the inner ($a<r$) or the outer ($a>r$) branches of these $\kappa(r,a)$ curves indeed have a point (shown with red diamonds) tangent to the $e(a)$ curve at some $a$. For example, $\asd$ peak at $7.33\ap$ (at $k=3$ eccentricity null) is bounded by two conspicuous jumps at $r_t=7.14\ap$ and $r_t=7.55\ap$. The former has $\kappa(r_t,a)$ tangent to $e(a)$ at $a_t\approx 7.56\ap$, while the latter has $\kappa(r_t,a)$ tangent to $e(a)$ at $a_t\approx 7.07\ap$. In this and other examples shown in Figure \ref{fig:features_ill}, one also notices that if the tangent point lies on the inner (outer) branch of  $\kappa(r,a)$, i.e. $a_t<r_t$ ($a_t>r_t$), the corresponding ASD jump is negative (positive) as $r$ increases. 

The first condition (\ref{eq:jump_cond}) implies that periastra (for $a_t>r_t$) or apoastra (for $a_t<r_t$) of particles at this $a_t$ end up at $r_t$, so that such particles spend a long time at this radius. The second condition implies that this condition is nearly fulfilled over an extended range of semi-major axes, implying that over a small but finite range of $a$ near $a_t$ particles have their turning points (apo- or periastra) coinciding at $r_t$, strongly modifying ASD there. Thus, features of ASD arising at these radii are essentially caustics. 

Conditions (\ref{eq:jump_cond}) can be easily manipulated to show that the semi-major axis $a_t$ of the tangent point is given by a solution of the equation 
\ba
\frac{\partial}{\partial a}\left(a e(a)\right)=\pm 1,~~~~~r\lessgtr a,
\label{eq:tan_cond}
\ea
where the upper/lower signs correspond to particle periastra/apoastra ending up at $r$. Once such $a=a_t$ is determined by solving (\ref{eq:tan_cond}), the radius $r_t$ at which the caustic feature of ASD is located is given by $r_t=a_t(1\mp e(a_t))$. Some useful results on $a_t,r_t$ are provided in Appendix \ref{sec:peaks_pedestals}, see equations (\ref{eq:step3})-(\ref{eq:gt}).

In Appendix \ref{sec:caustics} we show that the nature of an ASD feature appearing at a tangent radius $r_t$ is determined by the value of the dimensionless parameter $\chi$ defined as
\ba    
\chi=\frac{e^{\prime\prime}(a_t)-\kappa^{\prime\prime}(r_t,a_t)}{2}a_t^2,
\label{eq:chi}
\ea   
where primes imply differentiation w.r.t. semi-major axis $a$. A discontinuous but finite jump of $\asd$ appears at $r_t$ if $\chi<0$ at this point, see Appendix \ref{sec:causticsl0}; a situation with $\chi>0$ will be discussed in Section \ref{sec:free-ec}. In most cases the value of $\chi$ is set primarily by 
$e^{\prime\prime}(a_t)$, with $\kappa^{\prime\prime}(r_t,a_t)$ providing only a small contribution. Since a typical secular eccentricity profile with $\ffree=1$ depicted in Figures \ref{fig:overview}b \& \ref{fig:features_ill}a has $e^{\prime\prime}(a)<0$ everywhere, the condition $\chi<0$ needed for a finite jump to appear is fulfilled automatically at all tangent radii $r_t$. This explains the ubiquity of ASD jumps in Figures \ref{fig:overview}c \& \ref{fig:features_ill}b.

The amplitude of $\asd$ jump at a tangent radius $r_t$ is calculated in Appendix \ref{sec:causticsl0} and is given by equations (\ref{eq:dSig2}), (\ref{eq:dSig3}). The results of this calculation are shown as orange arrows in Figure \ref{fig:features_ill}b near the $\asd$ jumps at $r_t=8.18\ap$ and $8.31\ap$ (around the $k=2$ peak at $r\approx 8.23\ap$) and clearly provide a good estimate of the jump amplitude. Equation (\ref{eq:dSig3}) also specifies the direction of ASD jump at $r_t$, depending on the relative position of $a_t$ and $r_t$.

%%%%%%%%%%%%%%%%%%%%%%%%%%%%%%%%%%%%%%%%%
%%%%%%%%%%%%%%%%%%%%%%%%%%%%%%%%%%%%%%%%%

\section{Time evolution of ASD in a secularly-evolving disc}
\label{sec:sec-evolve}

%%%%%%%%%%%%%%%%%%%%%%%%%%%%%%%%%%%%%%%%%

Having understood the origin of the morphologies of the key features exhibited by ASD profiles, we are now in a position to examine the evolution of ASD as the disc evolves under secular perturbations due to a planet. As a practical example, we consider a system in which an inner planet has eccentricity  $\ep=0.2$ (lower than in Figure \ref{fig:overview}) and an external disc with surface density $\Sigma_a(a)$ in the form (\ref{eq:siga})-(\ref{eq:fout}). We also keep the free eccentricity of debris particles equal to the forced one throughout the whole evolution, i.e. $\ffree=1$ at all times. In Figure \ref{fig:time_ev} we show the snapshots of the secularly-driven $e(a)$ of the disc particles (top panels) and of the ASD profiles at nine different moments of time (in units of $\tsec$) shown in panels. The times shown range from $1\,\tsec$, when the secular evolution is just becoming important at the inner disc edge $a_\mathrm{in}$, to $40\,\tsec$ when the secular evolution has affected the disc out to $r=12\,\ap$. As in Figure \ref{fig:overview}c, we show not only the analytical $\asd(r)$ (black) but also the Monte Carlo sampled one (orange), to illustrate their perfect agreement aside from the shot noise associated with Monte Carlo sampling; dotted red curve shows the underlying $\Sigma_a(a)$.

Starting with eccentricity profiles, one can see a typical $e(a)$ evolution with radially-oscillatory $e(a)$ behavior reaching out further in the disc as $t$ increases. The number of eccentricity nulls within the shown radial range $r\in(3,12)\ap$ steadily increases, with the outermost one ($k=1$) marching out to larger $r$ (located at $\approx 4\,\ap$ at $t=1\,\tsec$ and $\approx 11.5\,\ap$ at $t=40\,\tsec$), in agreement with equation (\ref{eq:a_k}). At late times, $t\gg\tsec$, at a fixed radius $r$ in the inner disc the order of a nearby $e$-null increases as $k\propto t$, while the radial separation between two adjacent nulls goes down as $|a_{k+1}-a_k|\approx (2/7)r/k\propto t^{-1}$, see equation (\ref{eq:a_k}). In other words, the number of $e$-nulls per unit radial interval in the inner disc grows linearly with time. 

The ASD profiles shown in lower panels respond directly to the underlying evolution of $e(a)$. One can easily see conspicuous ASD peaks standing at most $e$-nulls during the full evolution span. At $t=1\,\tsec$ there is just one sharp peak of ASD located at $k=1$ null at $r=a_1\approx 4\,\ap$. At $t=2\,\tsec$ and $3\,\tsec$ there are two and three peaks, respectively, of somewhat different appearance. The $k=2$ peak is usually wider (as judged from the separation of the bounding ASD jumps) than the $k=1$ one, which can be understood by examining Figure \ref{fig:features_ill} and noticing that, quite generally, the ASD jumps having their tangent points $a_t$ closest to a given peak get more widely spaced the higher is the order $k$ of that peak (also see Appendix \ref{sec:peaks_pedestals}). And already at $k$ of several the ASD jumps corresponding to one peak may start affecting the ASD structure around the neighboring peaks\footnote{For example, in Figure \ref{fig:features_ill} the outer ASD jump of $k=5$ peak (at $r=a_5\approx 6.33\,\ap$) falls at $r=6.67\,\ap$, right next to $k=4$ peak (at $r=a_4\approx 6.75\,\ap$), and between its `own' jumps at $6.48\,\ap$ and $7.04\,\ap$).}. This greatly complicates the pattern of ASD around high-$k$ peaks, with the superposition of numerous ASD jumps and sharp peaks at $e$-nulls making ASD profile in the inner disc look rather chaotic as time increases (this is further discussed in Section \ref{sec:ep-dep} and Appendix \ref{sec:pedestal_overlap}). 

Focusing a bit more deeply on the vicinity of the first ($k=1$) $e$-null, we see that at $t=1\,\tsec$ this peak is sharp and narrow (bounded by two ASD jumps of quite different amplitude on each side), which agrees with the fact that $\zeta_1\approx 1.4>1$ at $r=a_1\approx 4\,\ap$ at this time. As $t$ increases, $k=1$ peak moves out, while decreasing in amplitude. By $t=3\,\tsec$ it reaches $r\approx 5.5\,\ap$ and the value of $\zeta_1$ crosses unity. Beyond that point in time, the ASD peak at $a_1$ is no longer discontinuous (although this may initially be difficult to see given the vertical scale in Figure \ref{fig:time_ev}) and its height steadily decreases in time. By $t=40\,\tsec$ this peak evolves into a mere ASD bump above $\Sigma_a(a)$ near $r=a_1\approx 11.5\,\ap$, and $\zeta_1\approx 0.5$ at this moment. Overall, the time evolution of ASD in the vicinity of $k=1$ $e$-null judged through the variation of $\zeta_1$ follows the expectations laid out in Section  \ref{sec:ASD-features-peaks} and Appendix \ref{sec:peaks}. Similar pattern can be noticed also for the ASD peak at $k=2$ $e$-null, which gradually narrows as $t$ increases, until it crosses $\zeta_2=1$ divide soon after $t=40\,\tsec$ and becomes non-singular.  

On a more quantitative level, one can use equations (\ref{eq:a_k}) and (\ref{eq:zeta1}) to express $k$ and the corresponding $a_k$ through a given value of $\zeta_k$:
\begin{align}
k(\zeta_k) &=\left(\frac{4}{35\pi}\zeta_k\frac{a_\mathrm{in}}{\ap}\ep^{-1}\right)^{7/9}\left(\frac{t}{\tsec}\right)^{2/9}\,,
\label{eq:k_viazeta}\\
%%%%%%
a_k(\zeta_k) &=a_\mathrm{in}\left(\frac{35\pi}{4}\zeta_k^{-1}\frac{\ap}{a_\mathrm{in}}\ep\frac{t}{\tsec}\right)^{2/9}\,.
\label{eq:ak_viazeta}
\end{align}
If we set $\zeta_k=1$ corresponding to the ASD sharp peak-to-bump transition, these equations predict the order $k$ of the peak and its location $a_k$ at which the peak-to-bump transition occurs at a given time $t$ and planetary eccentricity $\ep$. For example, equation (\ref{eq:ak_viazeta}) predicts that in Figure \ref{fig:time_ev}, at $t=3\,\tsec$, this transition occurs at $\approx 5.5\,\ap$, in agreement with what we found earlier.

%%%%%%%%%%%%%%%%%%%%%%%%%%%%%%%%%%%%%%%%%%%%%%%%%%%%%%%%%%%
\begin{figure*}
	\begin{center}
	\includegraphics[width=0.99\textwidth]{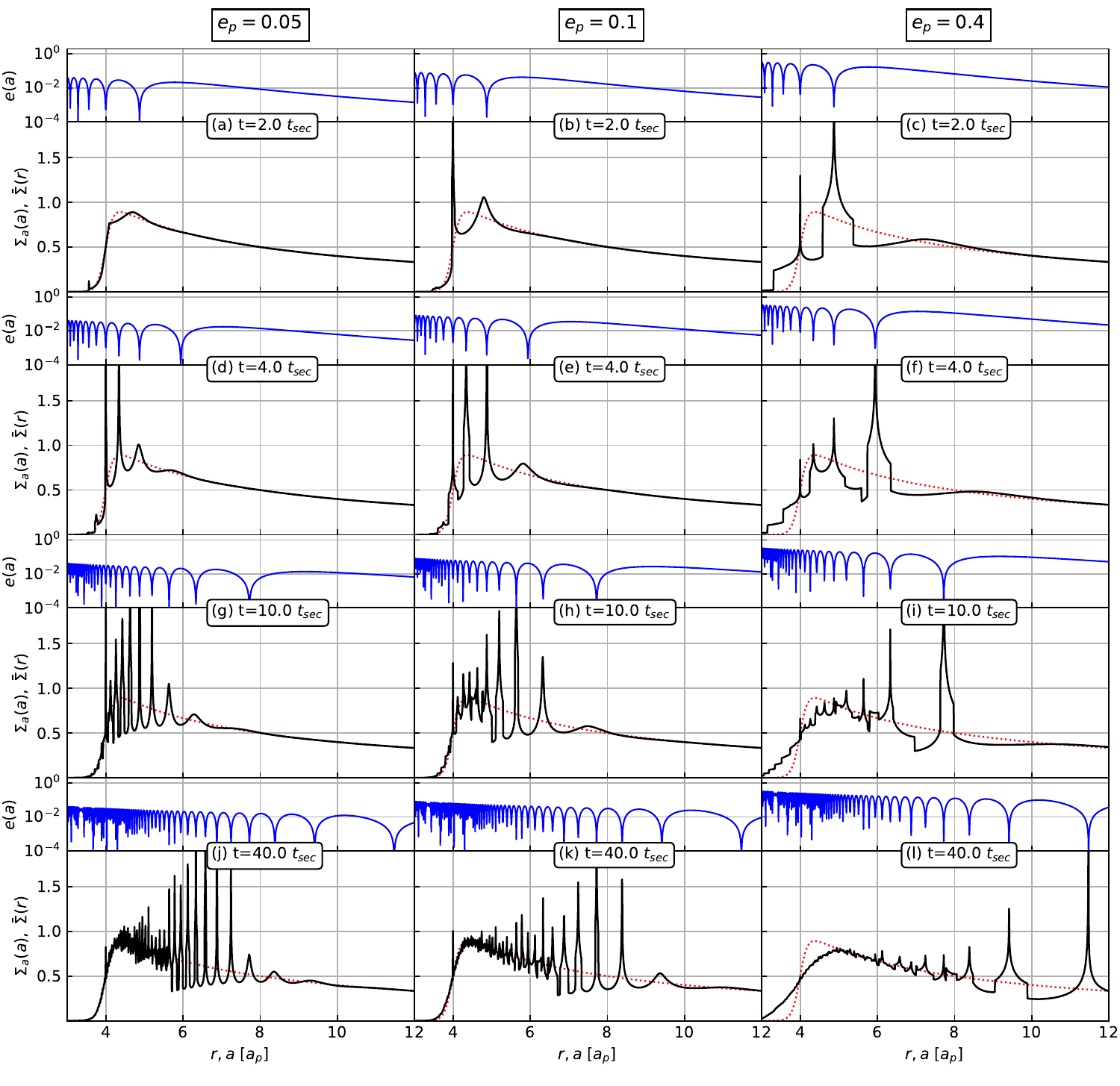}
	\caption{
    Similar to Figure \ref{fig:time_ev}, now illustrating the effect of varying $\ep$. Left, middle and right columns correspond to $\ep=0.05, 0.1$ and $0.4$, respectively. Snapshots of ASD are shown for 4 different moments of time (same for each $\ep$), indicated in every row. The outermost null of $e(a)$ at $t=40\,\tsec$ corresponds to $k=1$. See the text (Section \ref{sec:ep-dep}) for more details.  }
	\label{fig:ep_var}
	\end{center}
\end{figure*}
%%%%%%%%%%%%%%%%%%%%%%%%%%%%%%%%%%%%%%%%%%%%%%%%%%%%%%%%%%%

%%%%%%%%%%%%%%%%%%%%%%%%%%%%%%%%%%%%%%%%%

\section{Parameter variation and ASD evolution}
\label{sec:par-var}

%%%%%%%%%%%%%%%%%%%%%%%%%%%%%%%%%%%%%%%%%

Having studied a particular example in the previous section, we now examine how the variation of various parameters of the problem affects ASD evolution. In Section \ref{sec:ep-dep} we explore the effects of varying eccentricity of the planetary orbit $\ep$, in Section \ref{sec:free-ec} we look at the changes of ASD morphology brought about by the variation of free eccentricity fraction $\ffree$, and in Section \ref{sec:complex-discs} we consider variation of $\Sigma_a(a)$.

%%%%%%%%%%%%%%%%%%%%%%%%%%%%%%%%%%%%%%%%%

\subsection{Effects of varying planetary eccentricity}
\label{sec:ep-dep}

%%%%%%%%%%%%%%%%%%%%%%%%%%%%%%%%%%%%%%%%%

In Figure \ref{fig:ep_var} we show ASD snapshots at four different times $t=(2,4,10,40)\tsec$ (fixed in each row) for a debris disc-planet system in which we vary only $\ep=0.05,0.1,0.4$ (fixed in each column), everything else being the same (i.e. $\Sigma_a(a)$, $\tsec$, $\ap$, $\ffree=1$). We can also compare with the case of $\ep=0.2$, since the same moments of time are also shown in Figure \ref{fig:time_ev}b,d,g,i. Note that what really matters for the following comparison is not $\ep$ itself, but $\ep\ap$ since this is what sets $\ef$ and the normalization of $e(a)$ at every $a$, see equation (\ref{eq:e_forced}). But since we keep $\ap=1$ fixed throughout this work, this is equivalent to just varying $\ep$.

Because only $\ep$ changes and $\tsec$ is independent of it, the only change in $e(a)$ profile between the different $\ep$ cases at a given moment of time is the overall normalization of the $e(a)$ profile. In particular, one can easily see in Figure \ref{fig:ep_var} that the nulls of $e(a)$ at a given $t$ are the same regardless of $\ep$. For that reason, the ASD features associated with $e$-nulls appear at the same locations regardless of $\ep$, but their appearance differs substantially. 

%%%%%%%%%%%%%%%%%%%%%%%%%%%%%%%%%%%%%%%%%%%%%%%%%%%%%%%%%%%
\begin{figure*}
	\begin{center}
	\includegraphics[width=0.99\textwidth]{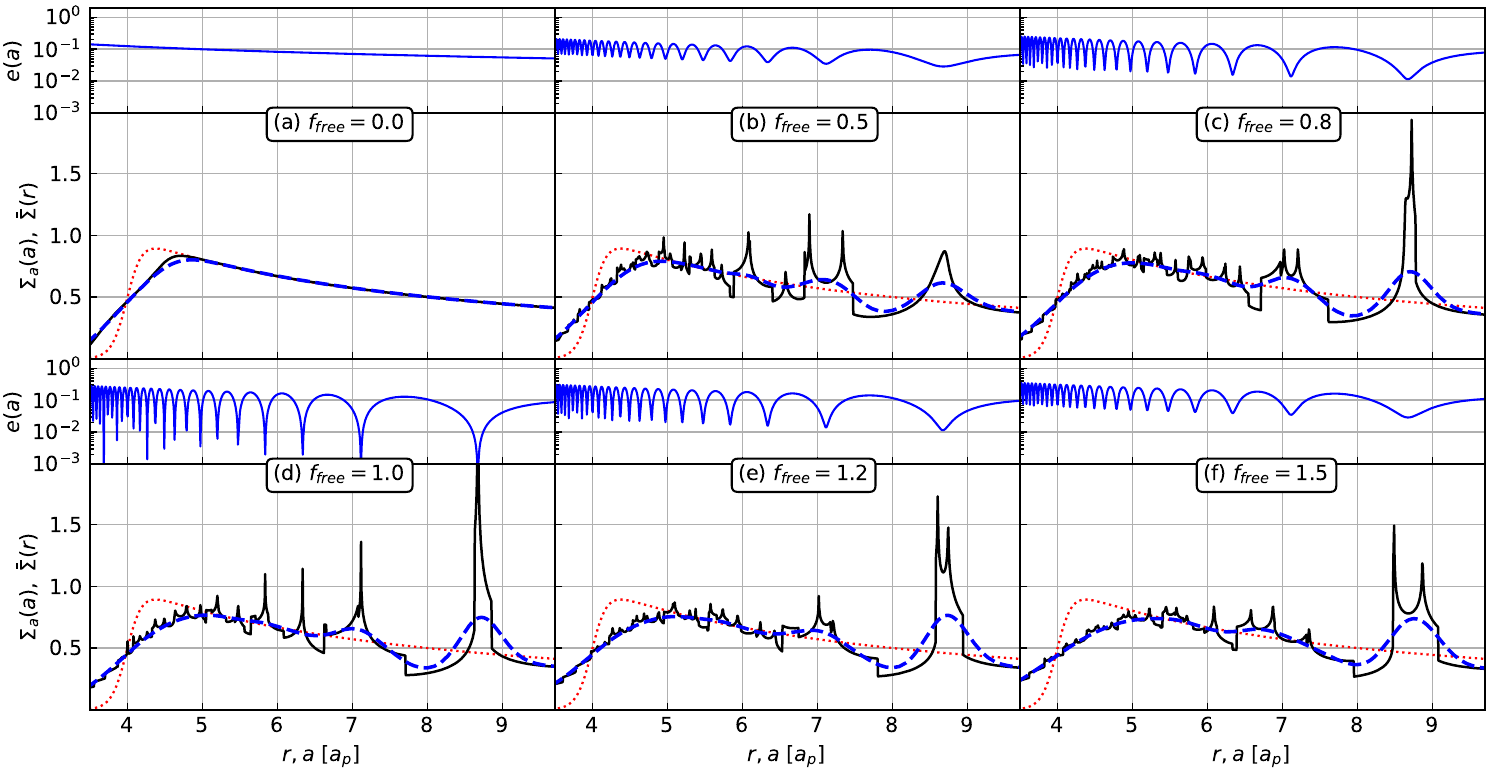}
	\caption{
    Similar to Figure \ref{fig:time_ev}, now illustrating the effect of varying $f_\mathrm{free}=\efree/\ef$. Snapshots of ASD for $\ep=0.4$ at $t=15t_\mathrm{sec}$ are shown with black solid curves for 6 different values of $f_\mathrm{free}$, indicated in each panel. Dashed blue curves show Gaussian convolution of ASD with $\sigma=0.25\ap$ to illustrate the effects of finite observational resolution. See text for discussion.}
	\label{fig:ffree_var}
	\end{center}
\end{figure*}
%%%%%%%%%%%%%%%%%%%%%%%%%%%%%%%%%%%%%%%%%%%%%%%%%%%%%%%%%%%

For example, at $t=2\,\tsec$ ASD calculation for $\ep=0.05$ shows no sharp peaks and only rather weak bumps at the first and second $e$-nulls (at $r=a_1\approx 4.9\,\ap$ and $r=a_2\approx 4\,\ap$, respectively). The $\ep=0.1$ calculation has a more pronounced bump at $a_1$ and a narrow, sharp peak at $a_2$, while $\ep=0.2$ features two sharp, narrow peaks at both locations, see Figure \ref{fig:time_ev}b. And the $\ep=0.4$ case really stands out by its massive, radially extended peak at $a_1$ (in addition to a narrower peak at $a_2$), rising high on a pedestal formed by two ASD jumps (at $r\approx 4.58\,\ap$ and $5.38\,\ap$). The effective width of this singular feature at $r=a_1$ is $0.8\,\ap$, if measured as the distance between the bounding ASD jumps; note that equation (\ref{eq:Delta_k}) in Appendix \ref{sec:peaks_pedestals} provides an analytical expression for the pedestal width. 

This difference in appearance of features at $k=1,2$ $e$-nulls at $t=2\,\tsec$ can again be easily traced to the phenomenology outlined in Section  \ref{sec:ASD-features-peaks} and Appendix \ref{sec:peaks} and the fact that $\zeta_k\propto \ep$ at fixed $k$ and $a_k$, see equation (\ref{eq:zeta1}). Given that at $k=1$ null $\zeta_1\approx 0.28,0.56,1.12,2.24$ for $\ep=0.05,0.1,0.2,0.4$, respectively, it is natural that ASD shows only a mild bump at $r=a_1$ for $\ep=0.05,0.1$ since $\zeta_1<1$, but a true singularity for $\ep=0.2,0.4$ (since $\zeta_1>1$). Similarly, at $k=2$ null of eccentricity $\zeta_2=0.69,1.37,2.75,5.5$ for the same set of $\ep$ values, and only the $\ep=0.05$ case features not a singular peak but a bump at $r=a_2$.  

Time evolution of the peak appearance can also be understood based on the variation of $\zeta_k$, just as in Figure \ref{fig:time_ev}. As $t$ increases, a peak corresponding to a particular $e$-null order $k$ gradually evolves towards a bump as its $a_k$ increases and $\zeta_k$ eventually crosses unity, see equation (\ref{eq:zeta1}). This process is particularly pronounced at low $\ep$: one can see that for $\ep=0.05$ even $k=3$ and $4$ features starting as singular peaks at $t=4\,\tsec$ evolve into bumps by $t=40\,\tsec$. These trends of peak-to-bump transition with $\ep$ and $t$ can be understood quantitatively using equations (\ref{eq:k_viazeta}), (\ref{eq:ak_viazeta}) with $\zeta_k=1$.

Another observation that one can make, in particular at $t=10\,\tsec$ and $40\,\tsec$, is that the apparent height of singular peaks at the same null order $k$ is larger for smaller $\ep$. This can be understood based on equation (\ref{eq:asd4}) characterizing asymptotic behavior of $\asd$ near the peak: one can see that the amplitude of the logarithmic term scales as $(\zeta_k^2-1)^{-1/2}$, and for the same $k$ and $t$ (i.e. same $a_k$) $\zeta_k$ is lower for smaller $\ep$.

Finally, if one focuses on the lower envelope of $\asd$ profile at $t=40\,\tsec$, one can see an interesting transition in its behavior around $5.7\ap$ for $\ep=0.05$ and around $6.5\ap$ for $\ep=0.1$: outside of this radius the minima of $\asd$ are quite regular and get deeper for higher $k$ (smaller $r$), while inside this radius the lower envelope of ASD becomes more chaotic and tends to follow $\Sigma_a$ profile. This transition can also be traced for $\ep=0.2,0.4$, as well as at earlier times, e.g. $t=10\,\tsec$.  

The explanation for this behavior lies in the phenomenon of a `pedestal overlap' --- overlap of the pedestals of neighboring ASD peaks, explored in more detail in Appendix \ref{sec:pedestal_overlap}. This overlap starts at high-$k$ peaks in discs that have evolved for a significant amount of time, so that the distance between the neighboring peaks (i.e. $e$-nulls) becomes smaller than the width of their pedestals. This overlap results in partial cancellation of nearby ASD jumps, leading to a chaotic appearance of an ASD profile, see Figure \ref{fig:features_ill}b. Together with a reduction of peak amplitudes at high $k$ (leading to higher $\zeta_k$ and lower peak amplitude, see equation (\ref{eq:asd4})), this eventually drives the ASD towards $\Sigma_a$. 

We show in Appendix \ref{sec:pedestal_overlap} that the overlap starts at a particular value of $\zeta_k=\zeta_\mathrm{po}\approx 2.95$, see equation (\ref{eq:zeta_po}). Plugging this value of $\zeta_k$ into equation (\ref{eq:ak_viazeta}) and setting $t=40\,\tsec$, we expect the overlap to start around $5.6\,\ap$ for $\ep=0.05$ and around $6.5\,\ap$ for $\ep=0.1$. These estimates are in excellent agreement with the radii at which the transition in the behavior of the lower envelope of ASD occurs in Figure \ref{fig:ep_var}j,k.

%%%%%%%%%%%%%%%%%%%%%%%%%%%%%%%%%%%%%%%%%

\subsection{Discs with $\efree$ different from $\ef$}
\label{sec:free-ec}

%%%%%%%%%%%%%%%%%%%%%%%%%%%%%%%%%%%%%%%%

So far, we explored ASD behavior only for eccentricity profile (\ref{eq:e_fr-fr=1}), which assumes the free eccentricity fraction $\ffree=1$ (Eq. \ref{eq:f_free}), typical for discs which start as dynamically cold. However, debris discs may also develop eccentricity profile with $\ffree$ different from unity, even if initially they had $\efree=\ef$. For example, it is possible that on timescales long compared to $\tsec$ the free eccentricity gets damped by some dissipative processes, e.g. gas drag caused by the leftover gaseous component \citep{Hughes2018}. In this case, one would end up with $\ffree<1$. If this process continues operating for a long time, it may completely damp $\efree$ resulting in $\ffree=0$ \citep[as assumed in e.g.][]{Lynch2022}, so that $e(a)=\ef(a)$ with no radial oscillations of $e(a)$ profile. 

There are also situations in which, starting with $\ffree=1$, the forced eccentricity of the disc could subsequently go down, making $\ffree>1$. According to equation (\ref{eq:e_forced}), this can be a result of an inward migration of the perturbing planet (lowering $\ap$), e.g. caused by its interaction with an inner planetesimal belt or other planets in the system \citep{Malhotra1993}. Alternatively, planetary eccentricity $\ep$ can go down, e.g. due to the phenomenon of resonant friction caused by its gravitational coupling to the debris disc \citep{Tremaine1998,Sefilian2023}, again lowering $\ef$ and raising $\ffree$.   

Moreover, it has been suggested \citep{Kennedy2020} that some debris discs may be born eccentric rather than circular as we assumed in Section \ref{sec:EccEvol}. In this case, $\ffree\neq 1$ is generally unavoidable and, depending on the magnitudes and orientation of the initial eccentricity vector ${\bf e}(t=0)$ and the forced eccentricity vector ${\bf e}_\mathrm{f}$, both $\ffree>1$ and $\ffree<1$ are possible.

To account for these possibilities, we now explore the effect of $\ffree$ on ASD profile. In Figure \ref{fig:ffree_var} we show $\asd(r)$ computed for $\ep=0.4$ at $t=15\,\tsec$ using the general $e(a)$ profile (\ref{eq:e_full}) with different values of $\ffree$. Panel (d) of that figure depicts a familiar case of $\ffree=1$ with sharp peaks at eccentricity nulls and ASD jumps forming pedestals. But once $\ffree$ starts deviating from unity even by a small amount, as in panels (c) and (e), $e(a)$ no longer reaches zero at  $a_k$; instead, $e(a)$ features minima at the locations very close to $a_k$ as follows from analyzing equation\footnote{The dependence of $\ef$ on $a$ leads to only a slight displacement of these minima from $a_k$.} (\ref{eq:e_full}). The ASD profile now exhibits not one (as in the $\ffree=1$ case) but two peaks around $a_k$, still with a pedestal around each pair of peaks. Radial separation between these two peaks generally gets wider the more $\ffree$ deviates from unity (see e.g. panel (f)), as long as the singularity exists. The latter is not always guaranteed, and one can see that a singular peak that is present around $8.7\ap$ for $\ffree=1$ turns into a bump for $\ffree=0.5$, see Figure \ref{fig:ffree_var}b,d. This and the narrowing of peak pedestals as $\ffree$ decreases below unity are caused by the dependence of the minimum and maximum eccentricities $e_\mathrm{min}$ and $e_\mathrm{max}$ on $\ffree$, which affects the locations and existence of the caustics as $\ffree$ is varied. Overall, it appears that the ASD peaks located at $e$-nulls in $\ffree=1$ case get split into two peaks once $\ffree$ starts deviating from unity.

%%%%%%%%%%%%%%%%%%%%%%%%%%%%%%%%%%%%%%%%%%%%%%%%%%%%%%%%%%%
\begin{figure} 
	\begin{center}
	\includegraphics[width=0.49\textwidth]{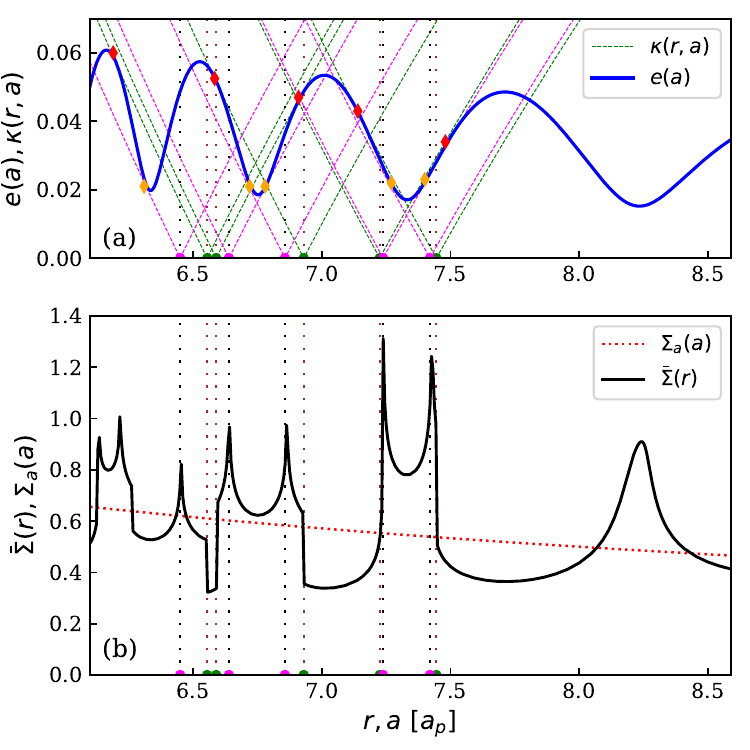}
	\caption{
    Similar to Figure \ref{fig:features_ill} but now illustrating the origin of peaks for $\ffree\neq 1$. All parameters of the calculation are the same as in Figure \ref{fig:features_ill}, except now $\ffree=0.5$, which results in a very different $\asd(a)$ in panel (b). The blue curve in panel (a) is $e(a)$ and it no longer reaches zero. Vertical dotted black lines again show the locations of the peaks in panel (b), with dashed magenta curves in panel (a) showing $\kappa(r,a)$ for these radii $r$. One can see that these $\kappa$ curves are tangent to $e(a)$ at points (shown with orange diamonds) where $e^{\prime\prime}>0$. As a result, $\chi>0$ at these locations (see equation (\ref{eq:chi})) leading to the emergence of sharp peaks at these caustic points, as discussed in Appendix \ref{sec:causticsg0}. Other curves are related to discontinuities at caustic points with $\chi<0$ (see Appendix \ref{sec:causticsl0}) and have the same meaning as in Figure \ref{fig:features_ill}. 
    } 
	\label{fig:features_ill1}
	\end{center}
\end{figure}
%%%%%%%%%%%%%%%%%%%%%%%%%%%%%%%%%%%%%%%%%%%%%%%%%%%%%%%%%%%

To understand the origin of this peak splitting, in Figure \ref{fig:features_ill1} we show ASD calculation identical to that in Figure \ref{fig:features_ill} except now we assume $\ffree=0.5$. Panel (a) shows a smooth $e(a)$, without a discontinuous derivative at $e$-nulls (which are no longer present). And in panel (b) one can see that sharp double peaks appear at radii $r$ such that $\kappa(r,a)$ is tangent to $e(a)$ at points shown as orange diamonds. There are also other tangent points, marked by red diamonds, which correspond to finite $\asd$ discontinuities. The key difference between the two sets of tangent points is the sign of the second derivative $e^{\prime\prime}(a)$: it is negative for the red tangent points corresponding to finite ASD jumps (just as in Figure \ref{fig:features_ill}) but positive for the orange tangent points corresponding to double peaks. 

This picture is fully in line with the phenomenology of tangent points or caustics described in Appendix \ref{sec:caustics}. There we show that if the value of the parameter $\chi$ defined by equation (\ref{eq:chi}) at the tangent point $a_t$ is positive, then a weak (logarithmic) singularity appears at the corresponding radius $r_t$, see Appendix \ref{sec:causticsg0}. This situation is opposite to the case of $\chi<0$ studied earlier, in which ASD experiences a finite jump, see Section \ref{sec:ASD-features-jumps} and Appendix \ref{sec:causticsg0}. A situation with $\chi>0$ is impossible for $e(a)$ profile with $\ffree=1$, but it appears naturally in the vicinity of the former $e$-nulls $a_k$ once $\ffree$ differs from unity, even by a small amount. To summarize, peak splitting arises because $\ffree\neq 1$ makes $\chi>0$ situation possible at some tangent points, leading to a singularity at the corresponding caustic.

In panel (a) of Figure \ref{fig:ffree_var}  we consider an extreme case of $\ffree=0$, when the free eccentricity is completely damped. In this case $e(a)=\ef(a)\propto a^{-1}$, see equation (\ref{eq:e_forced}), and neither nulls not caustic points (i.e. points tangent to $\kappa(r,a)$) are possible. As a result, ASD is also featureless and follows $\Sigma_a(r)$ very closely in the bulk of the disc, far from its edges. This behavior can be understood based on the results of \citet[][see their Appendix B]{Rafikov2023} for debris discs with $\Sigma_a(a)$ and $e(a)$ being power laws of $a$. In our case, $\Sigma_a\propto a^{-1}$ in the bulk of the disc, see Appendix \ref{sec:siga}, and $e(a)\propto a^{-1}$. In this case \citet{Rafikov2023} has shown that $\asd(r)\to \Sigma_a(r)[1-e^2(r)]^{-1/2}$. Due to the smallness of $e(a)$ in the upper panel of Figure \ref{fig:ffree_var}a , the correction factor in brackets does not deviate from unity by more than a per cent, in agreement with what $\asd(r)$ shows in the lower panel. In the opposite limit of $\ffree\to \infty$ (keeping $\ffree$ constant and $\ef\ffree$ below unity) one would find an identical outcome: smooth $e(a)$ and the resultant ASD.

%%%%%%%%%%%%%%%%%%%%%%%%%%%%%%%%%%%%%%%%%%%%%%%%%%%%%%%%%%%
\begin{figure*}
	\begin{center}
	\includegraphics[width=0.99\textwidth]{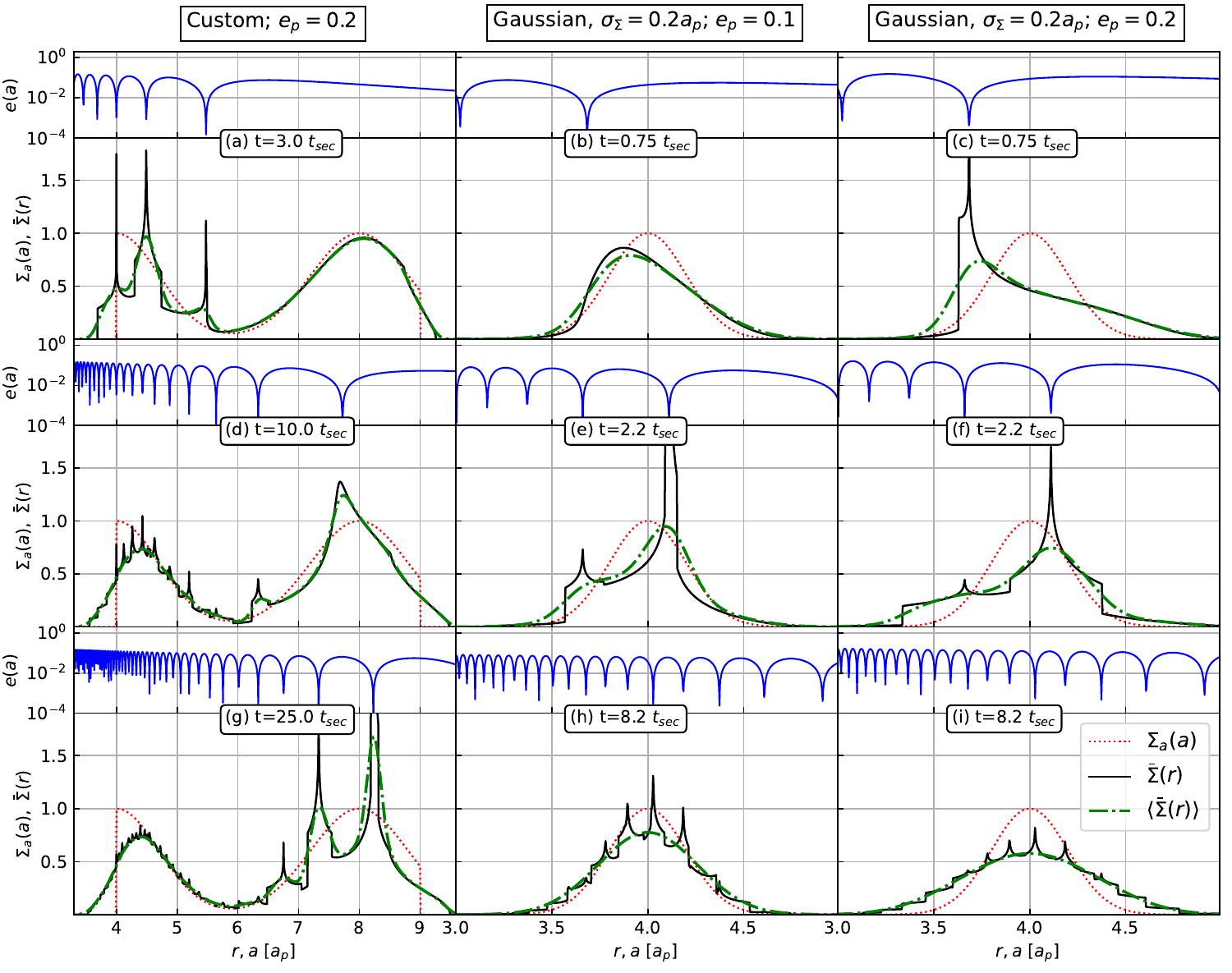}
	\caption{
    Similar to Figure \ref{fig:time_ev}, now illustrating the effect of varying $\Sigma_a(a)$ profile. Each column assumes a different model of $\Sigma_a(a)$ shown as dotted-red curves (including a narrow Gaussian ring in the second and third columns; note the different radial ranges), with the ASD shown at the times indicated in each panel as black solid curves. Dot-dashed green curves show Gaussian convolution of ASD with $\sigma=0.1\ap$ to illustrate the effects of finite observational resolution. See text for details. 
    }
	\label{fig:sigma_var}
	\end{center}
\end{figure*}
%%%%%%%%%%%%%%%%%%%%%%%%%%%%%%%%%%%%%%%%%%%%%%%%%%%%%%%%%%%

%%%%%%%%%%%%%%%%%%%%%%%%%%%%%%%%%%%%%%%%%

\subsection{Effect of varying $\Sigma_a(a)$}
\label{sec:complex-discs}

%%%%%%%%%%%%%%%%%%%%%%%%%%%%%%%%%%%%%%%%

So far, all our examples assumed a particular model of disc mass distribution in semi-major axis $a$, with $\Sigma_a(a)$ given by equations (\ref{eq:siga})-(\ref{eq:fout}) in Appendix \ref{sec:siga}. This profile has an extended power law segment, with smooth truncations at both inner and outer radii. It is natural to ask the question of how our results for the ASD behavior would change if we consider different, more structured profiles of $\Sigma_a(a)$. This question is additionally motivated by observations often revealing debris discs in the form of sharp and radially confined (narrow) rings \citep{Kennedy2020,ARKS1,ARKS2} and by theoretical works exploring the effect of a sharp (discontinuous) truncation of $\Sigma_a(a)$ on disc appearance \citep{Rafikov2023,Marino2021,Pearce2024}. 

To that effect, in Figure \ref{fig:sigma_var} we show several snapshots of $\asd(r)$ obtained for different $\Sigma_a(a)$ profiles (shown with dotted red curves). In the left column $\Sigma_a$ is represented by a sum of two broad Gaussians (with a deep minimum in between), sharply truncated on the inside and outside, see equation (\ref{eq:2G}) for an explicit expression. For this model we adopt $e(a)$ corresponding to $\ep=0.2$ and $\ffree=1$. In the other columns we adopt $\Sigma_a(a)$ in the form of a narrow Gaussian ring of radial width $\sigma_\Sigma=0.2\ap$ centered on $a_r=4\ap$, see equation (\ref{eq:ring}). For this $\Sigma_a$ profile we use $e(a)$ with $\ffree=1$ and two values of $\ep=0.1,0.2$ (center and right, respectively). Note that the times of snapshots are different in the left column and the other two.

Focusing first on the left column, we see that in the bulk of the disc the situation is qualitatively very similar to what we observed in our standard model of a radially extended disc: the ASD features multiple peaks that propagate out as time passes by following the outward motion of the $e$-nulls. Some peaks eventually evolve into bumps (as e.g. the one near $r=7.5\ap$ in panel (d)), closely replicating the patterns seen at the same moments of time in Figure \ref{fig:time_ev} which is made for the same $\ep$. The only noticeable difference compared to our standard $\Sigma_a$ disc model is the height of the peaks: compared to their analogues in Figure \ref{fig:time_ev}, the peaks are less pronounced in regions where $\Sigma_a$ is minimal, which can be easily understood by examining the asymptotic behavior (\ref{eq:asd4}) near the peak.

Near sharp disc edges the ASD exhibits a variety of behaviors. At the outer disc edge $e(a)$ varies weakly with $a$ (over the radial range where $\asd$ decays to zero) in the snapshots that are shown. As a result, the ASD exhibits a characteristic `asin’ profile near this edge, as predicted by  \citet[][see their Section 6.2.1]{Rafikov2023}. At the inner edge the situation is different, since $e(a)$ there varies on scales comparable to or smaller than the radial width over which ASD decays to zero. This leads to a variety of ASD behaviors, with several discontinuous jumps visible as $\asd$ goes to zero at early times, switching to a more continuous profile later on, as the spacing between $e$-nulls goes down. We defer the detailed exploration of the ASD behavior near the sharp edges of a secularly-evolving disc to a future study.  
   
Moving on to narrow Gaussian rings, we see that in the early stages of secular evolution, when the number of $e$-nulls in the ring vicinity is small (one or two), the ASD exhibits several individual features --- peaks, bumps, jumps --- which follow the phenomenology developed earlier for more extended discs. These features can be so significant that they substantially distort $\asd(r)$ away from the shape of $\Sigma_a(r)$, showing skewness that is dependent on whether the major $e$-nulls lie inside (e.g. at $t=0.75\tsec$) or outside (e.g. at $t=2.2\tsec$) of the $\Sigma_a(a)$ peak. At later times the density of $e$-nulls increases and numerous overlapping peaks start covering the radial extent of the ring. For low $\ep$, such that $\ef(a_r)$ is smaller than the typical relative ring width $2\sigma_\Sigma/a_r$ (close to the case shown in panel (h)), $\asd(r)$ eventually reduces to $\Sigma_a(r)$. But for larger $\ep$, such that $\ef(a_r)\gtrsim 2\sigma_\Sigma/a_r$, the radial width of the late-time ASD profile exceeds $\sigma_\Sigma$ and becomes more Lorentzian-like, with broader, more slowly decaying wings than a Gaussian; see panel (i).

%%%%%%%%%%%%%%%%%%%%%%%%%%%%%%%%%%%%%%%%%%%%%%%%%%%%%%%%%%%
\begin{figure*}
	\begin{center}
	\includegraphics[width=0.99\textwidth]{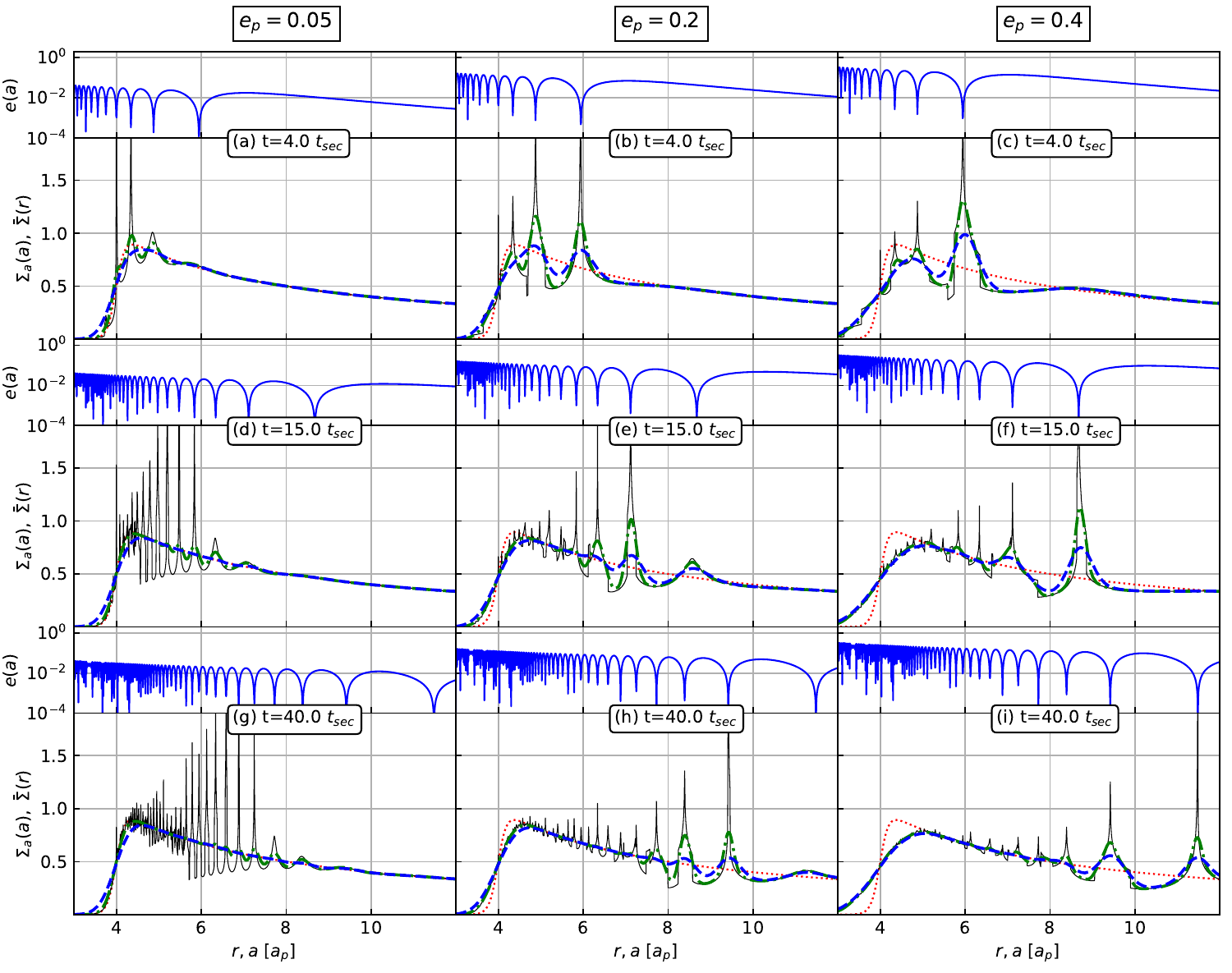}
	\caption{
    Snapshots of ASD (thin solid black curves) convolved (in radius) with a Gaussian to mimic the effects of finite resolution of observations. Shown are 3 snapshots at times $t=4, 15, 40 t_\mathrm{sec}$ for $\ep=0.05, 0.2, 0.4$ and $f_\mathrm{free}=1$, as labeled in individual panels. Smoothing is performed for the dispersion $\sigma=0.1 a_\mathrm{p}$ (dot-dashed green) and $0.25a_\mathrm{p}$ (dashed blue); dotted red curve is the underlying $\Sigma_a(a)$. }
	\label{fig:smooth}
	\end{center}
\end{figure*}
%%%%%%%%%%%%%%%%%%%%%%%%%%%%%%%%%%%%%%%%%%%%%%%%%%%%%%%%%%%

%%%%%%%%%%%%%%%%%%%%%%%%%%%%%%%%%%%%%%%%
%%%%%%%%%%%%%%%%%%%%%%%%%%%%%%%%%%%%%%%%%

\section{Observational appearance of ASD structures }
\label{sec:obs}

%%%%%%%%%%%%%%%%%%%%%%%%%%%%%%%%%%%%%%%%

We now turn our attention to the prospects of detection of the various ASD features described before in realistic observations of debris discs. The problem here is obvious: all features exhibited by the ASD --- peaks and jumps --- are radially very sharp, and it is not clear if this unique characteristic will survive in the observed debris disc images given their finite resolution. 

The true two-dimensional (2D) image of the disc in the plane of the sky gets convolved with a point-spread function (PSF) of the telescope, giving us the observed image at lower spatial resolution. 
Since in this study we are interested in ASD only, we adopted an approximate procedure to explore the impact of the finite resolution of observation on ASD structure: upon computing the exact $\asd(r)$ using {\sc debrispy}, we simply convolve this one-dimensional profile with a simple Gaussian of radial width $\sigma$. This method should provide a good qualitative assessment of resolution effects on the ASD inferred from observations. 
A detailed comparison with the results of direct 2D convolution is beyond the scope of this work and is left for future study.

In Figure \ref{fig:smooth} we show the results of applying this procedure to several ASD snapshots at three moments of time computed for three different values of $\ep$. We perform convolution using two different Gaussian widths, ``high-resolution'' $\sigma=0.1\ap$ and ``low-resolution'' $\sigma=0.25\ap$, referred to as such only in a relative sense, without explicit relation to any existing or future instrument. One can see that the sharp and tall peak features vanish upon convolution for either value of $\sigma$. The maximum contrasts of $\asd(a)$ relative to $\Sigma_a(a)$, formally divergent for unconvolved ASD, are about a factor of 2 at best for $\sigma=0.1\ap$ and lower for larger $\sigma$. The ASD jumps get smoothed and the peaks turn into bumps upon convolution. However, the pedestal structures formed by peaks and jumps still largely survive for $\sigma=0.1\ap$ and also to some degree in the low-resolution case.

An obvious trend is that the contrasts of the bumps of PSF-degraded ASD relative to $\Sigma_a(a)$ are a strong function of $\ep$. Indeed, for $\ep=0.05$ the convolved $\asd$ deviates from $\Sigma_a(r)$ only by $\sim 10-20\%$, whereas for $\ep=0.4$ it can easily reach a factor of 2. At higher $\ep$, including $\ep=0.2$, the set of reasonably sharp ASD bumps and troughs is quite pronounced on top of the smooth underlying $\Sigma_a$. We also note that the locations of the local maxima of convolved ASD remain very close to the locations of the sharp peaks of unconvolved $\asd$ at the $e$-nulls $a_k$.  

The actual value of resolution $\sigma$ needed to resolve and detect ASD features depends on the planetary eccentricity (which together with $\ap$ determines the local $\ef$ and $e(a)$, see equation (\ref{eq:e_forced})) and the signal-to-noise ratio (SNR) of observations. Indeed, looking at Figure \ref{fig:smooth}, one notices that for $\ep=0.05$ a convolution of a true ASD with $\sigma=0.25\ap$ Gaussian (dashed blue curves) leaves no apparent bumps of ASD even at $10\%$ level relative to the smooth continuum, and the convolved ASD is very close to $\Sigma_a(r)$ far from the disc edges. But if the PSF resolution improves to $\sigma=0.1\ap$ (dot-dashed magenta curves) and SNR becomes high enough to allow measurement of $\sim 10\%$ ASD contrasts, then 3-4 individual peaks can be measurable even when $\ep=0.05$. Note that the observable features do not necessarily reside at the lowest $k$ nulls of $e(a)$. For example, at $t=15\tsec$ (panel (d)) the (unconvolved) ASD bump at $a_1$ (and likely the one at $a_2$) has such a low amplitude, that it will not be picked up even in $\sigma=0.1\ap$ observations at $10\%$ contrast, but the features at $k=3$ and $4$ nulls of $e(a)$ should be above the noise. Similarly, at $t=40\tsec$ one would not see the features at $k=1,2$ nulls, but would have a chance of detecting the ones at $k=3,4,5$ nulls.

Things improve as $\ep$ increases and underlying ASD features naturally get wider. For example, for $\ep=0.2$ even at resolution of $\sigma=0.25\ap$ one can measure at $10\%$ contrast at least 2 (at $t=15\tsec$) and 3 (at $t=40\tsec$) bumps of ASD, starting at $k=1$ null. As the resolution improves to $0.1\ap$, these numbers change to 3 and 4. For $\ep=0.4$ the pedestals of peaks are so wide that at least 2 peaks can be easily detected for both values of $\sigma$, although the contrasts differ, of course. The number of detectable peaks is lower than in the $\ep=0.2$ case because of a more severe pedestal overlap in the $\ep=0.4$ case. 

Based on this discussion, it is clear that the detectability of an ASD bump depends on both the height of an (unconvolved) ASD peak and its radial extent set by the width of its pedestal --- the larger both are, the higher is the feature contrast in the convolved ASD. And to spatially resolve individual ASD peaks (at a fixed contrast) one needs resolution\footnote{For an ASD feature at $10\ap$ the assumed resolutions $\sigma=(0.1,0.25)\ap$ correspond to the PSF size of $1\%$ and $2.5\%$ of the radius of the feature. } $\sigma$ at the level of $\lesssim (0.2-0.3)$ of the peak separation, judging by Figure \ref{fig:smooth}. We leave a detailed exploration of a relation between the minimum SNR and $\sigma$ needed for a feature detection in realistic observations as a function of $\ep$ and $t$ to a future study. 

We also show a convolved ASD in Figure \ref{fig:ffree_var}, assuming a low resolution of $\sigma=0.25\ap$. The goal of that calculation was to see whether the peak splitting phenomenon discussed in Section \ref{sec:free-ec} can be used to measure the deviation of the free eccentricity fraction $\efree$ from unity. One can see that this is a difficult task, as the convolved ASD bumps look very similar to each other regardless of $\efree$ (as long as it is not too close to zero), e.g. see a bump at $8.8\ap$ in panels (b)-(f) of that figure. As a result, it is unlikely that the appearance of ASD inferred from observations can be used to measure the value of $\efree$, unless the PSF width $\sigma$ is very small. 

On the other hand, the maxima of the convolved ASD in Figure \ref{fig:ffree_var} are still located very close to the actual minima of $e(a)$, which themselves reside very near $a_k$ given by equation (\ref{eq:a_k}). This means that the radial locations of the ASD bumps can still be used to infer important details about the $e(a)$ structure and the values of $a_k$, regardless of the value of $\efree$ while it is not too different from unity.

In Figure \ref{fig:sigma_var} we mimic the effects of seeing on $\asd(r)$ obtained for different $\Sigma_a(a)$ profiles (see Section \ref{sec:complex-discs}) by showing its Gaussian convolution with $\sigma=0.1\ap$ (dot-dashed green curve). For the double-Gaussian profile (left column), this procedure may easily reduce the visibility of the already weak peaks near the minimum of $\Sigma_a$ below the detection threshold. For narrow Gaussian rings, the skewness of $\asd$ relative to $\Sigma_a$ remains obvious even after PSF convolution, with important implications for interpreting observations: seeing such skewed profiles may imply the presence of a low-$k$ ($k=1,2$) null of $e(a)$ profile on the shoulder of an underlying $\Sigma_a(a)$ profile. More generally, this exercise suggests that using pure Gaussians to fit $\asd$ profiles of narrow debris rings \citep{matra_REASONS, ARKS2} may be not the optimal strategy. 

Overall, we may conclude that despite a clear degradation of the sharpness of ASD structure in the presence of finite observational seeing, its key features still remain discernible in the convolved images and the locations of the main structures (bumps) can still be used to infer important information about the behavior of underlying $e(a)$.

%%%%%%%%%%%%%%%%%%%%%%%%%%%%%%%%%%%%%%%%%

\section{Effects of random free eccentricity}
\label{sec:free-ec-rnd}

%%%%%%%%%%%%%%%%%%%%%%%%%%%%%%%%%%%%%%%%

So far in this study we assumed eccentricity of debris particles to be a unique function $e(a)$ of  semi-major axis. However, various processes can produce an additional random component of the free eccentricity that is different from a deterministic $\efree$ set by the initial conditions and dissipative processes, as discussed in Section \ref{sec:free-ec}. Such a random eccentricity component adds to a deterministic $e(a)$ and leads to a distribution (rather than a unique value) of $e$ at a given value of $a$. Examples of processes leading to random velocity dispersion of debris particles are the gravitational self-stirring in massive discs \citep{Ida1990,Ida1992,Stewart2000,R2003a,R2003c}, gravitational stirring by a population of massive objects \citep{Kenyon2001,Ida1993,R2003b,Marino2021}, and mutual collisions \citep{Lohne2008,Pan2012}.

Non-zero random eccentricity component leads to a radial smearing of the various ASD structures, an effect similar to that of a finite resolution of observations. In this work we illustrate this effect approximately, in the same spirit as we modeled finite observational seeing in Section \ref{sec:obs}: we convolve  the exact ASD with a 1D Gaussian having a radial dispersion $\sigma_e(r) r$, where $\sigma_e(a)$ is the (dimensionless) dispersion of the random eccentricity component at a given $a$ and we neglect the difference between $a$ and $r$ in this calculation. We then also note that if we adopt the radial profile of eccentricity dispersion $\sigma_e(a)$ in the form 
\ba
\sigma_e(a)= \frac{\sigma_{e,0}}{a}
\label{eq:sig_e}
\ea
with $\sigma_{e,0}=$const, then the radial scale of the smoothing Gaussian is $\sigma_{e,0}$, i.e., radially constant, analogous to the dispersion $\sigma$ due to seeing that we used in producing smoothed curves in Figures  \ref{fig:ffree_var} \& \ref{fig:smooth}. Thus, we can use the same figures to illustrate the impact of random eccentricity with the dispersion in the form (\ref{eq:sig_e}) on the appearance of ASD; for example, the dot-dashed green curve in Figure \ref{fig:smooth} corresponds to $\sigma_e(a)$ with $\sigma_{e,0}=0.1\ap$. 

The conclusions we can draw from this exercise are the same as in Section \ref{sec:obs}: while the sharp features characteristic of the deterministic ASD get washed out by random free eccentricity, easily identifiable peaks of $\asd(r)$ are still detectable at locations very close to $a_k$ given by equation (\ref{eq:a_k}). Prominence of these features increases with the planetary eccentricity $\ep$, and typically only two or three peaks stand out in the ASD profile. In other words, one can still infer some key features of the underlying $e(a)$ profile even in the presence of a non-zero random eccentricity component. Note also that in the presence of both random eccentricity and finite resolution effects, the effective (radial) dispersion at which the ASD features need to be convolved for comparison with observations becomes $\left(\sigma^2+\sigma_e^2r^2\right)^{1/2}$, adding the two effects in quadratures.

%%%%%%%%%%%%%%%%%%%%%%%%%%%%%%%%%%%%%%%%
%%%%%%%%%%%%%%%%%%%%%%%%%%%%%%%%%%%%%%%%%

\section{Discussion}
\label{sec:disc}

%%%%%%%%%%%%%%%%%%%%%%%%%%%%%%%%%%%%%%%%

The main goal of this study was to develop analytical understanding of ASD structures arising in a secularly-evolving disc with a deterministic $e(a)$ profile in the form (\ref{eq:e_full}). This is a key step in building a forward modeling approach for interpreting debris disc observations, before adding complications such as the random eccentricity component in Section \ref{sec:free-ec-rnd}  and effects of observational seeing in Section \ref{sec:obs}. Comparisons between observations and simulated profiles, e.g. obtained using direct N-body simulations or derived from low-resolution Monte Carlo calculations, may be problematic if the latter do not adequately resolve the underlying radial structure of ASD. Since observations are inherently affected by finite angular resolution, meaningful comparisons require synthetic profiles that are first computed at sufficiently high radial resolution (which is what our study offers) and then convolved with an appropriate instrumental response. This may be especially important in large-scale or population-level studies, where subtle biases introduced by insufficient numerical resolution could systematically affect inferred planetary properties.

As we demonstrated in this work, a rather simple and physically-motivated secular eccentricity profile (\ref{eq:e_full}) can lead to a surprisingly rich phenomenology of behaviors exhibited by ASD. While we assumed a particular system architecture with a massless debris disc perturbed by a single inner planet, it is clear that analogous ASD structures should be exhibited in other setups in which eccentricity profile has a similar form with numerous extrema and (possibly) nulls, e.g. when a planet (or a star) is exterior to the disc \citep{Farhat2023}. 

More broadly, our analytical framework and intuition developed in this study allow us to predict ASD structures in discs with more general, non-secular $e(a)$ profiles. Based on the discussion in Sections \ref{sec:ASD-features}, \ref{sec:free-ec} and mathematical developments in Appendix \ref{sec:features_math}, one can predict that if an arbitrary $e(a)$ profile exhibits tangent points with respect to $\kappa(r,a)$, then ASD must feature a finite jump or a weakly divergent peak, depending on whether $e^{\prime\prime}$ (or, more precisely, $\chi$) is negative or positive at the tangent point. And if $e(a)$ exhibits a null, then either a weakly divergent peak or a bump will emerge at this location, depending on the value of $\zeta$ there. Our understanding of ASD behavior can also be employed to constrain planetary properties using the shape of the inner edge of the debris disc \citep{Marino2021,Rafikov2023,Pearce2024}. 

To a large degree, this improved understanding of ASD behavior has been made possible by the use of {\sc debrispy} (see Appendix \ref{sec:debrispy}), which turned the analytical framework of \citet{Rafikov2023} into a versatile and efficient numerical tool. Just to reiterate, for a given $e(a)$ profile, or more generally, for a given eccentricity distribution provided as a function of $a$, the results of \citet{Rafikov2023} are exact, as well as their implementation in {\sc debrispy}. An alternative method of computing ASD (and 2D images) by Monte Carlo sampling of orbits is inevitably noisy (e.g. see Figures \ref{fig:overview}c and \ref{fig:time_ev}) and requires very large number of sampled particles ($N\sim 10^7-10^8$) to provide reasonably accurate results for mass distribution \citep{Kennedy2020,Lovell2023,Chit2025}. Our analytical procedure is not only exact and much faster numerically, but it also allows one to compute $\asd$ at any single point of interest without computing the structure of the whole disc. This, in particular, allowed us to enable ASD calculation using an adaptive radial mesh, which focuses resolution at the locations where ASD gradients are highest, see Figure \ref{fig:mesh}. Based on all that, we recommend using {\sc debrispy} for general ASD calculations. 

On the observational side, depending on resolution, the ASD profiles of many debris disc systems seem to display multiple peaks separated by apparent density depletions \citep{ARKS2}. Especially for inclined systems, where the underlying disk morphology cannot be uniquely reconstructed from on-sky images (due to well-known degeneracies in fitting observations), such profiles are often interpreted as evidence for multiple belts of debris separated by gaps and sculpted by planet(s) embedded within the disc. However, our results clearly show that apparent depletions in observationally derived radial density profiles need not necessarily correspond to genuine gaps in the semi-major axis distribution of the parent bodies. Instead, a similar signature can be naturally produced in a radially smooth disk perturbed by an interior planet, a possibility that we explore in more detail next.

%%%%%%%%%%%%%%%%%%%%%%%%%%%%%%%%%%%%%%%%%

\subsection{Applications to interpretation of observations}
\label{sec:apps}

%%%%%%%%%%%%%%%%%%%%%%%%%%%%%%%%%%%%%%%%

Now we discuss how our understanding of the ASD behavior developed in this study can help us infer planetary properties based on observations of  secularly perturbed debris discs. Assuming that a low mass disc with a smooth underlying $\Sigma_a(a)$ is secularly perturbed by an unseen inner planet -- i.e., one that is not directly detectable because it is too faint and/or because it is difficult to detect via RV or transit methods --  we ask what the structure of the ASD inferred from observations can tell us about the planet. In doing this, it is important to remember that deducing an ASD profile from observations is not always trivial for several reasons. First, we need to remember that the observable is not the dust surface density but the intensity of emission at every point in the image. To derive the former from the latter using long-wavelength, sub-mm observations of cold dust, one needs to make some assumptions about the behavior of dust opacity and its temperature profile \citep[e.g.,][]{ARKS2}. Similarly, when dealing with short-wavelength, optical and near-IR scattered light observations of debris discs, one needs to account for the geometry of scattering and make assumptions about the scattering phase function \citep[e.g.,][]{Hom2024}. Second, de-projection of an observational image is a non-trivial task if the disc is warped and/or vertically extended \citep{ARKS_3}, such that dust at different radii contributes to emission at a given point on the sky. Third, a convolution of an image of an inclined disc with a non-trivial three-dimensional structure with a (generally non-circular) PSF can additionally impact the inferred ASD. Despite these complications, ASD retrieval from debris disk observations has been performed in a number of studies \citep[see e.g. Figs. 2 and 3 in][and references therein]{ARKS2}.

We also assume that disc observations have good enough resolution and SNR to enable detection of several individual bumps of ASD resulting from the convolution of a true ASD with the observational PSF. The resolution and sensitivity requirements needed for such a detection have been discussed in Section \ref{sec:obs}. 

Let us denote the radii at which the observed ASD reaches its maximum within each bump as $r_i$, $i=1,2,...$, with $i=1$ corresponding to the outermost bump and higher $i$ marking bumps at progressively smaller radii. As discussed in Sections \ref{sec:free-ec}, \ref{sec:obs}, \& \ref{sec:free-ec-rnd}, these locations should always be close to the null $a_k$ of $e(a)$ for $\ffree=1$ defined by equation (\ref{eq:a_k}), regardless of (i) observational resolution effects, as long as the bumps are detectable, (ii) deterministic free eccentricity fraction $\ffree$ and (iii) dispersion of the random eccentricity component $\sigma_e$. However, it is not always true that $r_i\approx a_i$ for the same $i$, simply because, as we discussed in Section \ref{sec:obs}, the low-order nulls of $e(a)$ may have low contrasts and be undetectable in observations, which is the case e.g. in Figure \ref{fig:smooth}a,d,g. What is true in general is that $r_i=a_{i+k_0}$, where $k_0$ is the highest null order $k$ that is not observable. If the ASD bump at $k=1$ null is observable, then $k_0=0$ and $i=k$, as in e.g. Figure \ref{fig:smooth}c,f,i. But if e.g. the bumps at the first two $e$-nulls are unobservable, as in Figure \ref{fig:smooth}g, then $k_0=2$ and $r_i=a_{i+2}$. 

This correspondence between $r_i$ and $a_k$ and equation (\ref{eq:a_k}) imply that the locations of the maxima of observed ASD should satisfy the relation 
\ba
\frac{r_i}{r_1}\approx \left(\frac{1+k_0}{i+k_0}\right)^{2/7},
\label{eq:rad_rat}
\ea  
where $r_1$ is the radius of the outermost detected bump, as long as the underlying $\Sigma_a(a)$ distribution is sufficiently smooth. For example, if the $k=1$ null of $e(a)$ is observable, then $k_0=0$, 
$r_i/r_1\approx i^{-2/7}$ and the maxima of the observed ASD would lie at
\ba
\{r_1,r_2,r_3,r_4,...\}\approx \{1,\,0.82,\,0.73,\,0.67,...\}r_1,~~~k_0=0.
\label{eq:rad_rat_0}
\ea  
But if the situation is as presented in Figure \ref{fig:smooth}g, with the bumps at the first two nulls undetected, then $k_0=2$ and 
\ba
\{r_1,r_2,r_3,r_4,...\}\approx \{1,\,0.92,\,0.86,\,0.82,...\}r_1,~~~ k_0=2.
\label{eq:rad_rat_2}
\ea  
Thus, if a debris disc features several ASD maxima at radii satisfying the relation (\ref{eq:rad_rat}) for some integer $k_0$, then this could indicate that the disc is indeed secularly perturbed by an inner planet. Of course, if there are only two maxima and this relation is obeyed, this could be a simple coincidence. But for a larger number of detected ASD peaks the probability of such a coincidence greatly diminishes, and one should consider the agreement with equation (\ref{eq:rad_rat}) as a strong evidence in favor of a secular planetary hypothesis. 

Once this possibility is established, we can use the radii of ASD maxima to yield an important constraint on a combination of the planetary mass $\Mp$ and its semi-major axis $\ap$. For that we need to know the distance to the system, giving us the actual values of $r_i$,
as well as the stellar mass $M_\star$ and the system age $t_{\rm age}$. If we now associate $r_i\approx a_{i+k_0}$, we can  re-write equation (\ref{eq:a_k}) in the form
\ba
\Mp \ap^2 \approx \frac{8\pi}{3 t_{\rm age}}\left(\frac{M_\star}{G}\right)^{1/2}  \left[(i+k_0)^2 r_i^7\right]^{1/2}, 
\label{eq:Ma2}
\ea  
which must hold for all values of $i$ corresponding to observed ASD peaks. As all quantities in the right-hand side of this relation are assumed to be known, we have a direct measurement of $\Mp\ap^2$.  

Going further, we can exploit the sensitivity of the shape of the ASD profile to planetary eccentricity $\ep$ to get further constraints. Figure \ref{fig:smooth} clearly demonstrates that the ASD bumps get narrower and have lower contrast as $\ep$ decreases. The sensitivity is actually not to $\ep$ alone but to the combination $\ep\ap$ as it determines the magnitude of $\ef$ at each radius (see equation \ref{eq:e_forced}) and the overall amplitude of the $e(a)$ profile, which in turns feeds into the observed ASD. One should also bear in mind the slight dependence of the observed ASD on $\ffree$ (see dashed blue curve in Figure \ref{fig:ffree_var}), resulting in some degeneracy of $\ep\ap$ with $\ffree$. Nevertheless, detailed modeling of ASD profile should make it possible to constrain the value of $\ep\ap$ with a reasonable accuracy.

%%%%%%%%%%%%%%%%%%%%%%%%%%%%%%%%%%%%%%%%%%%%%%%%%%%%%%%%%%%
\begin{figure}
	\begin{center}
	\includegraphics[width=0.48\textwidth]{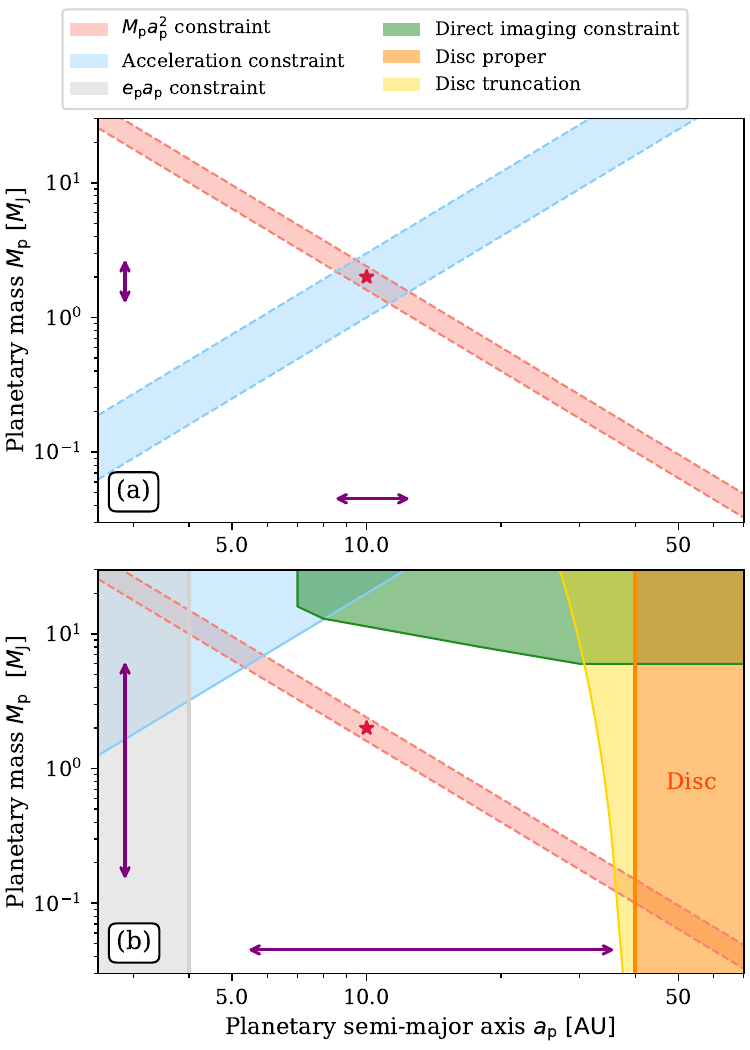}
	\caption{
    Various constraints on the planetary mass $\Mp$ and semi-major axis $\ap$ enabled by the detection of secular ASD structure in a debris disc (see text for details). In both panels the star marks the actual planetary parameters, and arrows show the individual ranges of $\ap$ and $\Mp$ allowed by observational measurements/constraints. Upper/lower limits are bounded by solid curves (colored region is excluded), while the actual measurements are bounded by dashed lines (colored region is allowed). 
    (a) $\Mp$ and $\ap$ determination possible when a secular measurement of $\Mp\ap^2$ is available (salmon band) together with the measurement of stellar acceleration yielding $\Mp\ap^{-2}$ (light blue band). As a result, $\Mp$ and $\ap$ are rather well constrained, see arrows.
    (b) Constraints on $\Mp$ and $\ap$ possible when only an upper limit on acceleration is available (light blue region) and combined with other constraints: direct imaging (green), lower limit on $\ap$ due to $\ep\ap$ measurement (grey), and an upper limit on $\ap$ due to dynamical truncation of the inner disc edge (yellow). Even in combination with the secular measurement of $\Mp\ap^2$ (salmon band), the resultant individual ranges of allowed $\ap$ and $\Mp$ (arrows) are wider than in panel (a).
    }
	\label{fig:meas}
	\end{center}
\end{figure}
%%%%%%%%%%%%%%%%%%%%%%%%%%%%%%%%%%%%%%%%%%%%%%%%%%%%%%%%%%%

The two constraints on $\Mp\ap$ and $\ep\ap$ alone do not uniquely determine $\Mp$, $\ap$, and $\ep$. That task requires additional constraints. The best one could come from measuring the acceleration of the parent star caused by planetary gravity either along the line of sight using the radial velocity method, or astrometrically e.g. using {\it Gaia}, or both in the best case scenario. Such a measurement would constrain a combination $\Mp\ap^{-2}$ (for a known $M_\star$), with some degeneracy with $\ep$ and planetary orbital orientation. When combined with $\Mp\ap^2$ constraint, this measurement would immediately yield $\Mp$ and $\ap$, with the value of $\ep$ then following from $\ep\ap$ constraint. 

This possibility is illustrated in Figure \ref{fig:meas}a, where we consider a hypothetical system with a $\Mp=2M_\mathrm{J}$ planet at $\ap=10$ AU and a disc with an inner edge at $a_\mathrm{in}=40$ AU. We assume that $\Mp\ap^2$ measurement is available to us from fitting the secular ASD features in the disc. It is shown as a salmon-colored band, with a finite width schematically reflecting the uncertainty in the stellar age and mass determination, see equation (\ref{eq:Ma2}). Also shown is the acceleration measurement (light blue band), which in this case would require acceleration to be measurable at the (rather optimistic) level of $\sim 1\,$m s$^{-1}$yr$^{-1}$. Finite width of this band reflects the uncertainty in the orientation and shape of the planetary orbit, although the secular measurement of $\ap\ap$ might greatly help in reducing it. The combination of these two measurements (including their uncertainties) constrains $\Mp$ and $\ap$ to lie within rather narrow ranges shown with arrows.

A non-detection of stellar acceleration would still provide an upper limit on $\Mp\ap^{-2}$, which can be combined with other constraints to limit the range of $\Mp$ and $\ap$ allowed by the secular measurement of $\Mp\ap^2$. This approach is illustrated in Figure \ref{fig:meas}b, where we show upper or lower limits set by various constraints as colored regions bounded by solid curves, while $\Mp\ap^2$ measurement is a salmon band bounded by dashed lines (as in panel (a) for both secular and acceleration measurements). Acceleration constraint, which in this case sets an upper limit on acceleration at the level of $\sim 10\,$m s$^{-1}$yr$^{-1}$, excludes the light blue corner of the $(\ap,\Mp)$ space. 

In addition, direct imaging should provide an $\ap$-dependent upper limit on $\Mp$, which we schematically show as a green region with the shape motivated by \citet{Apai2008}. Also, the secular measurement of $\ep\ap$ sets a direct lower limit on $\ap$ since $\ep<1$, shown as a grey constraint. Finally, the planet cannot orbit too close to the inner edge of the disc, as the particle density at the edge would be rapidly depleted \citep{Morrison2015,Pearce2024}. This sets an upper limit on $\ap$, for which we adopt a specific form given by equation (23) of \citet{Sefilian2021}, shown as a yellow region next to $\ap=a_\mathrm{in}$. Combination of these constraints leaves an extended white region of the parameter space unconstrained, however, when the secular measurement of $\Mp\ap^2$ is added (salmon), the allowed area in the $(\ap,\Mp)$ space shrinks dramatically.  Even though the individual ranges of allowed $\ap$ and $\Mp$ (shown by arrows) are much wider now compared to panel (a), such joint constraints should still be helpful in planning direct imaging campaigns aimed at detecting the putative planet perturbing the disc. 

Even if the perturbing planet has already been detected by direct imaging, secular constraints can still be very useful. First, a secular measurement of $\ep\ap$ would be very helpful for reducing the parameter space of planetary orbital characteristics consistent with the short orbital arc covered by initial observations. Second, a dynamical determination of $\Mp$ based on a secular measurement of $\Mp\ap^2$, in combination with the observed luminosity of the planet and the system's age, would provide important information about the initial properties of the planet \citep[i.e. its initial entropy, see][]{Spiegel2012} and its cooling history. 

We applied these ideas in practice by searching for possible secular signatures in the ASD profiles obtained for 24 debris discs by the ARKS project \citep{ARKS2}. Specifically, we searched for groups of ASD peaks or bumps in their Fig. 3 whose locations could be consistent with the relation given by Equation (\ref{eq:rad_rat}). Here we note that the ASD profiles in their Fig. 3 were derived from the {\sc CLEAN} images shown in their Fig. 2, which present the corresponding surface-brightness distributions. 
Although these quantities are distinct, sharp radial features appear at similar locations in the two figures (although their relative intensities differ). We did not find any obvious secular features in  the ARKS sample that are consistent with our theoretical expectations. This, however, is probably not too surprising given the finite resolution and SNR of these observations, the detrimental effects of  projection \citep[many of the ARKS systems are highly inclined; see][]{ARKS1}, and the radial narrowness of some of these discs. Future observations with, e.g., the proposed ALMA2040 upgrade\footnote{\href{https://www.euroalma2040.com/home}{https://www.euroalma2040.com/home}.}, offering substantially improved angular resolution and sensitivity, may help alleviate some of these limitations and enable such secular features to be identified more readily. The feasibility of this prospect can be assessed once the final specifications of ALMA2040 are established. For now, however, even this negative outcome can provide important information, as we show next.

%%%%%%%%%%%%%%%%%%%%%%%%%%%%%%%%%%%%%%%%%

\subsection{Implications of a non-detection of secular features}
\label{sec:non-detect}

%%%%%%%%%%%%%%%%%%%%%%%%%%%%%%%%%%%%%%%%

Now we briefly discuss what can be learned from a {\it non-detection} of secular structures in debris disc observations following the procedure described in Section \ref{sec:apps}. There are several possible reasons why a secular ASD structure can escape detection. 

The first (and most trivial) possibility is that the system simply does not harbor planets. Second, there may be a planet in the system, but its mass, or more precisely its $\Mp\ap^2$, is too low for the low-$k$ eccentricity nulls secularly induced by its gravity to have reached the inner edge of the disc within the lifetime of the system $t_\mathrm{age}$. For illustration, revealing the secular nature of the sharp ASD features requires at least two ASD peaks/bumps to reside within the disc (see Section \ref{sec:apps}), meaning that the secular structure would be missed if the second $e$-null lies interior to the inner disc edge $a_\mathrm{in}$ (even if the first null is within the disc proper). Using equation (\ref{eq:a_k}) with $k=2$, this condition for escaping a detection implies an upper limit on $\Mp\ap^2$:
\begin{align}
\Mp \ap^2 \lesssim \psi_\mathrm{low}~~~~~~\mathrm{with} 
\label{eq:Ma2-low}
\end{align}
\begin{align}
\psi_\mathrm{low} & =\frac{16\pi}{3}\left(\frac{M_\star}{G}\right)^{1/2}  \frac{a_\mathrm{in}^{7/2}}{t_{\rm age}}
\label{eq:Ma2-low1}\\
&\approx 
5.7\, M_\mathrm{J}\,\mathrm{AU}^2\left(\frac{M_\star}{M_\odot}\right)^{1/2}\left(\frac{a_\mathrm{in}}{40\,\mathrm{AU}}\right)^{7/2}\frac{200\,\mathrm{Myr}}{t_\mathrm{age}}. 
\end{align} 

Third, in the opposite limit the planetary $\Mp\ap^2$ may be so high that secular evolution has been fast and the low-$k$ eccentricity nulls are already located outside the outer disc edge $a_\mathrm{out}$. The ASD peaks at higher-$k$ nulls within the disc may be strongly affected by the pedestal overlap, making them invisible given the finite observational resolution. In this case the rough criterion for non-detection of secular ASD structures would be that $a_k(\zeta_k)$ given by equation (\ref{eq:ak_viazeta}) with $\zeta_k=\zeta_\mathrm{po}$ (see Appendix \ref{sec:pedestal_overlap}) is outside $a_\mathrm{out}$, which results in a lower limit on a combination of planetary parameters:
\begin{align}
\ep\ap\times \Mp \ap^2 \gtrsim \psi_\mathrm{high}~~~~~~\mathrm{with} 
\label{eq:Ma2-high}
\end{align}
\begin{align} 
\psi_\mathrm{high} &=\frac{32\zeta_\mathrm{po}}{105}\left(\frac{M_\star}{G}\right)^{1/2} 
\frac{a_\mathrm{out}^{9/2}}{t_{\rm age}} 
\label{eq:Ma2-high1}\\
&\approx 
276\, M_\mathrm{J}\,\mathrm{AU}^3\left(\frac{M_\star}{M_\odot}\right)^{1/2}\left(\frac{a_\mathrm{out}}{80\,\mathrm{AU}}\right)^{9/2}\frac{200\,\mathrm{Myr}}{t_\mathrm{age}}. 
\end{align} 
For illustration, \citet[][see their Fig. B1]{Pearce2024} present a $t_\mathrm{age}=16$ Myr snapshot of ASD from their direct N-body simulation of a debris disc with $a_\mathrm{out}=36$ AU perturbed by a $\Mp=2M_\mathrm{J}$, $\ep=0.2$, $\ap=10$ AU planet, with no apparent ASD peaks. This is natural since inequality (\ref{eq:Ma2-high}) is satisfied for the adopted system parameters, so pedestal overlap washes out secular structures across the whole disc making them indiscernible (given the number of simulated particles, $N=2\times10^4$, and the associated shot noise).  

Fourth, it is important to recall that in the presence of realistic observational resolution the ASD peaks/bumps are detectable even at low-$k$ $e$-nulls only if $\ep\ap$ is high enough, see Section \ref{sec:obs}. This suggests another possibility for a non-detection of secular ASD structures, namely that $\ep\ap$ is below some resolution- and SNR-dependent threshold, 
\begin{align}
\ep \ap \lesssim \psi_e(\sigma,\mathrm{SNR}),
\label{eq:epap_constr}
\end{align} 
in which case no constraint on $\Mp\ap^2$ is possible; the exact form of $\psi_e(\sigma,\mathrm{SNR})$ may also depend on $\Sigma_a(a)$ profile and should be determined for each system individually. But if $\ep\ap$ is above this threshold, then planetary parameters have to satisfy one of the constraints (\ref{eq:Ma2-low}) or (\ref{eq:Ma2-high}) for secular ASD structures to escape detection.    

There is also a fifth possibility: that even in the presence of a planet and its secular perturbations the eccentricity behavior of debris particles in the disc does not follow a simple prescription (\ref{eq:e_full}) adopted in this work. We will disregard this possibility for now but will comment on it in Section \ref{sec:future}.

%%%%%%%%%%%%%%%%%%%%%%%%%%%%%%%%%%%%%%%%%%%%%%%%%%%%%%%%%%%
\begin{figure}
	\begin{center}
	\includegraphics[width=0.49\textwidth]{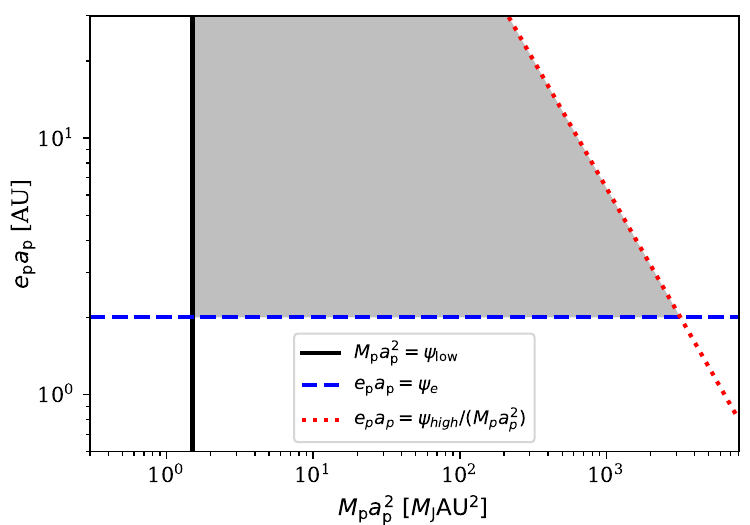}
	\caption{
    Constraints on combinations of planetary parameters $\Mp\ap^2$ and $\ep\ap$ provided by a non-detection of secular structures in disc ASD, illustrated using a system with properties similar to HD 107146 (see Section \ref{sec:non-detect} for details). Grey region of this parameter space is excluded if the system harbors an inner planet but no unambiguous secular structures are seen, while white region is still allowed. Different lines represent the thresholds following from equations (\ref{eq:Ma2-low})-(\ref{eq:epap_constr}).}
	\label{fig:non-det}
	\end{center}
\end{figure}
%%%%%%%%%%%%%%%%%%%%%%%%%%%%%%%%%%%%%%%%%%%%%%%%%%%%%%%%%%%

The three conditions (\ref{eq:Ma2-low})-(\ref{eq:epap_constr}) combined together set joint constraints on two combinations of planetary parameters --- $\ep\ap$ and $\Mp\ap^2$. These constraints are illustrated in Figure \ref{fig:non-det} for a system like HD 107146 \citep{Marino2018}, which harbors an extended, reasonably well-resolved debris disc \citep{ARKS1} around a $M_\star=1.04M_\odot$ star, which is $t_\mathrm{age}=150$ Myr old. Upon our examination, the disc extending from (roughly) $a_\mathrm{in}\approx 25$ AU to $a_\mathrm{out}\approx 150$ does not show obvious signs of secular structures consistent with the relation (\ref{eq:rad_rat}) in the CLEAN data presented in \citet{ARKS2}. System parameters allow us to immediately compute $\psi_\mathrm{low}\approx 1.5\, M_\mathrm{J}\,\mathrm{AU}^2$ and $\psi_\mathrm{high}\approx 6.4\times 10^3\, M_\mathrm{J}\,\mathrm{AU}^3$ for this system, and we show the corresponding constraints in Figure \ref{fig:non-det} via solid and dotted lines. 

Determining the value of $\psi_e$ is less trivial, as it requires detailed modeling of the disc ASD with the effects of seeing and SNR included. Instead, we simply assume for the purposes of illustration that secular structure in this disc becomes detectable at $\ep=0.2$ as shown in Figure \ref{fig:smooth}b,e,h. What matters for detectability is the combination $\ep\ap/a$ since it determines the value of $\ef$, and we evaluate it at $a\sim 10\ap$ in Figure \ref{fig:smooth}b,e,h, giving us $\ep\ap/a\sim 0.02$. In the disc of  HD 107146 at $a\sim 100$ AU we would have the same $\ep\ap/a$ if we set $\psi_e= \ep\ap\sim 2$ AU. Adopting this value (once again, chosen rather arbitrarily and only for illustration), we show the constraint (\ref{eq:epap_constr}) by a horizontal dashed line.   

One can see that a combination of the three constraints (\ref{eq:Ma2-low})-(\ref{eq:epap_constr}) excludes a certain (grey) region of the parameter space if (i) there is an inner planet in the system and (ii) no secular ASD structures are detected in the disc. Planets with parameters in this region should have revealed themselves via their secular ASD features. Planets with $(\Mp\ap^2, \ep\ap)$ outside this region (white areas) would not produce detectable secular ASD signatures given the finite observational resolution and SNR, meaning that their existence cannot be excluded by this method.
A joint constraint on planetary $\Mp\ap^2, \ep\ap$ represented by the grey region in Figure \ref{fig:non-det} can then be used to exclude a certain region of the more general three-dimensional parameter space of $\ep,\ap,\Mp$, which can help in planning subsequent observations of the system. 

Before moving on, we note that a particularly promising target for applying this method could be the Fomalhaut debris disc, which, to the best of our knowledge, currently provides the strongest and only observational evidence for a power-law forced eccentricity profile, with $d \log e / d \log a=  -1.75 \pm 0.16 $ as inferred based on ALMA observations \citep{Lovell2025Fom}, potentially indicative of secular perturbations by an unseen interior planet.

%%%%%%%%%%%%%%%%%%%%%%%%%%%%%%%%%%%%%%%%%

\subsection{Caveats and future extensions}
\label{sec:future}

%%%%%%%%%%%%%%%%%%%%%%%%%%%%%%%%%%%%%%%%

Our work necessarily employed some approximations. For example, we assumed a flat, razor-thin disc that is not warped, allowing us to avoid some ambiguity in defining the corresponding ASD. However, this approximation is expected to be reasonable for sufficiently vertically thin discs, as is the case for many observed systems, while its applicability to substantially thicker and/or warped discs remains uncertain
\citep[for instance, the ALMA ARKS survey finds debris disc aspect ratios of $\sim 0.002 - 0.2$; see][]{ARKS_3}. Relaxing this approximation would require generalizing the theoretical framework of \citet{Rafikov2023} to non-flat discs, explicitly accounting for the inclinations of debris particles and their resulting vertical structure due to e.g. secular planet--disk interactions \citep{Wyatt1999, Sefilian2025}.

A more serious assumption is that of a secular eccentricity profile in the form (\ref{eq:e_full}) with a single value of $e$ at any given semi-major axis $a$. This assumption should be reasonable for the parent objects of debris, which are only weakly affected by possible dissipative effects, and all our results should apply to the ASD of these objects. However, observations inform us about the spatial distribution of much smaller debris particles, which are the end product of a collisional cascade fed by the population of parent bodies. As the particles are broken down in collisions, their orbital parameters change as well \citep{Pan2012}, potentially driving their dynamics far from the simple prescription (\ref{eq:e_full}). Also, in discs with a substantial gaseous component, the free eccentricity of small particles can be strongly suppressed by gas drag, reducing their dynamics and associated ASD behavior to the situation shown in Figure \ref{fig:ffree_var}a.  

We leave a detailed exploration of these issues to  future work, but note for now that collisional evolution would likely turn a unique $e(a)$ profile (\ref{eq:e_full}) into a {\it distribution} of eccentricities at every $a$ and every particle size. Our discussion of the effect of random eccentricity component in Section \ref{sec:free-ec-rnd} should provide some idea of what impact this might have on the disc ASD. But it is also important to note that both the theoretical framework of \citet{Rafikov2023} and {\sc debrispy} can handle discs with a distribution of eccentricities with ease, as long as the eccentricity distribution and its dependence on $a$ are known. 

Also, ASD peaks and bumps represent locations where collisional evolution of debris particles is accelerated owing to their higher local density \citep{Mustill2009, Nesvold15}. This might cause preferential depletion of particles in the vicinity of peak locations (although this process would not be perfectly local), potentially reducing the contrasts of ASD peaks over their surroundings over time. The long-term disc evolution may therefore be more complicated than the purely dynamical picture presented here, with $\Sigma_a(a)$ changing in a non-trivial fashion simultaneously with the dynamical evolution. The steep dependence of the collisional timescale on the distance from the star would play an important role in shaping this evolution and the resultant appearance of ASD.

Future work should relax the simple prescription (\ref{eq:e_forced}) for the forced eccentricity behavior that is essentially scale-free.  One interesting option that we have not considered in this study is the possibility of a secular resonance located within the disc, at which the eccentricity formally diverges. Such a resonance can emerge if the debris disc is so massive that its self-gravity is secularly important \citep{R13a,R13,SR15A,RS15a,Sefilian2021, Sefilian2023}, there is more than one planet in the system \citep{Yelverton2018} or the orbit of the perturbing object is precessing \citep{R13a,SR15B,Pearce2015, Sefilian2021}. Relatedly, the self-gravity of a sufficiently massive debris disc can suppress secular eccentricity excitation by the planet and modify the radial profile of the forced eccentricity, generally making it steeper with increasing separation from the planet \citep{R13a, Sefilian2024}. We defer analytical investigation of ASD behavior in such systems to a future dedicated study (Rafikov \& Sefilian, in prep.).

%%%%%%%%%%%%%%%%%%%%%%%%%%%%%%%%%%%%%%%%
%%%%%%%%%%%%%%%%%%%%%%%%%%%%%%%%%%%%%%%%%

\section{Summary}
\label{sec:Summary}

%%%%%%%%%%%%%%%%%%%%%%%%%%%%%%%%%%%%%%%%

In this work we studied the behavior of axisymmetric surface density (ASD) of a debris disc subject to the secular gravitational perturbations of an inner planet. Starting with a smooth initial mass distribution in the disc, we find that secular evolution of the disc gives rise to a set of sharp, peculiar ASD features that may be detectable in disc observations. This work has been enabled by the development of a publicly available numerical framework {\sc debrispy} \citep[based on the analytical results of][]{Rafikov2023} that allows one to accurately and efficiently follow ASD evolution under a wide range of assumptions about the eccentricity behavior of disc particles. Our key findings are summarized below.

\begin{itemize}
    \item Secular evolution of debris particles results in an $e(a)$ profile that exhibits radial oscillations with numerous nulls (for $\ffree=1$, i.e. $\ef=\efree$) clustering towards the inner disc. As time goes by, the nulls of $e(a)$ move radially out, while also getting closer to each other at fixed $a$.
    \item Calculations assuming $\ffree=1$ reveal a set of sharp (weakly singular) peaks of ASD at the radial locations where the $e(a)$ profile exhibits nulls. These discontinuous peaks steadily evolve into finite bumps as the nulls move out; bumps are also prevalent at lower eccentricity of the perturbing planet.
    \item ASD profiles routinely exhibit numerous  discontinuous (but finite) jumps which emerge at the caustic locations where the periastra or apoastra of debris particles congregate.
    \item  Clustering of nulls eventually leads to overlap of the ASD structures in the inner disc (pedestal overlap phenomenon), washing out sharp ASD features.
    \item ASD structures become more pronounced and easier to detect when the perturbing planet has a higher eccentricity.
    \item Variation of $\ffree$ from unity leads to the splitting of an ASD peak into two, each of which is also weakly (logarithmically) singular. Discs with vanishing $\efree$ exhibit smooth ASD profiles.
    \item We provide quantitative analytical understanding for the origin of all ASD features based on the results of \citep{Rafikov2023} and support this theory with {\sc debrispy} calculations.
    \item Sharp secular ASD features get washed out by the effects of finite observational resolution and non-zero random component of free eccentricity. However, they should still be observable at well-defined locations relative to each other in high-resolution observations.
    \item Detection, as well as non-detection, of secular ASD features in debris discs can provide useful information about the properties of planets perturbing them.
\end{itemize}

We encourage the use of {\sc debrispy} for further studies of ASD behavior in discs with varied dynamical behaviors as well as for detailed forward modeling of ASD in observed debris discs. We also hope that our results will help motivate future debris disc observations and observing facilities with higher angular resolution and sensitivity.

%%%%%%%%%%%%%%%%%%%%%%%%%%%%%%%%%%%%%%%%%

\section*{Acknowledgements}

We thank the anonymous referee for useful suggestions.  R.R.R. acknowledges financial support through the Science and Technology Facilities Council (STFC) grant ST/T00049X/1 and the IAS. D.A. is supported by the STFC through an STFC PhD studentship (funder reference UKRI1731). A.A.S. is supported by the Heising-Simons Foundation through a 51 Pegasi b Fellowship.\\
\textit{Software:} \textsc{NumPy} \citep{harris_array_2020}, \textsc{SciPy} \citep{virtanen_scipy_2020} and \textsc{Matplotlib} \citep{thomas_a_caswell_matplotlibmatplotlib_2023}.

\section*{Data availability}
 The data underlying this article will be shared on reasonable request to the corresponding author. {\sc debrispy} is available on GitHub: https://github.com/DenizAkansoy/DebrisPy.

%%%%%%%%%%%%%%%%%%%%%%%%%%%%%%%%%%%%%%%%%%%%%%%%%%%%%%%%%%%%%%%%%%%%%%%%%%%%%%%%%%


\bibliographystyle{mnras}
\bibliography{Bibliography}

%%%%%%%%%%%%%%%%% APPENDICES %%%%%%%%%%%%%%%%%%%%%

\appendix

%%%%%%%%%%%%%%%%%%%%%%%%%%%%%%%%%%%%%%%%
%%%%%%%%%%%%%%%%%%%%%%%%%%%%%%%%%%%%%%%%%

\section{{\sc debrispy} software suite}
\label{sec:debrispy}

%%%%%%%%%%%%%%%%%%%%%%%%%%%%%%%%%%%%%%%%%

{\sc debrispy} is a modular Python package developed to numerically compute ASD profiles using the semi-analytic formalism of \citet{Rafikov2023}. In the most general case, for an eccentricity distribution $\psi_e(e,a)$ specified at every semi-major axis $a$, ASD is given by 
\begin{equation}
    \bar{\Sigma}(r) = \pi^{-1} \int_{r/2}^{\infty} a^{-1} \Sigma_a(a) \Phi_e(r,a) \md a,
\label{eq:debrispy-general-asd}
\end{equation}
where the eccentricity kernel $\Phi_e(r,a)$ is 
\begin{equation}
    \Phi_e(r,a) =\int_{\kappa}^{1} \frac{\psi_e(e,a)} {\sqrt{e^2-\kappa(r,a)^2}} \md e,
\label{eq:debrispy-general-kernel}
\end{equation}
and $\kappa=\kappa(r,a)$ is defined by equation~\ref{eq:kap}. Equations (\ref{eq:debrispy-general-asd}) and (\ref{eq:debrispy-general-kernel}) correspond to Equations~(16) and (28) of \citet{Rafikov2023}, respectively. The special case considered in this work, where the eccentricity is uniquely specified by a function $e(a)$, is recovered by taking $\psi_e(e,a)=\delta[e-e(a)]$, and is given by our Equation \ref{eq:asd-rel-gen} in Section \ref{sec:ASD}. 

The implementation is designed to be both flexible and computationally efficient. To this end, the package employs vectorised numerical operations, caches intermediate quantities for reuse, and offers CPU parallelisation. The calculation of ASD can be broken down into four sequential steps:

\begin{enumerate}
    \item specification of $\Sigma_a(a)$;
    \item specification of $e(a)$ or $\psi_e(e,a)$;
    \item construction of the corresponding eccentricity kernel (\ref{eq:debrispy-general-kernel}); 
    \item numerical evaluation of the ASD integral (\ref{eq:debrispy-general-asd}).
\end{enumerate}

\noindent In the {\sc debrispy} implementation, each of these stages is handled by a dedicated class, and we describe the associated numerical methodology in the following subsections.

%%%%%%%%%%%%%%%%%%%%%%%%%%%%%%%%%%%%%%%%%%%%%%%%%%%%%%%%%%%

\subsection{Input profiles}

The calculations require the specification of both the surface-density profile in semimajor axis space, $\Sigma_a(a)$, and the particle eccentricity distribution.

%%%%%%%%%%%%%%%%%%%%%%%%%%%%%%%%%%%%%%%%%%%%%%%%%%%%%%%%%%%

\subsubsection{$\Sigma_a(a)$ profile}

The implementation supports a broad class of physically motivated $\Sigma_a(a)$ profiles, including power-law, Gaussian, and step-function forms, as well as arbitrary user-defined functions. In all cases, the profile is defined over a finite domain $[a_{\rm min}, a_{\rm max}]$, outside of which the surface density is taken to vanish. 

%%%%%%%%%%%%%%%%%%%%%%%%%%%%%%%%%%%%%%%%%%%%%%%%%%%%%%%%%%%

\subsubsection{Eccentricity profiles}

Two separate eccentricity prescriptions are supported. In the first, the eccentricity is specified as a unique function of semi-major axis, $e(a)$, which is the prescription used for the calculations presented in this paper. The user can choose from built-in profiles, such as a power law of $a$, or supply an arbitrary function. In either case, the implementation verifies that the resulting eccentricities satisfy the physical constraint $0 \leq e(a) < 1$ throughout the domain.

Alternatively, the eccentricity at a given semi-major axis may be described by a full distribution, $\psi_e(e,a)$. Several built-in analytic distributions are supported, including a Rayleigh distribution of eccentricities, as well as other distributions considered in \citet{Rafikov2023}, for which closed-form expressions exist for the associated kernel $\Phi_e(r,a)$. More generally, arbitrary user-defined distributions $\psi_e(e,a)$ may be supplied. If the supplied function is already normalised, it is retained as a callable function and evaluated directly. Otherwise, the distribution is normalised numerically on a two-dimensional $(e,a)$ grid such that, at each semi-major axis, $\int_0^1 \psi_e(e,a) \md e = 1$.

%%%%%%%%%%%%%%%%%%%%%%%%%%%%%%%%%%%%%%%%%%%%%%%%%%%%%%%%%%%

\subsubsection{Adaptive gridding of eccentricity}

To ensure accurate numerical normalisation of user-supplied eccentricity distributions with sharp features or compact support, {\sc debrispy} can employ an adaptive gridding strategy based on gradient-driven remapping. The sampling grid is redistributed according to the local magnitude of $\nabla \psi_e$, concentrating resolution in regions of rapid variation of $\psi_e$ while maintaining coarser sampling elsewhere. This allows sharply peaked, truncated, or highly structured $\psi_e$ distributions to be resolved without requiring prohibitively fine uniform grids.

This procedure is only applied when the supplied $\psi_e$ distribution must be represented on a numerical grid (i.e. must be normalised). In this case, the normalised distribution is subsequently evaluated by interpolation, with nearest-neighbour, linear, and cubic interpolation schemes supported.

%%%%%%%%%%%%%%%%%%%%%%%%%%%%%%%%%%%%%%%%%%%%%%%%%%%%%%%%%%%

\subsection{Construction of the eccentricity kernel}

Prior to computing the ASD profile, {\sc debrispy} constructs the eccentricity kernel defined in Equation (\ref{eq:debrispy-general-kernel}). This kernel encapsulates the contribution of the eccentricity distribution to the ASD and is reused throughout the subsequent calculation, enabling fast evaluation of $\asd$ for multiple different $\Sigma_a(a)$ profiles.

When the eccentricity is specified uniquely as $e(a)$, the delta-function form of $\psi_e(e,a)$ allows the kernel to be evaluated directly, without numerical integration over eccentricity, see equation Equation (\ref{eq:asd-rel-gen}). 

As noted above, analytic kernel expressions are implemented for several commonly used eccentricity distributions. When these forms are used, the eccentricity integral in (\ref{eq:debrispy-general-kernel}) is evaluated analytically, so no numerical quadrature over $e$ is required.

For arbitrary eccentricity distributions, the kernel is evaluated numerically. {\sc debrispy} provides several integration schemes for this purpose, with Gauss--Legendre quadrature being recommended for most applications, as it provides a robust balance between accuracy, stability, and computational speed. The Gauss--Legendre scheme may be used either in fixed-order mode or in an adaptive-subdivision mode, in which the integration interval is subdivided into smaller intervals where additional resolution is required. Further details of the available integration schemes and configurable options are given in the package documentation \footnote{Package documentation available at: \href{https://debrispy.readthedocs.io/}{debrispy.readthedocs.io}.}.

The kernel itself may be tabulated either on a uniform $(r,a)$ grid or on an adaptively generated unstructured grid that concentrates samples in regions where the kernel varies rapidly. Once constructed, the tabulated kernel is reused in the ASD calculation through interpolation, with linear and cubic interpolation schemes currently supported.

%%%%%%%%%%%%%%%%%%%%%%%%%%%%%%%%%%%%%%%%%%%%%%%%%%%%%%%%%%%

\subsection{Computing the ASD}

Once the eccentricity kernel has been constructed, {\sc debrispy} evaluates the ASD integral over semi-major axis. In the general case, this corresponds to evaluating Equation~\ref{eq:debrispy-general-asd} on a user-specified radial grid. The calculation combines the input mass distribution in semi-major axis, $\Sigma_a(a)$, with the eccentricity kernel, $\Phi_e(r,a)$, and integrates over the finite region of semi-major-axis space for which particles can contribute to the chosen radius.

The ASD calculation may be performed either serially or with optional CPU parallelisation over independent radial grid points. Intermediate quantities, including the eccentricity kernel and interpolated input profiles, are cached where possible and reused in subsequent evaluations. In the calculations presented in this work, the semi-major-axis integral is evaluated using adaptive Gauss--Legendre quadrature.

%%%%%%%%%%%%%%%%%%%%%%%%%%%%%%%%%%%%%%%%%%%%%%%%%%%%%%%%%%%

\subsubsection{Adaptive grid refinement}

Sharp eccentricity structures can produce rapidly varying features in the final ASD profile. A uniformly spaced radial grid may therefore either miss such features or require a large number of points across the full domain. To address this, {\sc debrispy} includes an optional adaptive refinement procedure applied directly to the computed ASD profile.

\begin{figure}
	\begin{center}
	\includegraphics[width=0.49\textwidth]{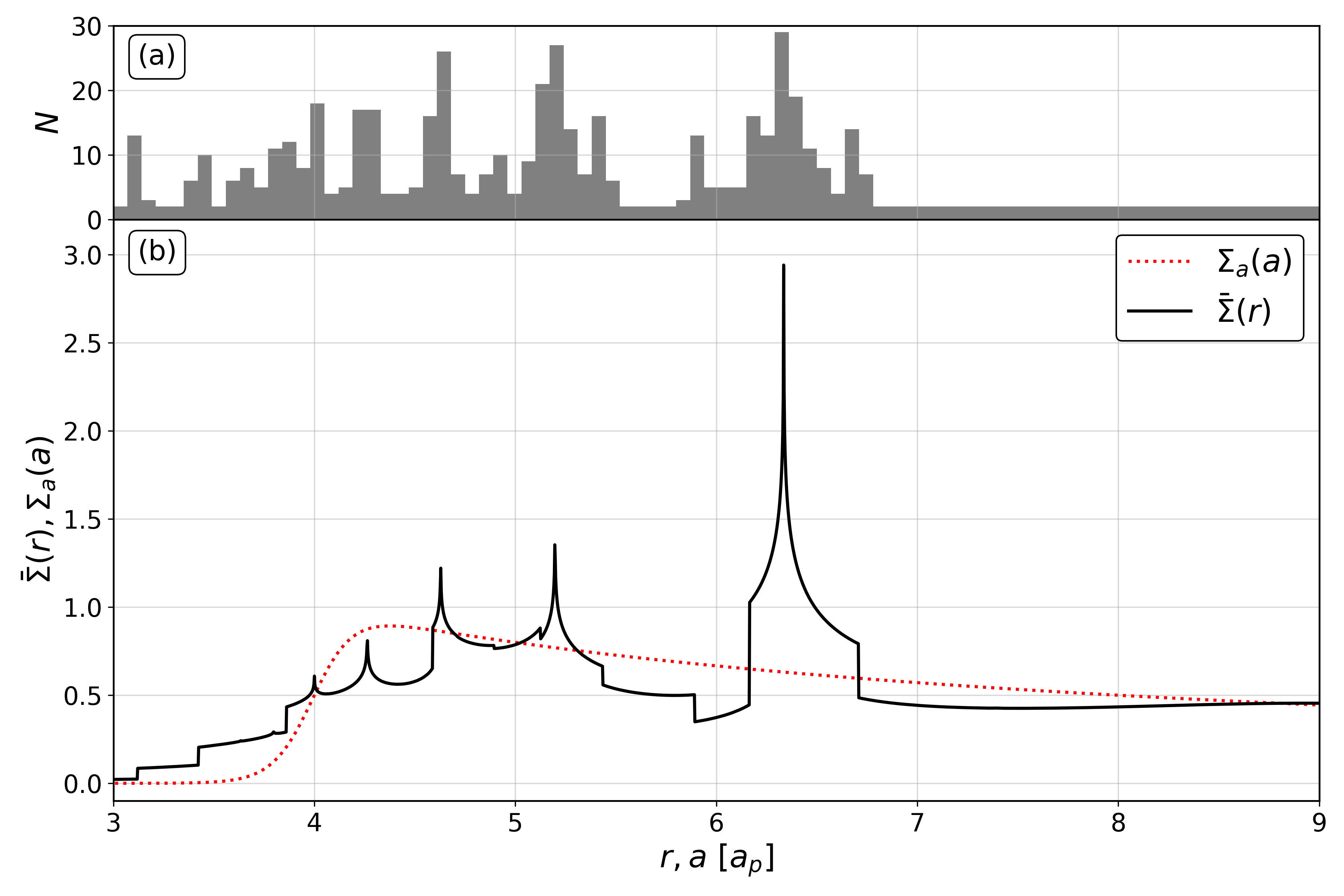}
	\caption{
    Figure illustrating the use of an adaptive grid in the ASD calculation. Panel (a) shows the density of grid points at which ASD shown in panel (b) is computed. One can see that our adaptive algorithm increases the density of grid points N (number of grid points per fixed radial interval) in regions where the ASD exhibits sharp features: peaks and discontinuous jumps. 
}
	\label{fig:mesh}
	\end{center}
\end{figure}

After an initial low-resolution profile has been evaluated, the local curvature of $\bar{\Sigma}(r)$ is estimated using finite differences. Regions exceeding a user-defined curvature threshold are refined by inserting additional radial points and recomputing the ASD at those locations. This procedure can be applied iteratively, concentrating resolution near sharp features while retaining coarser sampling in smooth regions. An example of this procedure is shown in Fig.~\ref{fig:mesh}, where panel (a) shows the resulting concentration of radial grid points, while panel (b) shows the corresponding refined ASD profile.

%%%%%%%%%%%%%%%%%%%%%%%%%%%%%%%%%%%%%%%%%%%%%%%%%%%%%%%%%%%
%%%%%%%%%%%%%%%%%%%%%%%%%%%%%%%%%%%%%%%%%%%%%%%%%%%%%%%%%%%

%%%%%%%%%%%%%%%%%%%%%%%%%%%%%%%%%%%%%%%%%%%%%%%%%%%%%%%%%%%

\subsection{Additional features}

In addition to the semi-analytic ASD calculation, {\sc debrispy} includes a Monte--Carlo sampling module. Semi-major axes are sampled from the input mass distribution using rejection sampling, with the appropriate Jacobian factor included when constructing surface-density profiles. For models with a unique eccentricity profile, eccentricities are evaluated directly as $e(a)$. For models with a full eccentricity distribution, eccentricities are sampled from $\psi_e(e,a)$ using rejection sampling at fixed semi-major axis. Orbital phases are sampled uniformly in mean anomaly, and Kepler's equation is then solved to obtain the corresponding particle positions, which are used to construct one-dimensional radial histograms of ASD or two-dimensional realisations of $\Sigma(r,\phi)$ in either Cartesian or polar coordinates.

The package also includes convolution tools for comparing high-resolution theoretical profiles with observationally resolved data. One-dimensional ASD profiles can be convolved with a Gaussian kernel, while two-dimensional Cartesian maps can be convolved with a Gaussian point-spread function, including elliptical and rotated beams.

%%%%%%%%%%%%%%%%%%%%%%%%%%%%%%%%%%%%%%%%
%%%%%%%%%%%%%%%%%%%%%%%%%%%%%%%%%%%%%%%%%

\section{\protect $\Sigma_a(a)$ profiles}
\label{sec:siga}

%%%%%%%%%%%%%%%%%%%%%%%%%%%%%%%%%%%%%%%%%

Throughout this work we mainly use the following $\Sigma_a(a)$ profile:
\begin{align}
\Sigma_a(a) &=\Sigma_0\left(\frac{a_\mathrm{in}}{a}\right)^{-\gamma}f_\mathrm{in}f_\mathrm{out},
\label{eq:siga}\\
f_\mathrm{in}&=\frac{1}{2}\left(1+\tanh\frac{a-a_\mathrm{in}}{w_\mathrm{in}}\right),
\label{eq:fin}\\
f_\mathrm{out} &= \frac{1}{2}\left(1+\tanh\frac{a_\mathrm{out}-a}{w_\mathrm{out}}\right).
\label{eq:fout}
\end{align}
We adopt the following parameter values: $\Sigma_0=1$, $\gamma=1$, $a_\mathrm{in}=4\ap$, $a_\mathrm{out}=15\ap$, $w_\mathrm{in}=w_\mathrm{out}=0.2\ap$. For a double Gaussian profile in Section \ref{sec:complex-discs} we use 
\begin{align}
\Sigma_a(a) = \sum\limits_{i=1,2}\Sigma_i\exp\left[-\frac{(a-a_i)^2}{2\sigma_i^2}\right]~~~\mathrm{if}~~~4\ap<a<9\ap, 
\label{eq:2G}
\end{align}
and zero otherwise, with $\Sigma_1=\Sigma_2=1$, $a_1=4\ap$, $a_2=8\ap$, $\sigma_1=0.7\ap$, $\sigma_2=0.8\ap$. For a narrow Gaussian ring in Section \ref{sec:complex-discs} we use
\begin{align}
\Sigma_a(a) = \exp\left[-\frac{(a-a_r)^2}{2\sigma_\Sigma^2}\right],
\label{eq:ring}
\end{align}
with $a_r=4\ap$ and varying $\sigma_\Sigma$.

%%%%%%%%%%%%%%%%%%%%%%%%%%%%%%%%%%%%%%%%
%%%%%%%%%%%%%%%%%%%%%%%%%%%%%%%%%%%%%%%%%

\section{Mathematical details on the origin and properties of ASD features}
\label{sec:features_math}

%%%%%%%%%%%%%%%%%%%%%%%%%%%%%%%%%%%%%%%%

%%%%%%%%%%%%%%%%%%%%%%%%%%%%%%%%%%%%%%%%%

\subsection{ASD features: peaks at nulls of $e(a)$}
\label{sec:peaks}

%%%%%%%%%%%%%%%%%%%%%%%%%%%%%%%%%%%%%%%%

As shown in Section \ref{sec:ASD-features-peaks}, for the secular eccentricity profile (\ref{eq:e_fr-fr=1}) the peaks of ASD can emerge at radii $r=a_k$, where $a_k$ is the semi-major axis of the $k$-th eccentricity null (i.e. $e(a_k,t)=0$) and is given by equation (\ref{eq:a_k}). We are interested in understanding the local asymptotic behavior of ASD as $r\to a_k$. To accomplish this, we need to know the local behavior of $\kappa(r,a)$ and $e(a)$ in the vicinity of $r=a_k$ as they determine the actual form of an ASD feature at this point, see equation (\ref{eq:asd-rel-gen}). 

For $r$, $a$ both close to $a_k$ we can approximate
\ba
\kappa(r,a)\approx \frac{|a-r|}{a_k}.
\label{eq:kappa_loc}
\ea
Also, using the expression (\ref{eq:e_fr-fr=1}) for $e(a,t)$ and expanding $\Ap$ in the vicinity of $a_k$ where the condition (\ref{eq:null_cond}) holds, we can approximate
\ba
e(a,t)\approx \ef(a_k) t\left|\frac{\partial \Ap(a_k)}{\partial a}(a-a_k)\right|=\zeta_k\frac{|a-a_k|}{a_k}.
\label{eq:eapprox}
\ea 
Here the coefficient $\zeta_k$ is defined as (also see equation (\ref{eq:zt_k})) 
\ba
\zeta_k = \ef(a_k)  a_k\left|\frac{\partial \Ap(a_k)}{\partial a}\right|t \approx \frac{7}{2}\ef(a_k) \Ap(a_k) t,
\label{eq:zeta}
\ea 
where we used equation (\ref{eq:Ap}). With equation (\ref{eq:e_forced}) and condition (\ref{eq:null_cond}) in mind, this expression for $\zeta_k$ can be reduced to equation (\ref{eq:zeta1}). The approximation (\ref{eq:eapprox}) requires
\ba   
\frac{t}{2}\left|\frac{\partial \Ap(a_k)}{\partial a}(a-a_k)\right|\lesssim \frac{\pi}{2}~~~~\mbox{or}~~~~
|a-a_k|\lesssim \frac{a_k}{7k},
\label{eq:loc_cond}
\ea  
where we again used equations (\ref{eq:Ap}) and (\ref{eq:null_cond}); this condition is always fulfilled.

%%%%%%%%%%%%%%%%%%%%%%%%%%%%%%%%%%%%%%%%%%%%%%%%%%%%%%%%%%%
\begin{figure} 
	\begin{center}
	\includegraphics[width=0.49\textwidth]{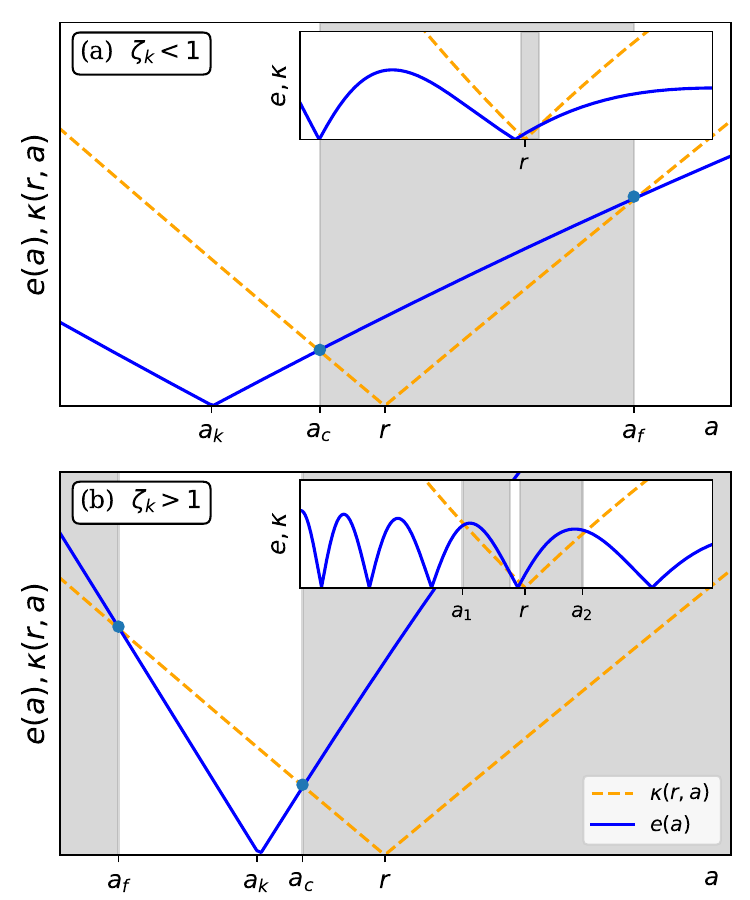}
	\caption{
    Illustration of different arrangements of $e(a)$ (solid orange) and $\kappa(r,a)$ (dashed blue) curves in a local vicinity of a null $a_k$ of $e(a)$, in the asymptotic limit $r\to a_k$. Two cases are shown: (a) $\zeta_k<1$ and (b) $\zeta_k>1$, see equations (\ref{eq:zeta}), (\ref{eq:zeta1}). Insets in each panel illustrate $e$ and $\kappa$ curves on a larger scale. Relative positions of various points ($r, a_c, a_f$, etc.) defined in the text are indicated. Grey regions correspond to semi-major axis intervals providing non-zero contribution to ASD, where $e(a)>\kappa(r,a)$. See Appendix \ref{sec:peaks} for details.
    }
	\label{fig:nulls_ill}
	\end{center}
\end{figure}
%%%%%%%%%%%%%%%%%%%%%%%%%%%%%%%%%%%%%%%%%%%%%%%%%%%%%%%%%%%

Equations (\ref{eq:kappa_loc}) and (\ref{eq:eapprox}) reveal the significance of $\zeta_k$: for $\zeta_k>1$ eccentricity increases with $|a-a_k|$ faster than the function $\kappa$ increases with $|a-r|$, and vice versa. As we show below, this dichotomy results in different outcomes of ASD calculation in the vicinity of $a_k$. 

Next we plug the local behaviors (\ref{eq:kappa_loc}) and (\ref{eq:eapprox}) into equation (\ref{eq:asd-rel-gen}). We also assume that the integral in the expression for ASD is determined locally, by the close vicinity of $a_k$, so that $a^{-1}\Sigma_a(a)$ can be evaluated at $a_k$ and taken outside the integral. As a result of this and some trivial algebraic manipulations, we find 
\ba  
\asd(r) \approx \frac{\Sigma_a(a_k)}{\pi}
\int_{r/2}^\infty \frac{\theta(r,a)~\md a}{\sqrt{(1-\zeta_k^2)(a_f-a)(a-a_c)}},
\label{eq:asd1}
\ea  
where
\ba
a_f=\frac{r-\zeta_k a_k}{1-\zeta_k},~~~~~
a_c=\frac{r+\zeta_k a_k}{1+\zeta_k},
\label{eq:afc}
\ea  
and $\theta=1$ when the expression inside the root is positive, and zero otherwise, setting the integration range. For future reference, it is useful to note that
\ba
a_f-a_c=\frac{2\zeta_k(r-a_k)}{1-\zeta_k^2},~~~
\frac{a_f+a_c}{2}=a_k+\frac{r-a_k}{1-\zeta_k^2},
\label{eq:afc1}
\ea  
so that $a_f,a_c\to a_k$ as $r\to a_k$.

We will now consider the cases $\zeta_k>1$ and $\zeta_k<1$ separately. To be specific, we will assume $r>a_k$, but the case of $r<a_k$ is treated analogously.

%%%%%%%%%%%%%%%%%%%%%%%%%%%%%%%%%%%%%%%%%

\subsubsection{Case of $\zeta_k>1$}
\label{sec:zg1}

In this case the relative arrangement of $e(a)$ and $\kappa(r,a)$ looks as shown in Figure \ref{fig:nulls_ill}b. One can see that when $\zeta_k>1$ these curves cross at four points, $a_1$, $a_f$, $a_c$, and $a_2$, arranged as follows:
\ba  
a_1<a_f<a_k<a_c<r<a_2.
\label{eq:orderg1}
\ea  
As a result, the integration range in (\ref{eq:asd1}) splits into two intervals: $(a_1,a_f)$ and $(a_c,a_2)$. It is important to note that crossings at $a_{1,2}$ would not arise if $e(a)$ was given by the local expression (\ref{eq:eapprox}). They emerge only when one considers the full eccentricity behavior (\ref{eq:e_fr-fr=1}), see inset, and the approximation (\ref{eq:eapprox}) is technically not applicable in their vicinity. 

\begin{figure} 
	\begin{center}
	\includegraphics[width=0.49\textwidth]{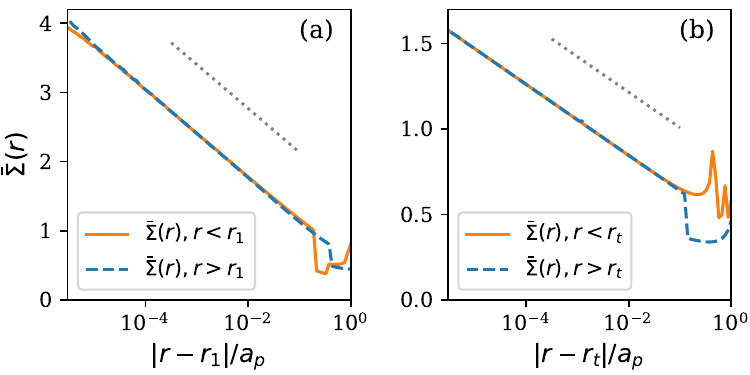}
	\caption{
    Illustration of ASD behavior near its peaks. Solid orange and dashed blue curves show $\asd(r)$ inside and outside the peak location, as a function of the distance from the peak. (a) Divergence near the $k=1$ null of $e(a)$ at $r_1\approx 5.944\ap$ in a calculation with $\ep=0.4$, $\ffree=1$ and $t=4\tsec$ shown in Figure \ref{fig:ep_var}f. See Appendix \ref{sec:zg1} for details. (b) Peak at the caustic point (with $\chi>0$) $r_t\approx 7.336\ap$ (with a corresponding tangent point $a_t\approx 7.057 \ap$) in a calculation with $\ep=0.4$, $\ffree=0.5$ and $t=15\tsec$ shown in Figure \ref{fig:ffree_var}b. See Appendix \ref{sec:causticsg0} for details. Dotted black lines show theoretically expected slopes based on equations (\ref{eq:asd4}) and (\ref{eq:cau2}) in panels (a) and (b), respectively. It is clear that $\asd$ diverges logarithmically with the distance from the peak.
    }
	\label{fig:log}
	\end{center}
\end{figure}

Nevertheless, when computing $\asd$ we can still use equation (\ref{eq:asd1}) since the leading-order (in $r-a_k$) contribution to ASD comes from around $a_c$ and $a_f$, so the actual behavior of $e(a)$ near $a_{1,2}$ is not important. With this in mind, for $\zeta_k>1$, equation (\ref{eq:asd1}) becomes
\ba    
\asd(r) \approx \frac{\Sigma_a(a_k)}{\pi\sqrt{\zeta_k^2-1}}
\left(\int_{a_1}^{a_f}+\int_{a_c}^{a_2}\right) \frac{\md a}{(a-a_f)(a-a_c)}.
\label{eq:asd2}
\ea 
Taking this integral, working in the limit $r\to a_k$ (so that $a_c,a_f\to a_k$), and using relations (\ref{eq:afc1}), one finds
\begin{align}    
\asd(r) &\approx  \frac{\Sigma_a(a_k)}{\pi\sqrt{\zeta_k^2-1}}
\nonumber\\
& \times \left(\ln\left|16(a_1-a_k)(a_2-a_k)\right|
-2\ln\left|a_f-a_c\right|\right).
\label{eq:asd3}
\end{align} 
In the limit $r\to a_k$ the integration end points $a_{1,2}$ become independent of $|r-a_k|$. Using this and equation (\ref{eq:afc1}) one obtains
\begin{align}    
\asd(r) \approx  \frac{\Sigma_a(a_k)}{\pi\sqrt{\zeta_k^2-1}}
\left[C(a_1,a_2,\zeta_k)-2\ln\left|r-a_k\right|\right]
\label{eq:asd4}
\end{align} 
as $r\to a_k$, where $C$ is a constant. Thus, $\asd(r)$ exhibits a weak (logarithmic) divergence as $r$ approaches the null $a_k$ of $e(a)$ in the form (\ref{eq:e_fr-fr=1}). We verify this expectation in Figure \ref{fig:log}a, where we show the $\asd(r)$ profile near its peak in log-linear coordinates. One can see that both the logarithmic behavior and the slope of the logarithmic term in equation (\ref{eq:asd4}) match the {\sc debrispy} results. 

%%%%%%%%%%%%%%%%%%%%%%%%%%%%%%%%%%%%%%%%%%%%%%%%%%%%%%%%%%%
%%%%%%%%%%%%%%%%%%%%%%%%%%%%%%%%%%%%%%%%%%%%%%%%%%%%%%%%%%%

%%%%%%%%%%%%%%%%%%%%%%%%%%%%%%%%%%%%%%%%%

\subsubsection{Case of $\zeta_k<1$}
\label{sec:zl1}

When $\zeta_k<1$, there are only two crossings of $e(a)$ and $\kappa(r,a)$, at $a=a_c,a_f$, see Figure \ref{fig:nulls_ill}a for an illustration. They are arranged as
\ba  
a_k<a_c<r<a_f,
\label{eq:orderl1}
\ea  
see equations (\ref{eq:afc}). The integration range in equation (\ref{eq:asd1}) is $(a_c,a_f)$, and ASD is given by
\ba  
\asd(r) \approx \frac{\Sigma_a(a_k)}{\pi\sqrt{1-\zeta_k^2}}
\int_{a_c}^{a_f} \frac{\md a}{(a_f-a)(a-a_c)}.
\label{eq:asd5}
\ea  
The integral in this expression evaluates to $\pi$ regardless of the values of $a_c$, $a_f$, so that
\ba  
\asd(r) \approx \frac{\Sigma_a(a_k)}{\sqrt{1-\zeta_k^2}}~~~~\mbox{as}~~~~r\to a_k.
\label{eq:asd6}
\ea  
One can see that, unlike the case of $\zeta_k>1$, ASD now features a non-singular behavior at the null of $e(a)$. Moreover, $\asd(a_k)$ is not very different from $\Sigma_a(a_k)$ unless $\zeta_k$ is close to unity. 

\begin{figure} 
	\begin{center}
	\includegraphics[width=0.49\textwidth]{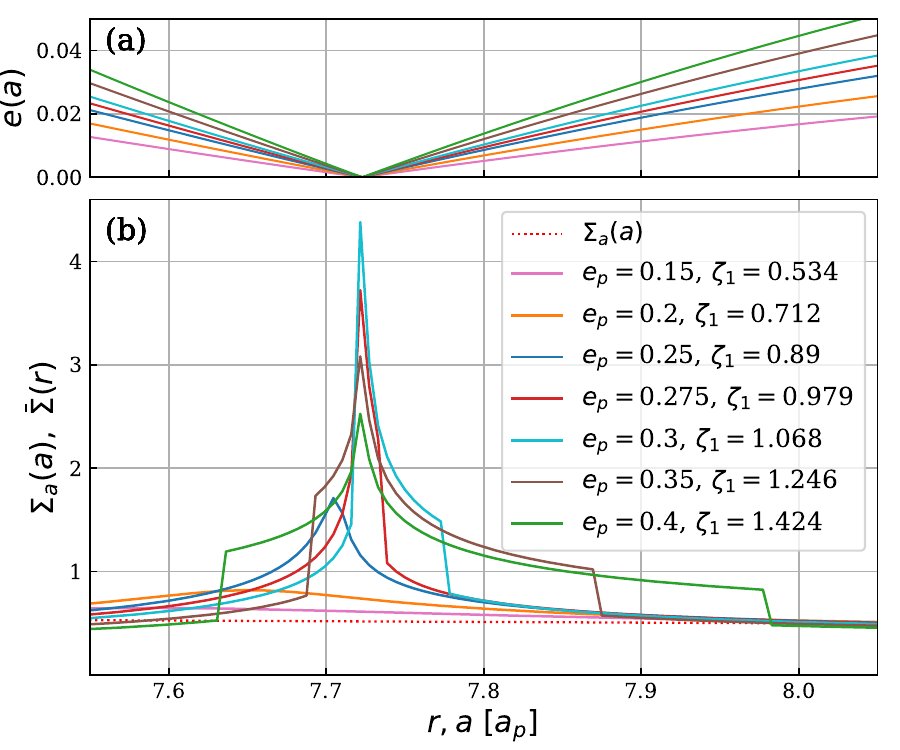}
	\caption{
    Evolution of an ASD peak at the eccentricity null as the parameter $\zeta_k$ is varied by changing $e_p$ (shown with curves of different color, labeled in the legend in panel (b)). Panels shows (a) $e(a)$ and (b) $\asd(r)$, $\Sigma_a(a)$. Shown is the local vicinity of $k=1$ eccentricity null at $t=10\tsec$ for $\ffree=1$. This null, visible in Figures \ref{fig:time_ev}g, \ref{fig:ep_var}g,h,i, is located at $r=a_1\approx 7.723\ap$. Note the characteristic changes of the peak height and shape as $\zeta_1$ varies across $\zeta_1=1$. See text for details.
    }
	\label{fig:peaks_zeta}
	\end{center}
\end{figure}

We illustrate the qualitative change in the appearance of ASD near a peak at $r=a_k$ as $\zeta_k$ varies crossing unity in Figure \ref{fig:peaks_zeta}, focusing on a $k=1$ null of $e(a)$. Variation of $\zeta_1$ at a fixed peak location (i.e. semi-major axis $a_1$ of the eccentricity null) is enabled by varying planetary eccentricity, as per equation (\ref{eq:zeta1}). One can see that for low $\ep$, when $\zeta_1<1$, there is only a mild enhancement of $\asd$ around $a_1\approx 7.723\ap$. As $\ep$ increases towards unity, this enhancement increases in amplitude and narrows in radial extent, until, around $\zeta_1=1$ (which happens at $\ep$ slightly below 0.3), a sharp, narrow peak appears at $a_k$. As $\zeta_1$ increases further, the apparent height of the peak goes down (it still remains a singularity), while the peak broadens radially (the width can be judged from the locations of two $\asd$ jumps on each side of the peak, creating a `pedestal' for this feature). Overall, this progression fully confirms our understanding of $\asd$ behavior near the peak. Further examples illustrating these theoretical predictions are highlighted when discussing Figures \ref{fig:time_ev} \& \ref{fig:ep_var}.

%%%%%%%%%%%%%%%%%%%%%%%%%%%%%%%%%%%%%%%%%%%%%%%%%%%%%%%%%%%
%%%%%%%%%%%%%%%%%%%%%%%%%%%%%%%%%%%%%%%%%%%%%%%%%%%%%%%%%%%

%%%%%%%%%%%%%%%%%%%%%%%%%%%%%%%%%%%%%%%%
%%%%%%%%%%%%%%%%%%%%%%%%%%%%%%%%%%%%%%%%%

\subsection{ASD features: caustics at tangent points}
\label{sec:caustics}

%%%%%%%%%%%%%%%%%%%%%%%%%%%%%%%%%%%%%%%%

Another notable type of features exhibited by the ASD are 
caustics arising when $e(a)$ and $\kappa(r,a)$ are tangent to each other. Depending on the situation, they result in either the discontinuous but finite jumps of $\asd$ (see Section \ref{sec:ASD-features-jumps}) or in weak singularities (see Section \ref{sec:free-ec}). 

%%%%%%%%%%%%%%%%%%%%%%%%%%%%%%%%%%%%%%%%%%%%%%%%%%%%%%%%%%%
\begin{figure} 
	\begin{center}
	\includegraphics[width=0.49\textwidth]{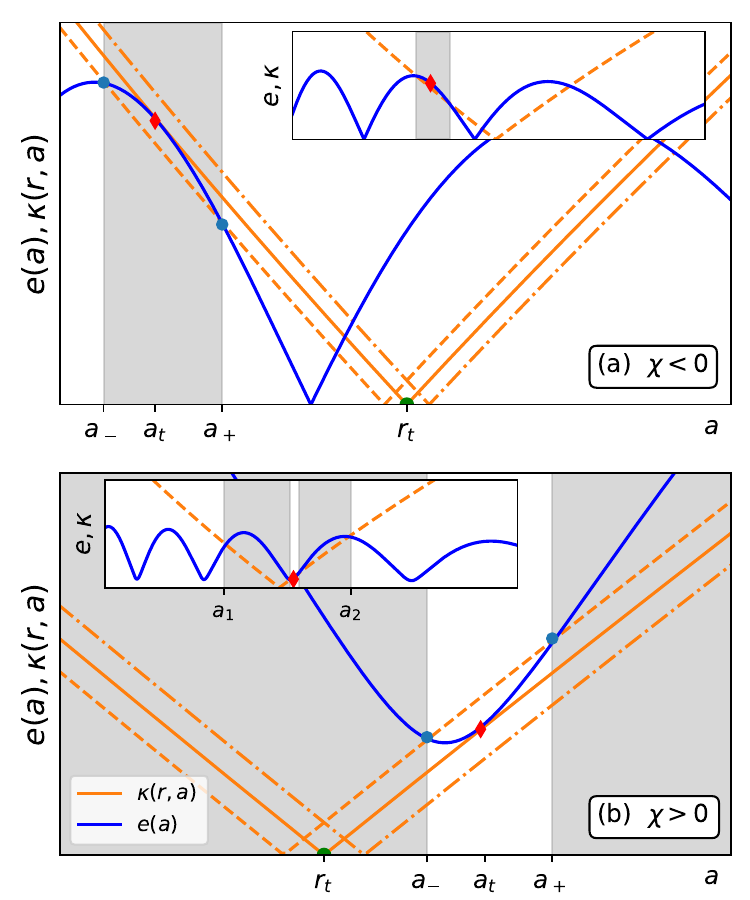}
	\caption{
    Schematic illustration of the ASD calculation near caustic points, as $r\to r_t$, for (a) $\chi<0$ and (b) $\chi>0$, with $\chi$ defined by equation (\ref{eq:chi}). Main panels show the local vicinity of a tangent point $a=a_t$ (marked with a red diamond), where the eccentricity curve $e(a)$ (solid orange) and $\kappa(r_t,a)$ curve (solid blue) are tangent; insets illustrate the geometry on a more global scale. Dashed ornge curve shows $\kappa(r,a)$ for $r<r_t$ ($\delta r<0$), while dot-dashed orange curve shows $\kappa(r,a)$ for $r>r_t$ ($\delta r>0$). Shaded regions show the integration regions in the ASD calculation for $\delta r<0$ (dashed orange curve); crossings of $e(a)$ and $\kappa(r,a)$ in this case are shown by blue dots. See Appendix \ref{sec:caustics} for details.}
	\label{fig:cau_ill}
	\end{center}
\end{figure}
%%%%%%%%%%%%%%%%%%%%%%%%%%%%%%%%%%%%%%%%%%%%%%%%%%%%%%%%%%%

We want to understand the behavior of $\asd$ near such caustic points, and we now carry out the local asymptotic analysis of the ASD behavior as $r\to r_t$. This means considering different situations illustrated in Figure \ref{fig:cau_ill} and evaluating the integral in the ASD definition (\ref{eq:asd-rel-gen}) in the vicinity of $a_t$. To this goal we first manipulate the expression for $e^2-\kappa^2$ entering the equation (\ref{eq:asd-rel-gen}), when $r\to r_t$ and $a\to a_t$. In this limit, keeping in mind the condition (\ref{eq:jump_cond}) and carrying out Taylor expansion in $a-a_t$, we find
\begin{align}
& e^2(a)-\kappa^2(r,a)\approx 2e(a_t)[e(a)-\kappa(r,a)]
\nonumber\\
& \approx 
2e(a_t)\Big[e(a_t)-\kappa(r,a_t)
+\left(e^\prime(a_t)-\kappa^\prime(r,a_t)\right)(a-a_t)
\nonumber\\
&+\frac{e^{\prime\prime}(a_t)-\kappa^{\prime\prime}(r,a_t)}{2}(a-a_t)^2\Big],
\label{eq:rad}
\end{align}  
where primes denote differentiation with respect to $a$. We next perform an expansion in $\delta r=r-r_t$, which needs to be done only in terms involving $\kappa$ and its derivatives. One can easily see that, according to the definition (\ref{eq:kap}),
\ba
\kappa(r,a)=s_t\frac{\delta r}{a}+\kappa(r_t,a), ~~~~s_t=\mathrm{sgn}(r_t-a).
\ea  
Plugging this expression into (\ref{eq:rad}), evaluating at $a=a_t$, and using the conditions (\ref{eq:jump_cond}), we obtain
\begin{align}
e^2(a)-\kappa^2(r,a)\approx 
2e(a_t)\Big[& -s_t\frac{\delta r}{a_t}+s_t\frac{\delta r}{a_t^2}(a-a_t)
\nonumber\\
&+\chi\frac{(a-a_t)^2}{a_t^2}\Big].
\label{eq:rad1}
\end{align}  
where $\chi$ is defined by equation (\ref{eq:chi}) and we approximated $\kappa^{\prime\prime}(r,a_t)\to \kappa^{\prime\prime}(r_t,a_t)$, which is sufficient for our purposes near the tangent radius $r_t$. Parameter $\chi$ is very important for deciding the behavior of ASD, as we will see later. Note that $\kappa^{\prime\prime}(r_t,a_t)=2 s_t r_t/a_t^3=2 s_t [1+s_t e(a_t)]/a_t^2$, i.e. looks differently depending on whether $r_t>a_t$ or vice versa. But in situations where $e(a)$ rapidly varies with $a$ (which is true for a secularly-driven $e(a)$ given by equation (\ref{eq:e_fr-fr=1}) at late times), the $\kappa^{\prime\prime}$ term is likely to be subdominant compared to $e^{\prime\prime}$.

Introducing a new variable $x=(a-a_t)/a_t$ and dropping the terms quadratic in $\delta r$ as needed, we re-write expression (\ref{eq:rad1}) as  
\begin{align}
e^2(a)-\kappa^2(r,a)\approx 
2e(a_t)\chi\left[\left(x+\frac{s_t\delta r}{2\chi a_t}\right)^2-\frac{s_t\delta r}{\chi a_t}\right].
\label{eq:rad2}
\end{align}  
This form reveals the significance of the sign of $s_t\delta r/\chi$. When $s_t\delta r/\chi>0$, there are two crossings of $e(a)$ and $\kappa(r,a)$ in the vicinity of $a_t$ (i.e. near $x=0$), at
\ba
x_{\pm}=-\frac{s_t\delta r}{2\chi a_t}\pm\sqrt{\frac{s_t\delta r}{\chi a_t}}\approx \pm\sqrt{\frac{s_t\delta r}{\chi a_t}}~~~\mathrm{as}~~\delta r\to 0,
\label{eq:xpm}
\ea  
or at $a=a_\pm=a_t(1+x_\pm)$. But if $s_t\delta r/\chi<0$, then $e$ and $\kappa$ curves do not cross at all near $a_t$.

We will use the expression (\ref{eq:rad2}) in equation (\ref{eq:asd-rel-gen}) to analyze the ASD behavior around the caustics as $r\to r_t$, i.e. as $\delta r\to 0$. There are two qualitatively different outcomes depending on whether $\chi<0$ (Section \ref{sec:causticsl0}) or  $\chi>0$ (Section \ref{sec:causticsg0}); we consider both cases below.

%%%%%%%%%%%%%%%%%%%%%%%%%%%%%%%%%%%%%%%%%

\subsubsection{Caustics for $\chi<0$: discontinuous but finite jumps}
\label{sec:causticsl0}

The case of $\chi<0$, which is very typical for $e(a)$ in the form (\ref{eq:e_fr-fr=1}), is illustrated in Figure \ref{fig:cau_ill}a. There we depict a situation when $a_t<r_t$, so that $s_t=1$; the case $a_t>r_t$ is discussed later. 

If we look at the $\kappa(r,a)$ curve for $r<r_t$, i.e. when $\delta r<0$, one can see that it has two crossings with $e(a)$, at $a_\pm$, consistent with the equation (\ref{eq:xpm}) since $s_t\delta r/\chi>0$. The interval $a\in (a_-,a_+)$ provides a non-zero contribution to $\asd$ since $e>\kappa$ there. We can easily calculate this contribution by changing integration variable in (\ref{eq:asd-rel-gen}) from $a$ to $x$ and using equation (\ref{eq:rad2}) to write
\ba  
\asd(r) \approx \frac{\Sigma_a(a_t)}{\pi\sqrt{2 e(a_t) |\chi|}}
\int_{x_-}^{x_+} \frac{\md x}{(x_+ - x)(x-x_-)}.
\label{eq:dSig1}
\ea  
The integral in this expression is equal to $\pi$, see equations (\ref{eq:asd5}), (\ref{eq:asd6}), and the whole contribution is independent of $\delta r$.

As $r\to r_t$, $\delta r$ shrinks and $a_-\to a_+$. When $r$ passes through $r_t$, $\delta r$ changes sign, and $e(a)$ and $\kappa(r,a)$ no longer cross each other (as they switch into $s_t\delta r/\chi<0$ regime), the contribution (\ref{eq:dSig1}) suddenly vanishes. As a result, at $r=r_t$ ASD exhibits a sudden downward jump (as $r$ increases) by
\ba  
\Delta\asd(r) \approx -\frac{\Sigma_a(a_t)}{\sqrt{2 e(a_t) |\chi|}}.
\label{eq:dSig2}
\ea  

If we were to consider a situation when $a_t>r_t$ (so that $s_t=-1$), we would find a similar situation, but reversed: an extra contribution to ASD would emerge as $r$ crossed $r_t$ from below, implying a sudden upward jump of ASD at $r_t$. The calculation of this contribution is the same as in (\ref{eq:dSig2}) with the same amplitude of the jump as in (\ref{eq:dSig2}). Thus, a general expression for the ASD jump at $r_t$ is 
\ba  
\Delta\asd(r) \approx -\frac{s_t\, \Sigma_a(a_t)}{a_t\sqrt{e(a_t) \left(\kappa^{\prime\prime}(r_t,a_t)-e^{\prime\prime}(a_t)\right)}},
\label{eq:dSig3}
\ea  
where we also used the definition (\ref{eq:chi}). 

This calculation explains the sharp but finite discontinuities of $\asd$ emerging at tangent points $r_t$ in all our calculations since $\chi<0$ is very common for secularly-induced $e(a)$.

%%%%%%%%%%%%%%%%%%%%%%%%%%%%%%%%%%%%%%%%%

\subsubsection{Caustics for $\chi>0$: singularities}
\label{sec:causticsg0}

Under certain circumstances, e.g., when the free eccentricity $\efree$ is not equal to $\ef$ (see Section \ref{sec:free-ec}), the possibility of $\chi>0$ may also arise. This situation is schematically depicted in Figure \ref{fig:cau_ill}b, where for illustration we have chosen the case of $r_t<a_t$ (and $s_t=-1$) for the $\kappa$ curve tangent to the eccentricity profile. The analysis of $\asd$ behavior near the caustic at $r_t$ is in some ways similar to the calculations in Section \ref{sec:zg1}. 

Starting at $r<r_t$, when $\delta r<0$, one can see that $s_t\delta r/\chi>0$, so that $\kappa$ and $e$ curves cross at two points given by equation (\ref{eq:xpm}). Unlike the case studied in Section \ref{sec:causticsl0}, now $e>\kappa$ for $x<x_-$ and $x>x_+$. In practice, without loss of generality, we should also limit $x$ from below and above by some distant points $x_1<0$ and $x_2>0$, $|x_{1,2}|\sim 1$, as we did in Section \ref{sec:zg1}. While the behavior of $e^2-\kappa^2$ will not obey the local approximation (\ref{eq:rad2}), this will not be important for the outcome, as we see next. 

With this in mind, using (\ref{eq:rad2}) and changing the integration variable from $a$ to $x$, equation (\ref{eq:asd-rel-gen}) becomes
\ba  
\asd(r) \approx \frac{\Sigma_a(a_t)}{\pi\sqrt{2 e(a_t) \chi}}\left(\int_{x_1}^{x_-}+\int_{x_+}^{x_2}\right) \frac{\md x}{(x-x_-)(x-x_+)}.
\label{eq:cau}
\ea  
We can evaluate this integral similarly to equations (\ref{eq:asd2}), (\ref{eq:asd3}), finding in the limit of $\delta r\to 0$ (when $x_\pm\to 0$)
\begin{align}    
\asd(r) \approx  \frac{\Sigma_a(a_t)}{\pi\sqrt{2 e(a_t) \chi}}
\left(\ln\left|16x_1x_2\right|
-2\ln\left|x_+-x_-\right|\right).
\label{eq:cau1}
\end{align} 
Using equation (\ref{eq:xpm}), we finally obtain the asymptotic behavior of ASD near $r_t$ (for $r<r_t$) in the form
\begin{align}    
\asd(r) \approx  \frac{\Sigma_a(a_t)}{\pi\sqrt{2 e(a_t) \chi}}
\left[C(x_1,x_2,a_t,\chi)
-\ln\left|r-r_t\right|\right],
\label{eq:cau2}
\end{align} 
where $C$ is a constant. 

Once $r$ exceeds $r_t$, $\delta r>0$ and $s_t\delta r/\chi$ becomes negative, meaning that there are no intersections of $e$ and $\kappa$ curves near $a_t$. Despite that, there is still a contribution to ASD from the vicinity of $a_t$ since $e>\kappa$ there. Upon the variable change $a\to x$, the integration range in (\ref{eq:asd-rel-gen}) is $(x_1,x_2)$, and, plugging equation (\ref{eq:rad2}) into (\ref{eq:asd-rel-gen}) and integrating, one arrives at the same expression (\ref{eq:cau2}) also for $r>r_t$.

Thus, for $\chi>0$ the ASD shows a logarithmic divergence in the form (\ref{eq:cau2}) at the caustic point $r_t$. This behavior is illustrated in Figure \ref{fig:log}b for one such $\chi>0$ caustic, clearly confirming the analytical picture presented here.

%%%%%%%%%%%%%%%%%%%%%%%%%%%%%%%%%%%%%%%%%=1

\subsection{Peaks and their pedestals}
\label{sec:peaks_pedestals}

%%%%%%%%%%%%%%%%%%%%%%%%%%%%%%%%%%%%%%%%

In many ASD plots shown in this study one can see that in the case of $\ffree=1$ the $\asd$ peaks at $e$-nulls often have an underlying `pedestal' structure, formed by ASD jumps on each side of the peak. Figure \ref{fig:features_ill} shows that these bounding jumps correspond to the tangent points $a_t$ closest to a particular peak location $a_k$, sitting on the two nearest $e(a)$ cycles, see the discussion in Section \ref{sec:ASD-features}.  Given the ubiquity of these pedestals, we explore some of their properties below.

%%%%%%%%%%%%%%%%%%%%%%%%%%%%%%%%%%%%%%%%%%%%%%%%%%%%%%%%%%%
\begin{figure} 
	\begin{center}
	\includegraphics[width=0.49\textwidth]{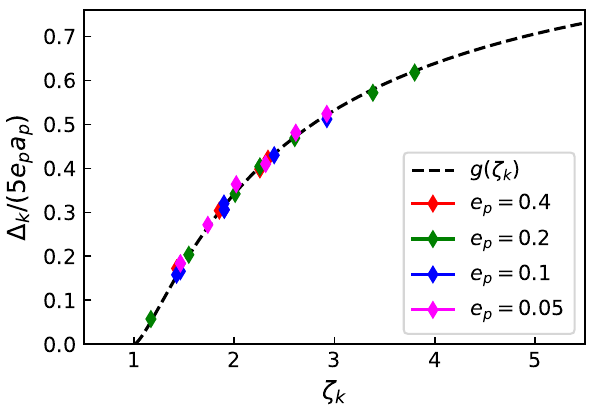}
	\caption{
    Peak widths $\Delta_k$ plotted for several values of $\ep$, measured at different moments of time $t$ and null order $k$. The dashed curve shows the analytical prediction for $\Delta_k$ given by equation (\ref{eq:Delta_k}).}
	\label{fig:width_zeta}
	\end{center}
\end{figure}
%%%%%%%%%%%%%%%%%%%%%%%%%%%%%%%%%%%%%%%%%%%%%%%%%%%%%%%%%%%

First, we estimate the radial width of a pedestal for a peak at the $e$-null of order $k$ by finding the locations of its bounding jumps. To find $a_t$ of the corresponding tangent points we first plug $e(a)$ in the form (\ref{eq:e_forced}), (\ref{eq:e_fr-fr=1}) into equation (\ref{eq:tan_cond}): 
\begin{align}    
\frac{5}{4}\ap\ep t\left|\frac{\partial \Ap}{\partial a}\right|_{a_t}
\cos\frac{\Ap(a_t)t}{2}=\pm 1\,.
\label{eq:step1}
\end{align} 
Working in the vicinity of $a_k$, we can approximate $|\partial \Ap/\partial a|_{a_t}\approx (7/2)\Ap(a_k)/a_k$ (see equation \ref{eq:Ap}) and then use the condition (\ref{eq:null_cond}) and definition (\ref{eq:zeta1}) to obtain
\begin{align}    
\cos\frac{\Ap(a_t)t}{2}=\pm \zeta_k^{-1}\,.
\label{eq:step2}
\end{align} 
Taylor expanding the argument of cosine near $a_k$ and using the condition (\ref{eq:null_cond}) again, we find that the locations $a_t$ of the tangent points are given by 
\begin{align}    
a_t-a_k\approx \pm\frac{2}{7\pi k}a_k \,\mathrm{acos}\left(\zeta_k^{-1}\right)\,
\label{eq:step3}
\end{align} 
for the inner/outer jump. The separation of the jump from the peak is $r_t-a_k=a_t-a_k\mp a_t e(a_t)$. Evaluating $e(a_t)$ using equations (\ref{eq:e_forced}), (\ref{eq:e_fr-fr=1}), and (\ref{eq:step2}), and manipulating the result using definition (\ref{eq:zeta1}) we find 
\begin{align}    
r_t-a_k\approx \mp\frac{5}{2}\ap\ep \, g(\zeta_k)
\label{eq:step4}
\end{align} 
for the inner/outer ASD jump, where we defined a function
\begin{align}    
g(t)=\sqrt{1-t^{-2}}-t^{-1}\mathrm{acos}\left(t^{-1}\right)\,.
\label{eq:gt}
\end{align} 
The full width $\Delta_k$ of the pedestal of a $k$-th peak is the difference of $r_t$ for each jump, meaning that
\begin{align}    
\Delta_k \approx  5\ap\ep \, g(\zeta_k)\,.
\label{eq:Delta_k}
\end{align} 
Note that non-zero $\Delta_k$ requires $\zeta_k>1$ (so that $g(\zeta_k)>0$), which is also the condition for a singular peak to exist at $a_k$ in the first place, see Appendix \ref{sec:peaks}.

We illustrate the prediction (\ref{eq:Delta_k}) in Figure \ref{fig:width_zeta} with a dashed curve. We also show the data on pedestal widths (normalized by $5\ep\ap$ to remove the $\ep$ dependence) measured using our ASD profiles for several values of $\ep$ (diamonds), for different $k$ and at different times. One can see that $\Delta_k(\zeta_k)$ derived from ASD profiles agree with the analytical estimate (\ref{eq:Delta_k}) very well. 

We can use this result to understand some trends noticeable in peak width behavior in Figures \ref{fig:time_ev}, \ref{fig:ep_var}. In particular, for a given $k$ and fixed $\ep$, as time increases and $a_k$ grows, both $\zeta_k$ and $\Delta_k$ decrease. Similarly, at a fixed $t$, the peak width increases with $\ep$ for a fixed $k$, and also with $k$ for a fixed $\ep$ (as $k/a_k\propto k^{9/7}$ in $\zeta_k$). 

%%%%%%%%%%%%%%%%%%%%%%%%%%%%%%%%%%%%%%%%%=1

\subsection{Pedestal overlap}
\label{sec:pedestal_overlap}

%%%%%%%%%%%%%%%%%%%%%%%%%%%%%%%%%%%%%%%%

As the disc undergoes a prolonged secular evolution with $t\gg\tsec$, the number of $e$-nulls and corresponding ASD peaks increases, while the distance between the consecutive peaks decreases as $a_k-a_{k+1}\approx (2/7)a_k/k$, since $k$ grows with $t$ at fixed $a_k$, see equation (\ref{eq:a_k}). On the other hand, as we look at peaks with increasing $k$ (at fixed $t$), the width of a peak pedestal $\Delta_k$ increases. This means that at some $k=k_\mathrm{po}$ a pedestal of the $k$-th peaks would start overlapping with the pedestal of $k+1$ peak. In other words, for $k<k_\mathrm{po}$ the peaks stand isolated from each other, but for $k>k_\mathrm{po}$ they start overlapping. This makes the ASD profile look erratic as the ASD jumps of neighboring peaks start partly `canceling' each other and steadily driving $\asd(r)$ to $\Sigma_a(r)$ at high $k$. This pedestal overlap also leads to a transition in the behavior of the lower envelope of ASD noted in the end of Section \ref{sec:ep-dep} and Figure \ref{fig:ep_var}.

We can determine the condition for pedestal overlap to occur by setting $a_k-a_{k+1}\approx \Delta_k$, which, with the help of equations (\ref{eq:zeta1}) and (\ref{eq:Delta_k}), reduces to 
\begin{align}
\zeta_k g(\zeta_k)\approx \pi/2. 
\label{eq:po_cond}
\end{align}
In other words, there is a critical value of $\zeta_k=\zeta_\mathrm{po}$, at which the pedestal overlap starts. By solving transcendental equation (\ref{eq:po_cond}), we find this critical value to be 
\begin{align}    
\zeta_\mathrm{po}\approx 2.95\,.
\label{eq:zeta_po}
\end{align} 
Pedestal overlap occurs for peaks with $\zeta_k>\zeta_\mathrm{po}$, whereas for $\zeta_k<\zeta_\mathrm{po}$ ASD peaks remain isolated from each other. This may have important implications for observations, see Section \ref{sec:non-detect}.

%%%%%%%%%%%%%%%%%%%%%%%%%%%%%%%%%%%%%%%%%%%
%%%%%%%%%%%%%%%%%%%%%%%%%%%%%%%%%%%%%%%%%%%
%%%%%%%%%%%%%%%%%%%%%%%%%%%%%%%%%%%%%%%%%%%

% Don't change these lines
\bsp	% typesetting comment
\label{lastpage}

\end{document}